\documentclass[preprint2]{aastex}
\usepackage{natbib}

\usepackage{lineno}
\usepackage{array,multirow,makecell}
\usepackage{amsmath}
 \usepackage{color}
\usepackage{epsfig} 
  \usepackage{graphicx}
\usepackage[top=1.5cm,bottom=1.5cm, left=2.0cm, right=2.0cm]{geometry}
\usepackage{amssymb}

\slugcomment{Not to appear in Nonlearned J., 45.}

\shorttitle{Cloud modelling for brown dwarfs and young giant exoplanets}
\shortauthors{Charnay et al.}

\begin{document}


\title{A self-consistent cloud model for brown dwarfs and young giant exoplanets: comparison with photometric and spectroscopic observations}


\author{B. Charnay\altaffilmark{1}, B. B\'ezard \altaffilmark{1}, J.-L. Baudino\altaffilmark{2}, M. Bonnefoy\altaffilmark{3}, A. Boccaletti\altaffilmark{1} and R. Galicher \altaffilmark{1}}


\email{benjamin.charnay@obspm.fr}


\altaffiltext{1}{LESIA, Observatoire de Paris, PSL Research University, CNRS, Sorbonne Universit\'es, UPMC Univ. Paris 06, Univ. Paris Diderot, Sorbonne Paris Cit\'e, 5 Place Jules Janssen, 92195 Meudon, France}
\altaffiltext{2}{Department of Physics, University of Oxford, Oxford, UK}
\altaffiltext{3}{Univ. Grenoble Alpes, IPAG; CNRS, IPAG, 38000 Grenoble, France}


\begin{abstract}
We developed a simple, physical and self-consistent cloud model for brown dwarfs and young giant exoplanets. 
We compared different parametrisations for the cloud particle size, by either fixing particle radii, or fixing the mixing efficiency (parameter $f_{sed}$) or estimating particle radii from simple microphysics. 
The cloud scheme with simple microphysics appears as the best parametrisation by successfully reproducing the observed photometry and spectra of brown dwarfs and young giant exoplanets. 
In particular, it reproduces the L-T transition, due to the condensation of silicate and iron clouds below the visible/near-IR photosphere. 
It also reproduces the reddening observed for low-gravity objects, due to an increase of cloud optical depth for low gravity. 
In addition, we found that the cloud greenhouse effect shifts chemical equilibriums, increasing the abundances of species stable at high temperature. This effect should significantly contribute to the strong variation of methane abundance at the L-T transition and to the methane depletion observed on young exoplanets.
Finally, we predict the existence of a continuum of brown dwarfs and exoplanets for absolute J magnitude=15-18 and J-K color=0-3, due to the evolution of the L-T transition with gravity.
This self-consistent model therefore provides a general framework to understand the effects of clouds and appears well-suited for atmospheric retrievals.

\end{abstract}



\section{Introduction}

More than 40 planetary mass companions have been detected by direct imaging, which is currently limited to young and warm giant exoplanets. The latest generation of exoplanet imagers, instruments SPHERE \citep{beuzit08} and GPI \citep{macintosh14}, allows us to detect and characterize fainter objects than previously, such as 51 Eri b \citep{macintosh15, samland17} and HIP65426b \citep{chauvin17}. This technique can provide disk-averaged emission spectra of exoplanets, giving information about the composition and thermal structure of their atmospheres, and about the formation and evolution of planetary systems.
The study of brown dwarfs is of high interest to understand the properties of young giant exoplanets. Firstly, the physics and chemistry occurring in their atmospheres is expected to be very similar, the main difference being the surface gravity (typically log(g)$\sim$5 for brown dwarfs and log(g)$\sim$3.5-4 for exoplanets, with g in cm s$^{-2}$). Secondly, detected brown dwarfs cover a wide range of effective temperature (300-2500 K). Finally, high quality spectra of isolated brown dwarfs have been obtained, and many brown dwarfs have been observed, providing statistical information about their atmospheres. 

\begin{figure}[!h] 
\begin{center} 
	\includegraphics[width=8.5cm]{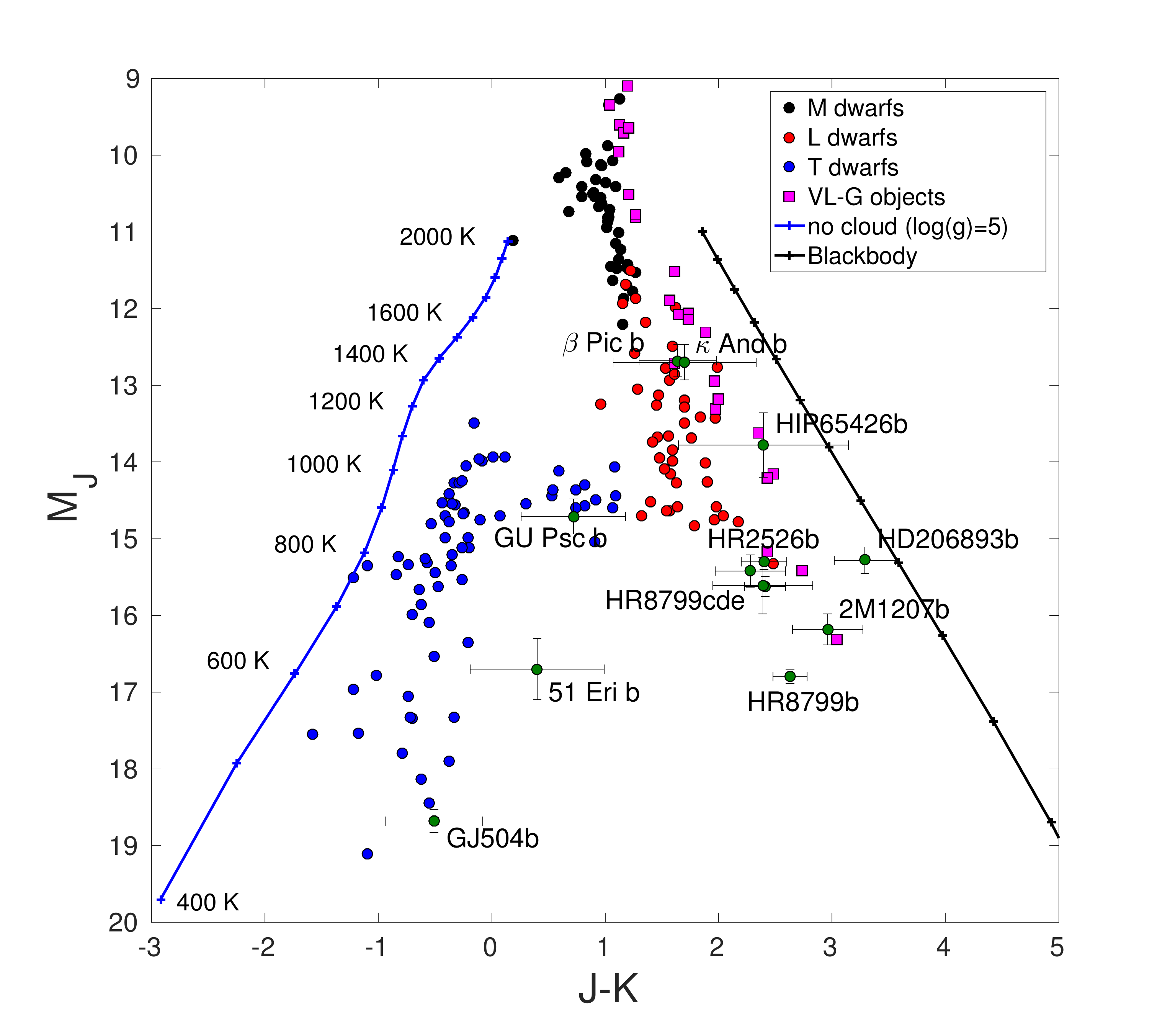}
\end{center}  
\caption{Color-magnitude diagram of M, L and T dwarfs with J-K colors plotted against the absolute J magnitude (MKO). M dwarfs are plotted as black dots, L dwarfs as red dots, T dwarfs as blue dots, low-gravity brown dwarfs as purple squares and directly imaged substellar companions as green dots.
The blue solid line was computed with spectra from Exo-REM assuming no cloud, log(g)=4, with T$_{eff}$ evolving from 400 to 2000 K and with object radii from the Ames-Dusty evolution model. The black solid line was computed assuming a blackbody for the spectrum.
Data for M, L, T field dwarfs are from \cite{dupuy12}. Data for low-gravity brown dwarfs (VL-G objects) are from \cite{liu16} and data for substellar companions are from \cite{rajan17, kuzuhara13, zurlo16, currie13, mohanty07, naud14, carson13, chauvin17, delorme17, konopacky16}.
}
\label{figure_1}
\end{figure}

\begin{figure*}[!h] 
\begin{center} 
	\includegraphics[width=8.5cm]{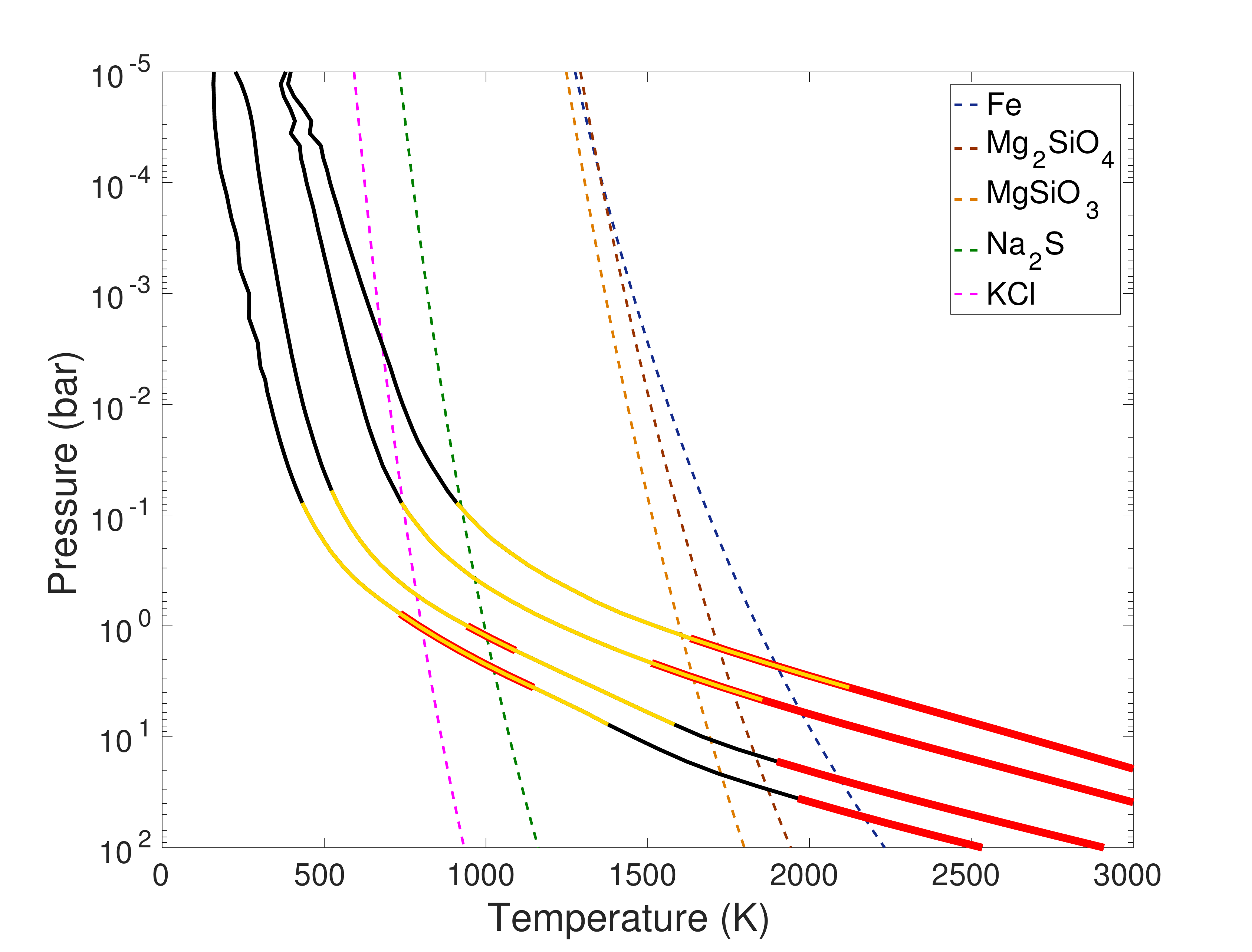}
	\includegraphics[width=8.5cm]{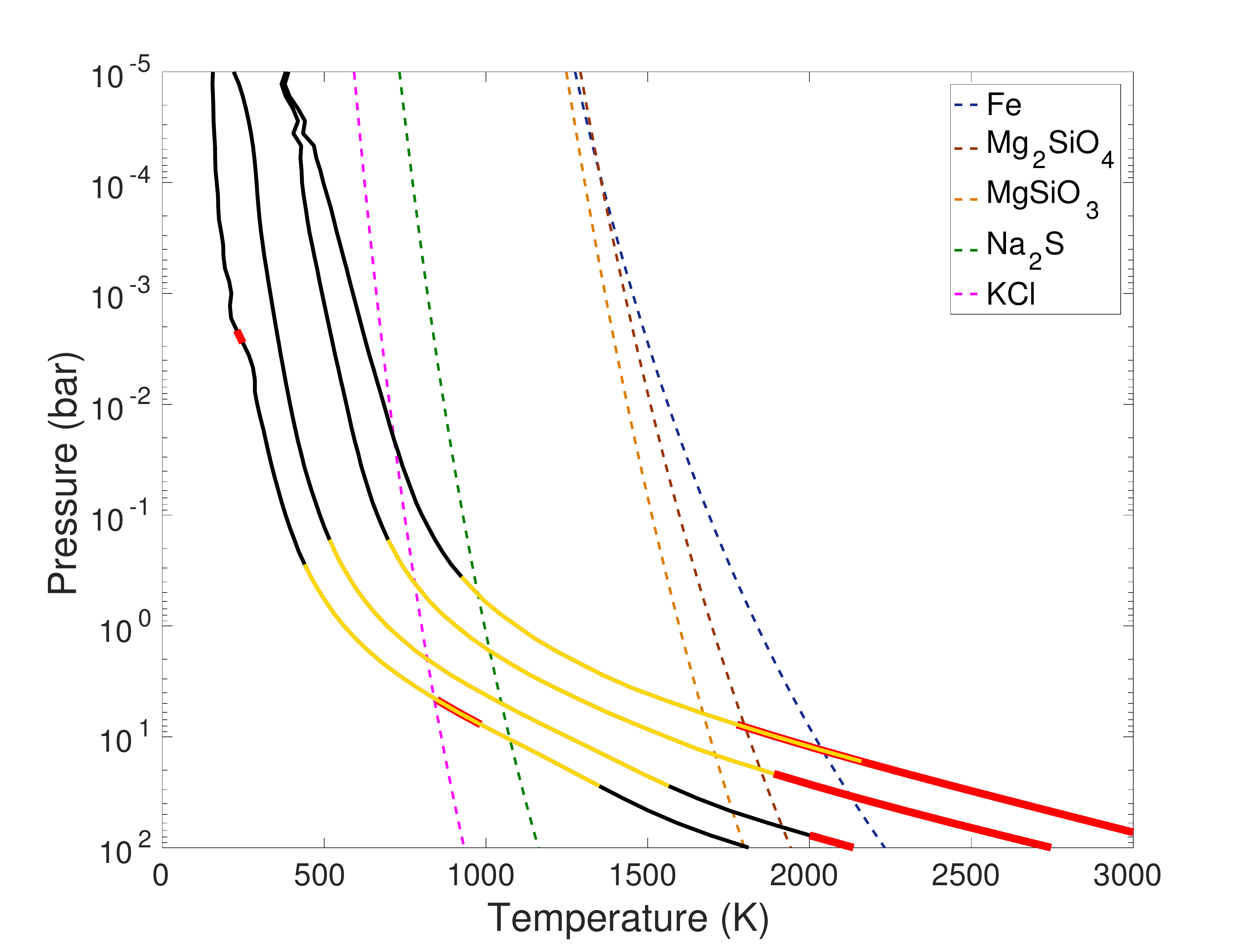}
\end{center}  
\caption{Temperature profiles and cloud condensation curves. For both panels, black curves are temperature profiles calculated with Exo-REM without cloud. Red curves represent the convective region and yellow curves represent the photosphere (region where most of the thermal emission comes from) computed for wavelengths from 0.625 to 5 $\mu$m. Dashed curves are condensation temperature curves for Fe, Mg$_2$SiO$_4$, Na$_2$S and KCl clouds assuming solar atmospheric metallicity. The left panel shows temperature profiles for log(g)=4 and T$_{eff}$=700, 900, 1300 and 1600 K (from left to right). The right panel shows temperature profiles for the same effective temperatures and log(g)=5.}
\label{figure_2}
\end{figure*}

As brown dwarfs cool, emission spectra and colors evolve. A strong transition, named the L-T transition, occurs for effective temperatures of 1100-1400 K. Late-L dwarfs appear red in color-magnitude diagram, with CO absorption in spectra, while early-T dwarfs appear much bluer with stronger CH$_4$ features in spectra (see reviews by \cite{kirkpatrick05} and \cite{cushing14}). The chemical change is attributed to the cooling of the atmosphere, shifting the CO-CH$_4$ chemical equilibrium. The favoured hypothesis to explain the L-T transition is linked to the formation of iron and silicate clouds in the photosphere of L dwarfs, making them redder and closer to a blackbody emission in color-magnitude diagram \citep{tsuji96, allard01, ackerman01, burrows06}. Figure \ref{figure_1} shows a color-magnitude diagram in J-K bands with color curves from a cloud-free model (Exo-REM, detailed en section 2) and from a blackbody. For colder objects like T dwarfs, clouds are believed to form deeper in the atmosphere and pass below the photosphere. Figure \ref{figure_2} illustrates this effect by showing temperature profiles obtained with the model Exo-REM, without cloud and for different effective temperatures. The L-T transition could be strengthened by the formation of holes in the cloud cover \citep{burgasser02, marley10}. The presence of clouds is also suggested by the temporal variability of brown dwarfs, particularly at the L-T transition \citep{apai13, crossfield14, yang16, biller17a}. T dwarfs and Y dwarfs appear slightly redder than what is expected for cloud-free objects. This reddening could be due to other condensates such as sulfide, alkali salt and water clouds forming at lower temperatures \citep{morley12, morley14}.

\begin{figure}[!h] 
\begin{center} 
	\includegraphics[width=8.5cm]{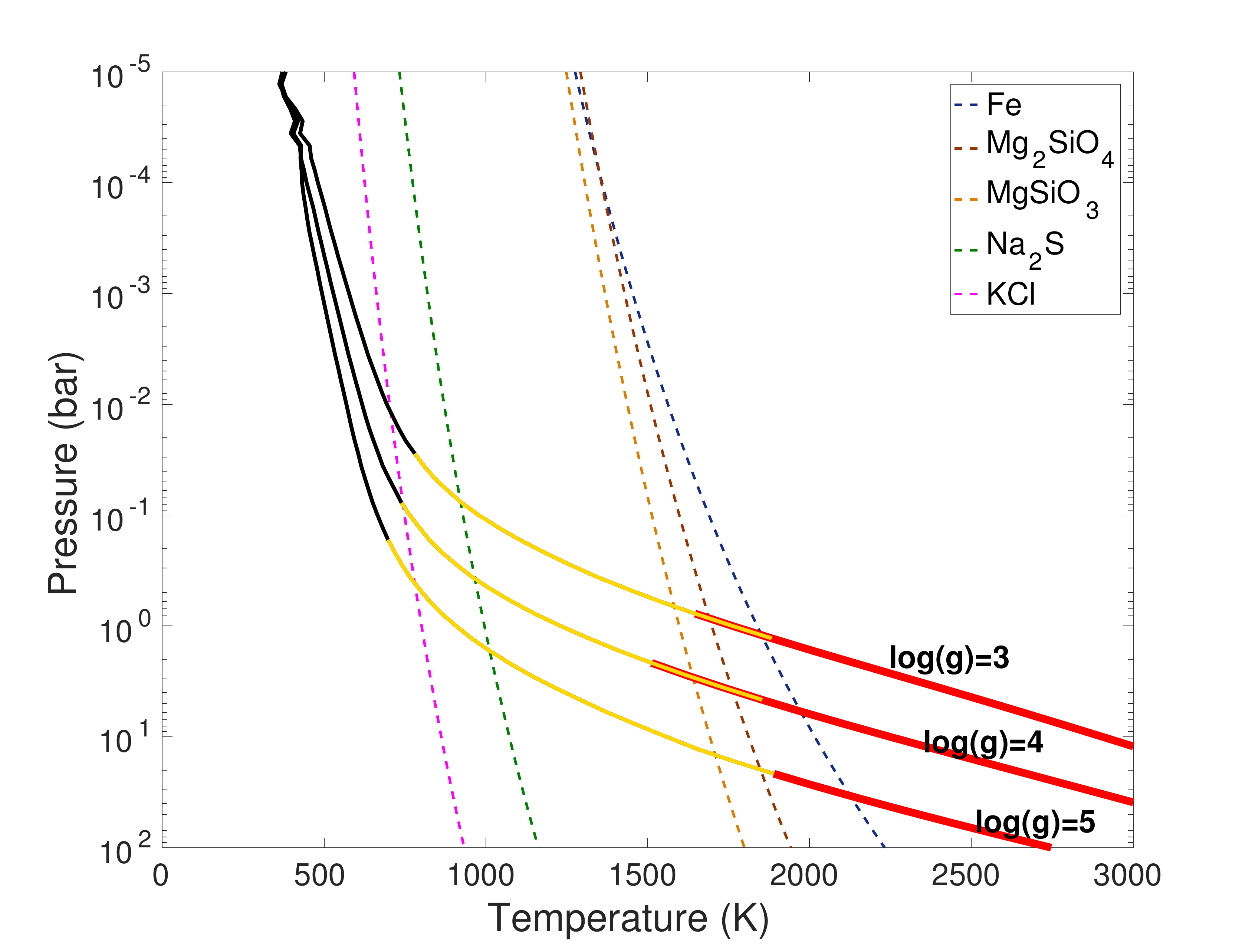}
\end{center}  
\caption{Temperature profiles as in Figure 2 but with T$_{eff}$=1300 K and log(g)=3, 4 and 5 (from top to bottom).}
\label{figure_3}
\end{figure}

Exoplanets discovered by direct imaging seem to have colors quite similar to field brown dwarfs, and so seem to have spectra shaped by clouds. The formation of clouds on young giant exoplanets is not surprising, since clouds and hazes are present in every Solar System atmosphere. Moreover, transit spectra and visible light curves provide strong indications for the presence of clouds and hazes on highly irradiated exoplanets \citep{demory13, sing16}. However, most young giant exoplanets seem to occupy a slightly distinct space compared to field brown dwarfs in color-magnitude diagrams. They generally appear redder than brown dwarfs of similar effective temperature. It is particularly remarkable for HR8799 bcde \citep{marois10}, 2M1207b \citep{chauvin04} and VHS1256b \citep{gauza15} which are shifted toward bottom right of the M$_J$ J-K diagram compared to late-L dwarfs (see Figure \ref{figure_1}). A similar reddening is observed for isolated young brown dwarfs (\cite{gagne15, faherty16, liu16}, and see Figure \ref{figure_1}).
It seems that there is a strong reddening and a delayed (i.e. happening at lower temperature) L-T transition for low-gravity objects, as if they were more cloudy. The delayed L-T transition could be explained by the effect of gravity on the thermal structure \citep{marley12}. If gravity decreases, the temperature in the deep atmosphere increases and the passage of the cloud condensation level below the photosphere occurs for a lower effective temperature (see temperature profiles in Figure \ref{figure_3}). The reddening for low-gravity objects is not well understood. It could suggest that the cloud settling is less efficient for low gravity, leading to optically thicker clouds \citep{madhusudhan11}.

Previous modelling studies of clouds in the atmospheres of brown dwarfs and young giant exoplanets are mostly based on 1D radiative-convective models. Up to now, there are two main classes of cloud models. The first class corresponds to sophisticated models computing cloud self-consistently with microphysics. These include: 
\begin{itemize}
\item Drift-Phoenix \citep{helling08}, which includes a full microphysics scheme with cloud formation on TiO$_2$-seed for 11 solids from 60 chemical reactions. 
\item BT-Settl \citep{allard12b}, which includes 40 condensable species whose particle size are computed by comparing the characteristic timescales of condensation, sedimentation and mixing. 
\end{itemize}
The second class corresponds to parametric cloud models which do not include microphysics and use fixed parameters to describe the cloud particle distribution such as: 
\begin{itemize}
\item  The model by \cite{tsuji02}, which includes corundum, iron and silicate clouds, fixing particle sizes and with clouds forming with no sedimentation from the condensation level to the top of the photosphere.
\item The model by \cite{ackerman01}, with following updates by \cite{marley10,morley12, morley14}, which includes iron, silicate, sulfide, salt and water clouds and determines cloud distribution and particles radii assuming a fixed value of the ratio $f_{sed}$ of sedimentation velocity by vertical mixing velocity.
\item  The model by \cite{burrows06} and \cite{madhusudhan11}, which includes iron and silicate clouds, fixing particle sizes and the cloud profile with 3 free parameters.
\end{itemize}

All these models have difficulties to produce a sharp L-T transition and the reddening for low-gravity objects. Alternatively, cloud-free models have been developed to explain the L-T transition, photometry and spectra of brown dwarfs assuming changes in the temperature gradients caused by fingering convection and thermochemical instabilities of CO/CH$_4$ and N$_2$/NH$_3$ \citep{tremblin15, tremblin16, tremblin17}. However, in these models, the atmosphere has to be quasi-isothermal in the fingering convective part, what is quite drastic and has not been demonstrated by 2D/3D simulations yet.  Moreover, it is difficult to argue for no cloud effect on the spectra of brown dwarfs and exoplanets, as discussed before (e.g. the clear detection of clouds on irradiated exoplanets). It is also possible that fingering convection and cloud formation occur in parallel.

In this paper, we present a new self-consistent cloud model able to reproduce and to explain the photometry and spectra of brown dwarfs and giant exoplanets. We wanted to avoid a complex cloud model with full microphysics, because they are difficult to interpret, they may be based on many unknown parameters and they are not always efficient for atmospheric characterization because of the absence of free parameters for clouds. We favoured instead a simple model but capturing the dominant physical processes and with the minimum number of free parameters. 
In section 2, we describe the model and different hypotheses for cloud particle size. 
In section 3, we describe the results of the cloud model and compare it to the observed photometry and spectra of brown dwarfs and exoplanets. Finally, we discuss implications and perspectives for future work in section 4 and 5.

\section{Methods}
\subsection{Radiative-convective equilibrium model: Exo-REM}

Exo-REM is a 1-D radiative-equilibrium model, developed to simulate the atmospheres and spectra of young giant exoplanets \citep{baudino15, baudino17}. Exo-REM solves for radiative-convective equilibrium, assuming that the net flux (radiative+convective) is conservative and neglecting stellar heating, which is verified for the long orbit exoplanets that were imaged nowadays. 
The basic input parameters of the model are the effective temperature of the planet, the acceleration of gravity at 1 bar and the elemental abundances. Conservation of flux over the grid (64 pressure levels) is solved iteratively using a constrained linear inversion method. Sources of opacity include the H$_2$-H$_2$ and H$_2$-He collision-induced absorption, ro-vibrational bands from 8 molecules (H$_2$O, CH$_4$, CO, CO$_2$, NH$_3,$ PH$_3$, TiO and VO) and resonant lines from Na and K (see Baudino et al. 2015, 2017 for references). For VO, we used a more recent linelist from \cite{mckemmish16}, which is available in the ExoMol database (http://exomol.com/). Molecular and atom line absorption is represented through a k-correlated distribution method and the net fluxes are calculated from 0.625 to 500 $\mu$m. The truncation at 0.625 $\mu$m is a limitation for high temperatures but it should be acceptable for objects with T$_{eff}$ $\leqslant$ 1700 K.
The vertical profiles of the different molecules and atoms are calculated for a given temperature profile either assuming thermochemical equilibrium or allowing for some non-equilibrium chemistry between C-, O- and N- bearing compounds. For the latter case, we use an analytical formulation based on a comparison of chemical time constants  with vertical mixing time from \cite{zahnle14}. For this calculation, vertical mixing is parametrized by an eddy mixing coefficient Kzz (see next section). Chemical gas abundances are modified accordingly when condensates form. Figure \ref{figure_annexe1} in the Appendix shows abundance profiles assuming equilibrium (dashed lines) or non-equilibrium (solid lines) chemistry. By default, simulations are performed with non-equilibrium chemistry.

The first version of Exo-REM only treated absorption of iron and silicate particles, fixing the cloud vertical profile as in \cite{burrows06} and with a single free parameter, which is the optical depth at some wavelength.
We have upgraded the model by accounting for both absorption and scattering of thermal radiation by clouds. Fluxes are computed using the two-stream approximation assuming the hemispheric closure, as suggested for exoplanets by \cite{heng14}, using an average cosine of the zenith angle equal to 2/3 for the upward flux and -2/3 for the downward flux. For non-cloudy atmospheres, we obtain almost identical results between this version of Exo-REM and the previous one, for which fluxes were properly integrated over all zenithal angles.
The model is essentially valid for planets and brown dwarfs with effective temperatures between 300 and 1700 K. The limitation at $\sim$1700 K is due to the truncation at 0.625 $\mu$m and to the lack of some condensates (e.g. Al$_2$O$_3$) and other absorbing species (FeH, ions,...) forming at high temperatures.

\subsection{Cloud scheme}

\subsubsection{Cloud condensation chemistry and optical properties}

Iron and silicate clouds are supposed to strongly affect the colors and spectra of L brown dwarfs \citep{marley10}. Sulfide and alkali salt (Cr, MnS, Na$_2$S, ZnS, and KCl) clouds may play a significant role for T dwarfs \citep{morley12}, while water clouds should be the dominant condensate on Y dwarfs \citep{morley14}.
Our model includes the formation of iron, silicate, Na$_2$S, KCl and water clouds. 
We did not include all sulfide and salt condensates for simplicity and because Na$_2$S is expected to be the dominant one for T dwarfs \citep{morley12}.
For simulating cloud condensation, we used saturation vapour pressures and abundances of condensate-forming species from \cite{visscher06}, \cite{visscher10} and \cite{morley12} (see also Appendix A). 
We computed optical properties (extinction coefficient, single scattering albedo and assymetry factor) assuming spherical particles following a log-normal size distribution with an effective variance of 0.3. We computed these values for a grid of wavelengths and mean particle radii (from 0.1 to 100 $\mu m$), using optical indexes from \cite{baudino15}, \cite{morley12} and \cite{querry87}. In Exo-REM, optical properties are interpolated from these precomputed tables for a given wavelength and particle radius.

\subsubsection{Vertical mixing and sedimentation}

At equilibrium, the downward flux of falling cloud particles is balanced by the upward flux of cloud particles and vapour due to advection and turbulent mixing. This balance can be simply expressed as \citep{ackerman01}:

\begin{equation} 
\frac{\partial q_c}{\partial z} = -\frac{\partial q_v}{\partial z} -\frac{v_{sed}}{K_{zz}} q_c
\end{equation} 

where $q_c$ is the mass mixing ratio of a condensate, $q_v$ is the mass mixing ratio of vapour, $v_{sed}$ is the sedimentation velocity of cloud particles and $K_{zz}$ is the eddy diffusion coefficient representing the vertical mixing due to convection, waves or turbulence.
We make the assumption that the supersaturation is weak. In that case, $q_v$=$q_s$ above the cloud condensation level, where $q_s$ is mass mixing ratio of vapour at saturation ($q_s$=$p_{sat}/p$ where $p$ is the pressure and $p_{sat}$ is the saturation vapour pressure).

The mass mixing ratio of condensate at a level n+1 ($q_c(z_{n+1})$) can be expressed by solving equation (1) analytically with the method of variation of parameters:

\begin{equation}
\begin{split}
q_c(z_{n+1})= exp\left(-\int_{z_n}^{z_{n+1}} \frac{v_{sed}(z)}{K_{zz}(z)}dz \right)  \\
 \times \left[q_c(z_n) -  \int_{z_n}^{z_{n+1}} \frac{\partial q_s(z)}{\partial z} exp\left(\int_{z_n}^{z} \frac{v_{sed}(z')}{K_{zz}(z')}dz' \right) dz\right]
\end{split}
\end{equation}

This analytical expression provides a high numerical stability and accuracy for the integration.
The mixing ratio at level $z_{n+1}$ is thus the sum of term $A$ corresponding to the advection of cloud particles from level $z_{n}$:
\begin{equation}
A=exp\left(-\int_{z_n}^{z_{n+1}} \frac{v_{sed}(z)}{K_{zz}(z)}dz \right) q_c(z_n) 
\end{equation} 
and term $B$ corresponding to the condensation between levels $z_{n}$ and $z_{n+1}$:
\begin{equation}
\begin{split}
B=-  exp\left(-\int_{z_n}^{z_{n+1}} \frac{v_{sed}(z)}{K_{zz}(z)}dz \right)  \\
\times \int_{z_n}^{z_{n+1}} \frac{\partial q_s(z)}{\partial z} exp\left(\int_{z_n}^{z} \frac{v_{sed}(z')}{K_{zz}(z')}dz' \right) dz
\end{split}
\end{equation}

We computed the sedimentation of particles assuming that they fall at the terminal velocity given by
\cite{fuchs64} and \cite{rossow78}:
\begin{equation} 
v_{sed}=\frac{2 \beta r^2 g (\rho_p-\rho_a)}{9\eta}
\end{equation} 
valid for low Reynolds numbers, and where $r$ is the particle radius, $g$ is the gravitational acceleration, $\rho_p$ is the particle density, $\rho_a$ is the atmosphere density, $\eta$ is the viscosity of the atmospheric gas and $\beta$ is the Cunnigham slip factor, which describes non-continuum effects. An experimental expression of the Cunnigham slip factor is \citep{rossow78}:

\begin{equation} 
\beta= 1+ \frac{4}{3} K_n 
\label{eq_beta}
\end{equation}

where $K_n$ is the Knudsen number, equal to the ratio of the mean free path $\lambda$ to the particle radius:

\begin{equation} 
K_n= \frac{\lambda}{r} = \frac{k_B T}{\sqrt{2} \pi d^2} \frac{1}{p r} 
\end{equation} 

with $k_B$ the Boltzmann constant, $T$ the temperature and $d$ the gas molecular diameter.

\begin{figure}[!h] 
\begin{center} 
	\includegraphics[width=8.5cm]{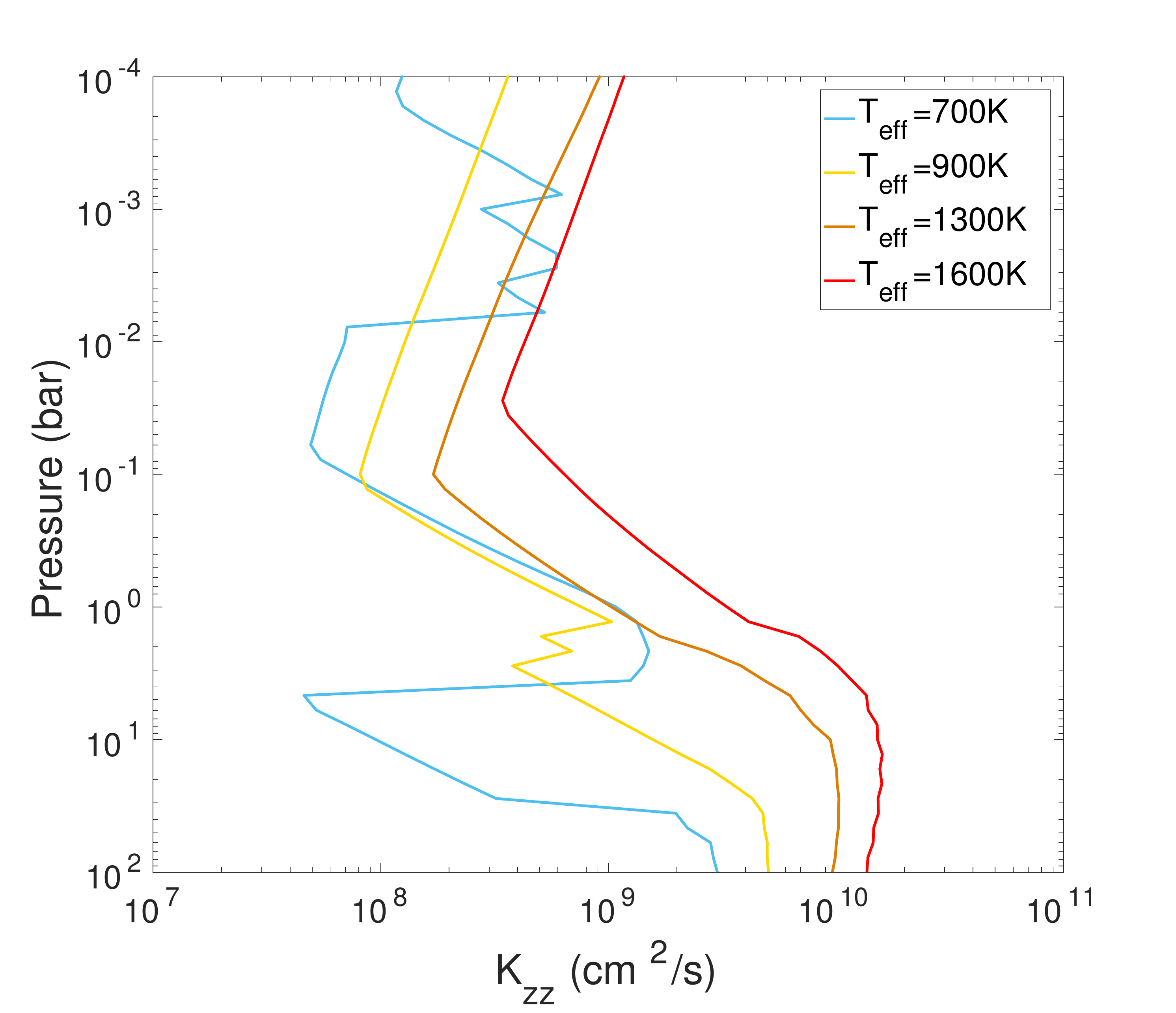}
	\includegraphics[width=8.5cm]{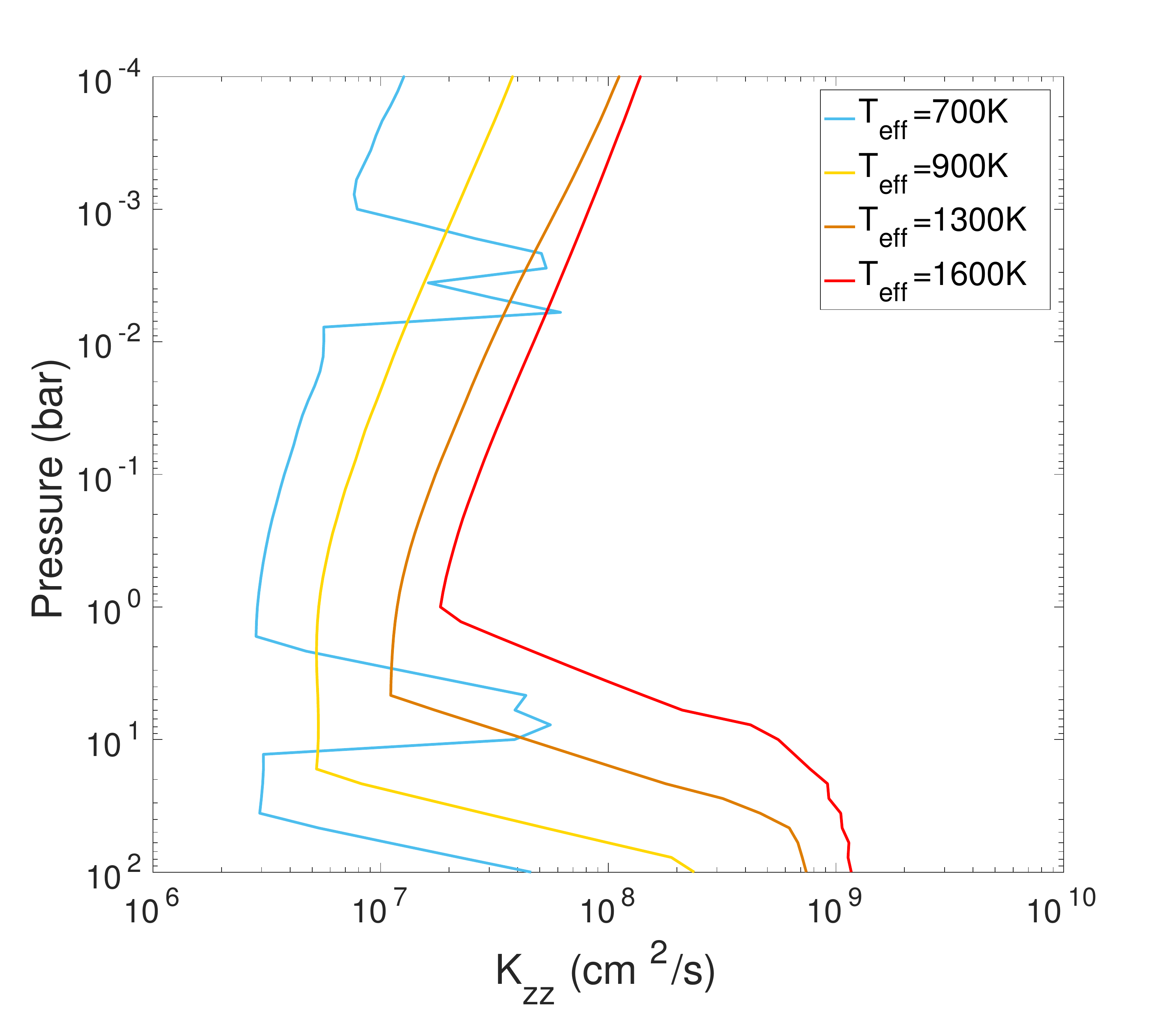}
\end{center}  
\caption{K$_{zz}$ profiles derived from our parametrization for T$_{eff}$=700, 900, 1300 and 1600 K and log(g)=4 (top) and log(g)=5 (bottom).}
\label{figure_4}
\end{figure}

For the eddy mixing coefficient, we used a formula from \cite{gierasch85} and \cite{ackerman01}, based on the mixing length theory:

\begin{equation} 
K_{zz} = \frac{H}{3} \left(\frac{L}{H}\right)^{4/3} \left(\frac{R F_{conv}}{\mu \rho_a c_p}\right)^{1/3}
\end{equation} 

where $H$ is the atmospheric height scale, $L$ is the mixing length, $R$ is the universal gas constant, $\mu$ is the mean molecular weight, $c_p$ is the specific heat and $F_{conv}$ is the convective heat flux. The factor 1/3 on the left is arbitrary, used to match observations of giant planets. We assumed that the mixing length is equal to $H$, even if it is generally expected to be smaller. Contrary to \cite{ackerman01} who fixed $F_{conv}=\sigma T_{eff}^4$ for simplicity, we used the convective flux derived by Exo-REM in convective regions. Outside convective region, the convective heat flux is null, however there is still mixing produced by gravity waves, whose amplitude grows with altitude. According to 2D simulations of brown dwarfs by \cite{freytag10}, the vertical mixing and vertical winds fall by 1-2 orders of magnitude above the convective region. We parametrize $K_{zz}$ in radiative regions using formula (9) with $F_{conv}=10^{-6}\sigma T_{eff}^4$ and assuming overshooting as described in \cite{ludwig02, ludwig03} and \cite{helling08}. With this parametrization, $K_{zz}$ decreases by a factor of $\sim$50 above the convective region and increases above with altitude (approximately $\propto P^{-1/3}$, see Figure \ref{figure_4}). For effective temperatures lower than 900 K, our model predicts the formation of a second convective layer at 1-10 bar, also noticed by \cite{morley12}. Convection also appears at 10$^{-2}-10^{-3}$ bar for low temperature but has limited impact on the results here.
The magnitude and vertical evolution of $K_{zz}$ with our parametrization seems compatible with the 2D simulations by \cite{freytag10}.

\subsubsection{Hypotheses for computing cloud radii}

We assumed that cloud particles follow a log-normal size distribution with a constant effective variance $\nu_{eff}$=0.3. The effective radius $r_{eff}$ (i.e. the area weighted radius of the particle distribution) is slightly lower than the sedimentation radius $r_{sed}$, which is the equivalent radius of mass-weighted sedimentation flux. The effective radius and the sedimentation radius are linked together by \citep{ackerman01}:

\begin{equation} 
r_{eff}=  r_{sed} \  exp\left( -\frac{\alpha+1}{2} ln(1+\nu_{eff})\right)
\end{equation}

where $\alpha$ represents the exponent of the radius dependence of $v_{sed}$. $\alpha$ should vary between 1 and 2. For simplicity, we use an average value $\alpha$=1.4, as \cite{ackerman01}. The error for the calculation of $r_{eff}$ is lower than 7$\%$.

For computing the mass mixing ratio (equation 2), we need to make assumptions about the sedimentation radii of cloud particles. We included three options for computing cloud particles radii: 1) fixing the radii, 2) fixing the ratio of sedimentation velocity by vertical mixing velocity (parameter $f_{sed}$), 3) computing the radii with simple microphysics by comparing characteristic timescales.

\paragraph{1) Fixed radii\\}

The simplest possibility is to fix cloud particle radii as a free parameter. In that case, we consider that all particles form with the same size, which does not evolve. This assumption represents a quite realistic description of the vertical distribution of cloud above the condensation region. However, there is no indication about the cloud particle size, which could depend on many parameters such as the kind of condensate, the temperature, the pressure, the gravity and the atmospheric metallicity.

\paragraph{2) Fixed $f_{sed}$\\}

\cite{ackerman01} proposed that the vertical distribution of cloud particles is qualitatively well represented by fixing the ratio of the particle sedimentation velocity by the characteristic vertical mixing velocity. 
This ratio is defined by parameter $f_{sed}$ as: $f_{sed}=  \frac{v_{sed}}{w^{*}}$, where $w^*=\frac{K_{zz}}{H}$ and $v_{sed}$ is given by equation (5). 
This parametrization assumed that at each vertical level, the size of particles evolves so that the characteristic timescale of mixing ($\tau_{mixing}=\frac{H}{w^*}$) is similar to the characteristic timescale of sedimentation ($\tau_{sed}=\frac{H}{v_{sed}}$).
The two extreme values for $f_{sed}$ are $f_{sed}$=0, which would correspond to clouds with no sedimentation as in AMES-Dusty \citep{chabrier00}, and $f_{sed}$=$+\infty$, which would correspond to clouds removed instantaneously by sedimentation as in AMES-Cond \citep{allard01}.
$f_{sed}$ generally is $\sim$ 1-5 for clouds in the solar system, and photometry of brown dwarfs can be reproduced with $f_{sed} \sim$ 3 \citep{ackerman01, marley10}. 

\begin{figure}[!h] 
\begin{center} 
	\includegraphics[width=8.5cm]{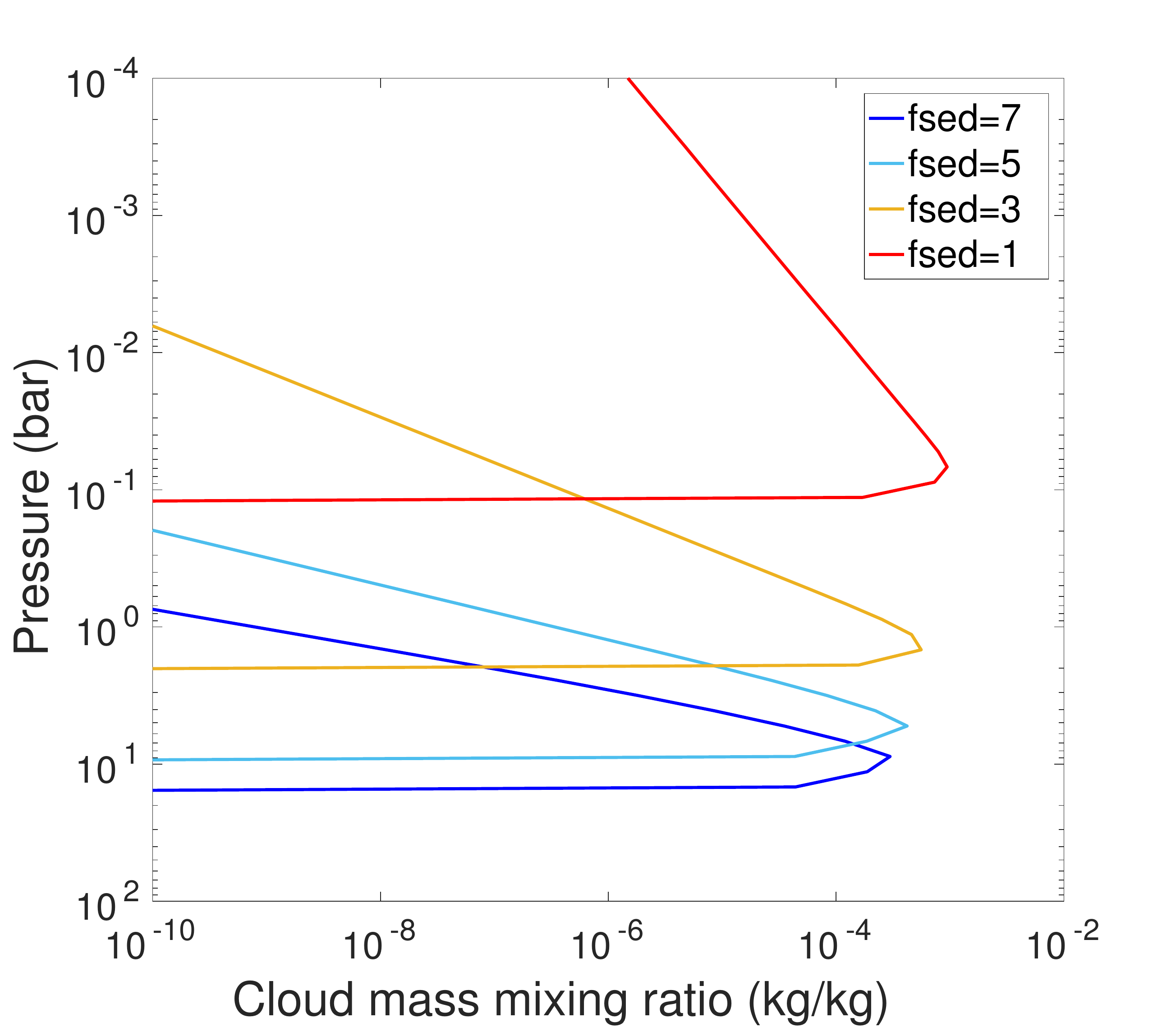}
	\includegraphics[width=8.5cm]{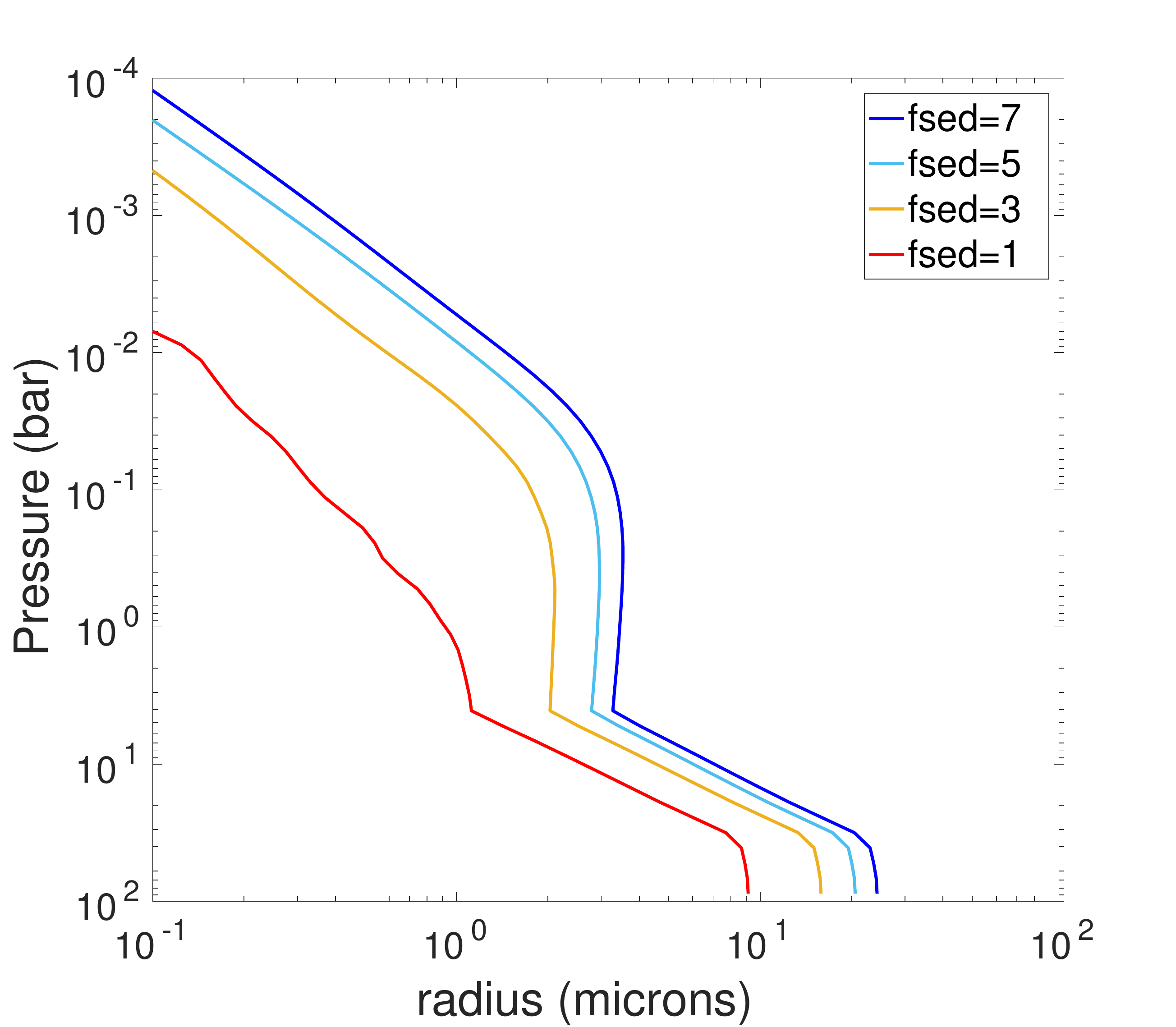}
\end{center}  
\caption{Evolution with pressure of condensate mixing ratio (top panel) and cloud particle radii (bottom panel)  for iron clouds, with T$_{eff}$=1300 K and log(g)=5, for $f_{sed}$=1-5.}
\label{figure_5}
\end{figure}

By assuming a fixed $f_{sed}$, the cloud vertical distribution can easily be computed from equation (2). In particular, in a cloud layer where condensation is negligible compared to vertical mixing (i.e. $B\ll A$ in equations 3 and 4, what is generally the case for one atmospheric scale height above condensation), the cloud mixing ratio simply evolves as $q_c \propto p^{f_{sed}}$ (see Figure \ref{figure_5}). Cloud particle sedimentation radii are directly computed from formula (5-8):

\begin{equation} 
r_{sed}=  \frac{2}{3}\times\lambda\left(\sqrt{1+10.125\frac{ \eta w^* f_{sed}}{g (\rho_p-\rho_a)\lambda^2}}-1\right)
\end{equation}

This parametrization with fixed $f_{sed}$  is widely used in the brown dwarf and exoplanet modeling community. It represents a simple and efficient way of computing cloud distribution with quite realistic particle sizes and with only one free parameter. However, this parametrization assumes that cloud particles radii change as a function of altitude to maintain a constant efficiency of vertical mixing. The cloud mass mixing ratio profile is in that case completely independent of the vertical mixing.
When condensation is negligible (almost no new particles are formed), there is no physical reason for such an efficient evolution of cloud particle radii. In particular, that may lead to an overestimation of cloud mass mixing ratio and optical thickness when $K_{zz}$ decreases with altitude, like above the convective region. 

\paragraph{3) Simple microphysics\\}
Cloud microphysics is a complex research area that still lacks a global understanding, and information about clouds in brown dwarf and exoplanets is very limited. It is therefore preferable to limit at maximum the level of complexity in a cloud microphysics model for such objects.
\cite{rossow78} developed a simple method for deducing the dominant processes controlling cloud microphysics in planetary atmospheres and for estimating the radii of cloud particles. This method is based on the comparison of the timescales of the main physical processes involved in the formation and growth of cloud particles. Processes having the shortest timescales are supposed to dominate, driving the particle size, while processes with longer timescales are neglected. In this approach, the mean particle size in steady state is obtained by equalizing the timescale of the dominant particle growth process with the  timescale of the dominant particle removal process.
This method succeeds in reproducing particle size and density for water clouds on Earth and Mars, sulphuric clouds on Venus and ammonia icy clouds on Jupiter \citep{rossow78}.
We implemented it in our cloud model, considering only condensation growth, coalescence, vertical mixing and sedimentation as physical processes driving the distribution of cloud particles. Growth by coagulation is an important process for high altitude aerosols, as sulphuric clouds on Venus and organic haze on Titan. We found that coagulation should be negligible compared to condensation for the conditions considered here.
Cloud formation starts by the formation of stable cloud embryos by nucleation. These cloud embryos are small condensate particles whose size exceeds a critical value allowing growth by condensation.
Heterogeneous nucleation generally is the most efficient nucleation process for planetary atmospheres \citep{rossow78} and is expected to control cloud formation on brown dwarfs \citep{helling06}.
Embryos produced by heterogeneous nucleation form on the surface of various small aerosols, called cloud condensation nuclei (CCN), which decrease the level of supersaturation required for nucleation.
For a brown dwarf or a giant exoplanet, these CCN could be micrometeorits or small particles produced by homogeneous or chemical nucleation, as for instance TiO$_2$ particles \citep{helling06}.
Heterogeneous nucleation is a very fast process, much faster than the subsequent growth phase. Once nucleation is finished, the particle number density becomes relatively constant and particles grow by condensation with a characteristic timescale $\tau_{cond}=\left(\frac{1}{r}\frac{dr}{dt}\right)^{-1}$ given by \citep{rossow78}:

\begin{equation} 
\tau_{cond}=  \left\{ 
\begin{array}{rcl}
\frac{r^2 \rho_p}{2 \gamma \eta q_s S} & \mbox{for} & K_n \ll 1 \\
\frac{2 r \rho_p \rho_a}{3 \gamma q_s S} \frac{1}{\sqrt{2 k T /\pi m}} & \mbox{for} & K_n \gg 1 
\end{array} \right\}
\end{equation}

where $K_n$ is given by equation (7), $\gamma$ is a constant taken equal to 2 (i.e. the value for water clouds on Earth), $m$ is the atmospheric molecular mass and $S= \frac{q_v-q_s}{q_s}$ is the supersaturation. As \cite{rossow78}, we assumed that the supersaturation is constant in the cloud. We used it as a free parameter with $S=10^{-2 \pm 1}$, the range of water nucleation in Earth's atmosphere.
In the cloud forming region, the condensation growth timescale is generally proportional to $r^2$ ($K_n \ll 1$). The time for a cloud embryo having an arbitrary size to grow up to a radius $r$ is approximately $\tau_{cond}$. The growth is then limited by removal of particles by sedimentation or vertical mixing. We consider that it is stopped once $\tau_{cond}$ equal the minimum value between the mixing timescale ($\tau_{mixing}$) and the sedimentation timescale ($\tau_{sed}$), given by:

\begin{equation} 
\tau_{mixing}= \frac{H^2}{K_{zz}} 
\end{equation} 

and:

\begin{equation} 
\tau_{sed}=  \frac{H}{v_{sed}}
\end{equation} 

This equilibrium between particle growth and removal fixes the size of particles ($r_{cond}$). Since small particles grow very fast, the final size distribution should be quite narrow and insensitive to the initial size distribution. This allows us to only use one radius to describe the size distribution, which is assumed to follow the log-normal distribution given in paragraph 2.2.3. We also assume that particle effective radii cannot be lower than 0.1 $\mu m$ and larger than 100 $\mu m$, which are the limits of our grid of cloud optical properties.

A particle larger than the others would fall faster and can grow during its descent by collecting the smaller particles. This process, called gravitational coalescence, may efficiently limit the mean size of particles by quickly removing particles larger than a certain size.
The characteristic timescale of coalescence is \citep{boucher95, rossow78}:

\begin{equation} 
\tau_{coal}=  \frac{4 r \rho_p}{3 v_{sed} \rho_a q_c}
\end{equation} 

However this process is supposed to be efficient only for liquid or icy particles, which have a high sticking efficiency, and only in the deep atmosphere. It should not occur for silicate, sulfide and salt clouds. We therefore limit coalescence to water and liquid iron clouds.
$\tau_{coal}$ is generally proportional to $1/r$ and decreases more slowly than $\tau_{sed}$. It is therefore dominant in the deep atmosphere for small particles. When coalescence is efficient ($\tau_{coal} < \tau_{sed}$ and $\tau_{coal} <\tau_{mixing}$) , we fix the radius of particles with $\tau_{cond}=\tau_{coal}$ and modify the sedimentation velocity used in equation (2) as $v_{sed}'=v_{sed}(1+\tau_{sed}/\tau_{coal})$ to take into account the removal of particles by coalescence.

The comparison of these different timescales provides the size of new particles formed at a given level. However, some particles are also transported from below, leading to several particle populations. For simplicity, we assumed that vapour exceeding saturation only condenses to produce new particles rather than making grow the already formed particles coming from adjacent levels. We also assumed that there is only one particle population at each level and we fix the mean size in order that the sedimentation mass flux of this particle population is equal to the sum of sedimentation mass fluxes from new condensed particle and advected particles. Therefore, we compute the mean radius $r(z_{n+1})$ at level n+1, verifying:

\begin{equation}
\begin{split} 
q_c(z_{n+1})\times v_{sed}(r(z_{n+1}))= A\times v_{sed}(r(z_{n})) \\
+ B\times v_{sed}(r_{cond}(z_{n+1}))
\end{split}
\end{equation} 

where the first term in the left hand is the sedimentation mass flux at level n+1, the first term in the right hand is the sedimentation mass flux of particles advected from level n and the second term represents the sedimentation mass flux of particles condensed at level n+1.
Our simplified approach allows us to properly simulate cases where advection or condensation dominates. 
In fact, such situations occur in most of the atmosphere because saturation pressures vary quickly with altitude.

\begin{figure}[!h] 
\begin{center} 
	\includegraphics[width=8.5cm]{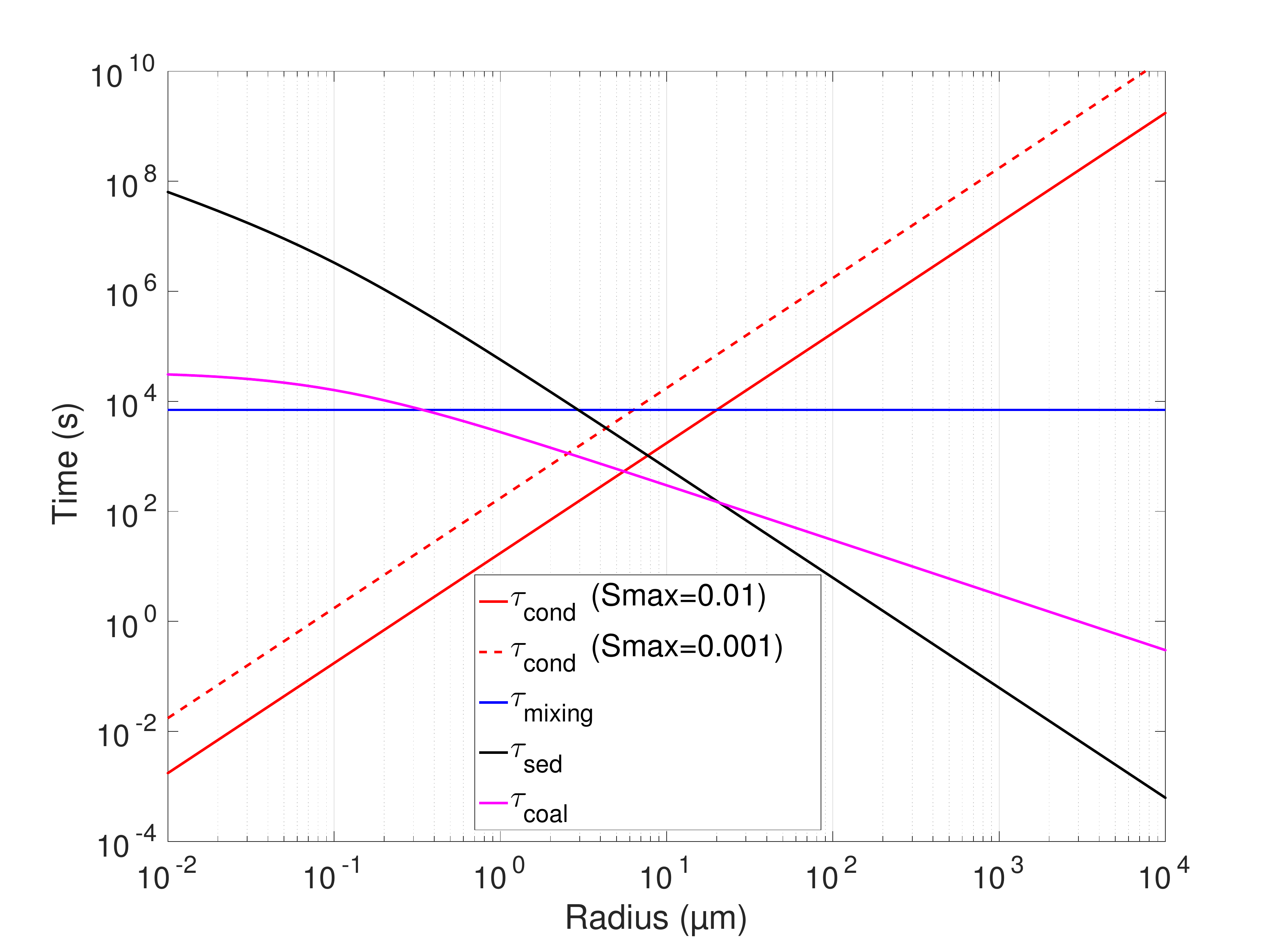}
\end{center}  
\caption{Characteristic timescales for condensation growth (red), sedimentation (black), vertical mixing (blue) and coalescence (purple) as a function of particle radius. Timescales were computed for iron at the condensation level ($\sim$10 bars) for T$_{eff}$=1300 K and log(g)=5. The condensation growth timescale was computed with a supersaturation S=0.01 (solid line) and S=0.001 (dashed line).}
\label{figure_6}
\end{figure}

Figure \ref{figure_6} shows the characteristic timescales of condensation growth, coalescence, vertical mixing and sedimentation as a function of radius at the bottom of iron cloud (p=10 bars) for T$_{eff}$=1300 K and log(g)=5. At this pressure and with $S=10^{-2}$, coalescence limits particle growth to around 5 $\mu m$. Without coalescence, particle growth would be limited to around 8 $\mu m$. In the case with coalescence, particle size is quite close to the one obtained by fixing $f_{sed} \sim$ 3 ($f_{sed}$=3 corresponds to $\frac{\tau_{mixing}}{\tau_{sed}}=3$).

\begin{figure*}[!h] 
\begin{center} 
	\includegraphics[width=8.5cm]{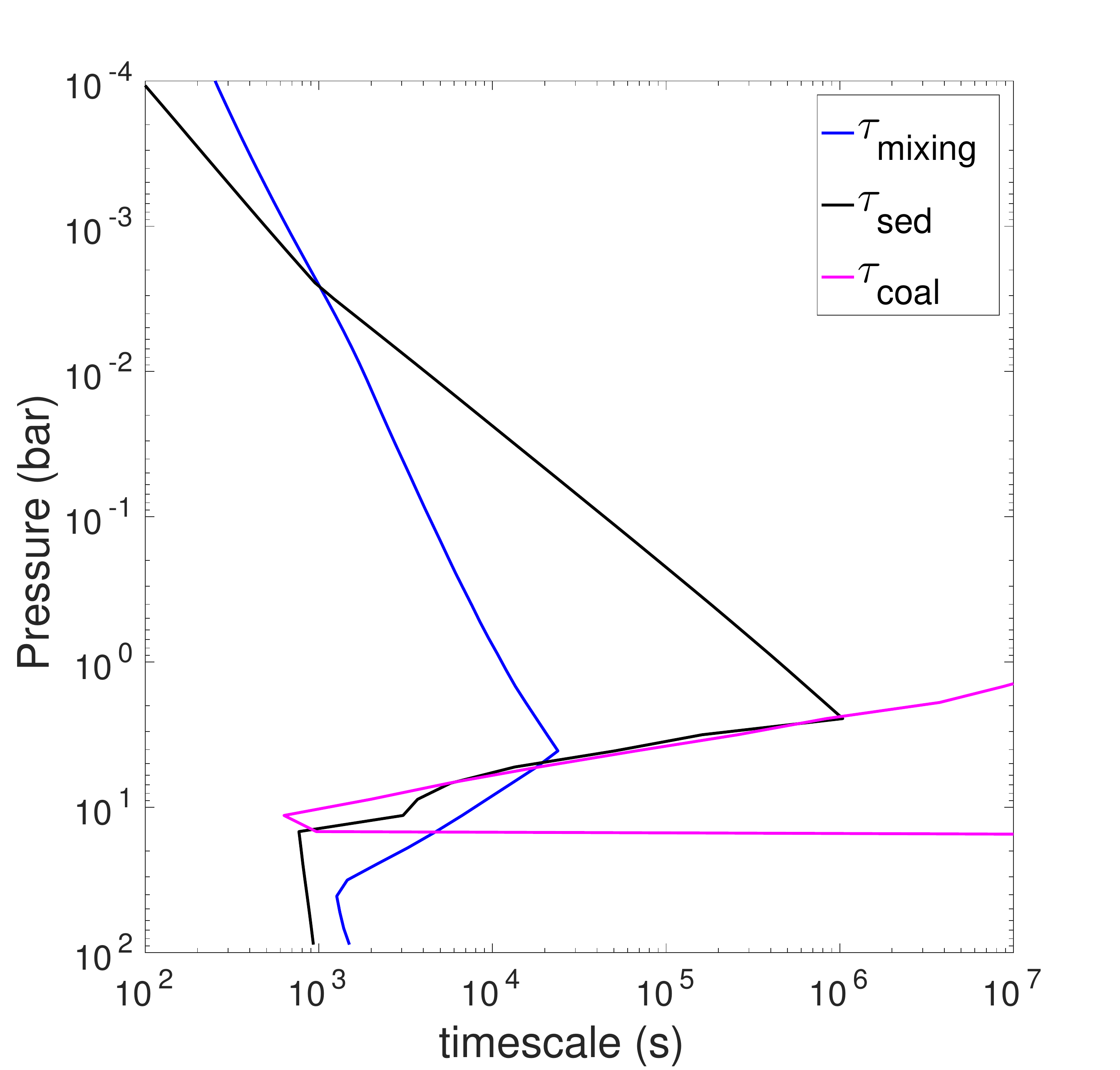}
	\includegraphics[width=8.5cm]{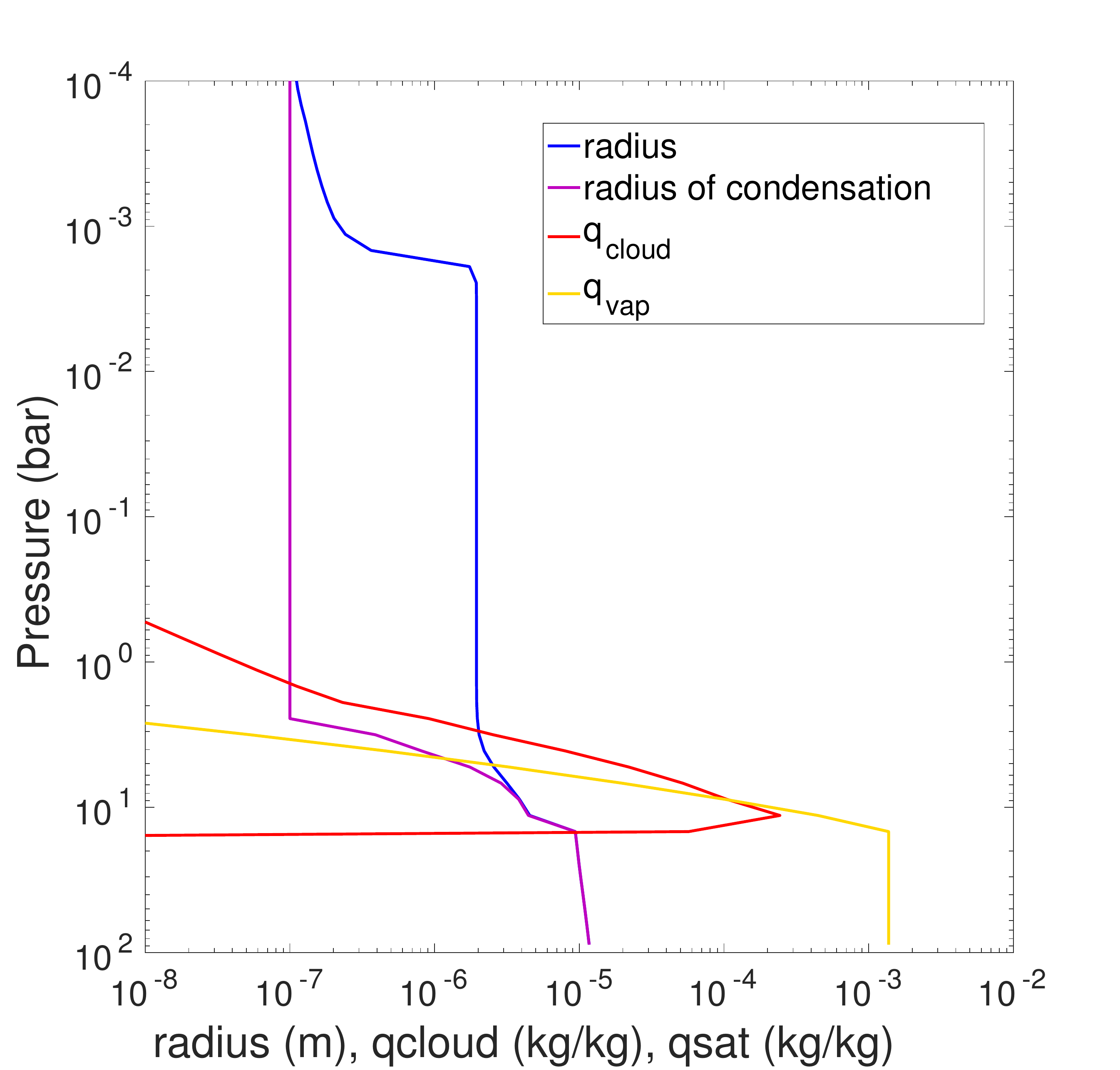}
\end{center}  
\caption{Evolution with pressure of characteristic timescales, cloud particle radii and condensate mixing ratio for iron clouds, for T$_{eff}$=1300 K and log(g)=5. The left panel shows the timescales of vertical mixing (blue), sedimentation (green) and coalescence (purple). The right panel shows the evolution of the mean radius (blue, in m), the radius of new condensed particles (purple, in m), the mass mixing ratio of condensate (red, in kg/kg) and vapour (yellow, in kg/kg). 
}
\label{figure_7}
\end{figure*}

Figure \ref{figure_7} shows the evolution of timescales (left panel) and cloud radius (right panel) with altitude for the same example.
Condensation growth is first limited by coalescence. It is after limited by vertical mixing for pressures lower than 4 bar down to 3$\times$10$^{-3}$ bar. Cloud particle size (blue curve in the right panel) initially corresponds to the size of new condensed particles (purple line). Progressively, advection dominates over condensation (i.e. $q_c \gg q_s$) and the particle radius evolves much less than the condensation radius. It becomes constant for pressure lower than 4 bars where advection dominates and the condensation radius reaches our limit value.

Finally, this model with simple microphysics is intermediate between the first two options (fixed radius and fixed $f_{sed}$). It computes cloud particle radii that are not arbitrarily fixed but estimated with physical arguments, and particle radii evolve with altitude only when condensation is not negligible.

\section{Results}

\subsection{Effects of clouds on temperature profiles, atmospheric composition and spectra}

\subsubsection{Temperature profiles}

\begin{figure}[!h] 
\begin{center} 
	\includegraphics[width=8.5cm]{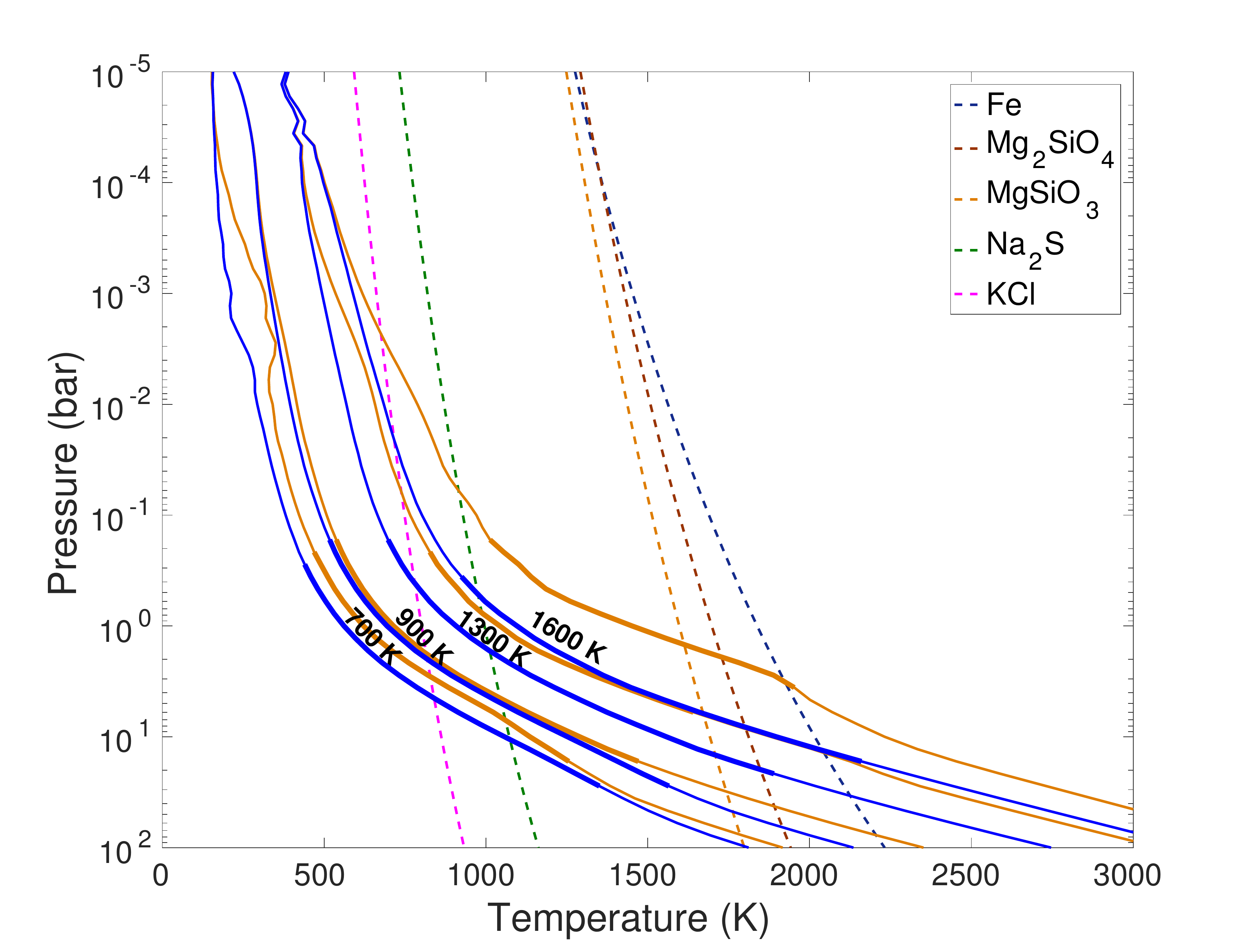}
\end{center}  
\caption{Temperature profiles with (orange) and without clouds (blue) computed for log(g)=5 and T$_{eff}$=700, 900, 1300 and 1600 K. Here, we used the cloud model with simple microphysics, with iron and silicate clouds, and also sulfide/salt clouds for T$_{eff}$=700 K. We assumed a supersaturation S=0.01. The thick curves represent the photosphere computed for wavelengths between 0.6 and 5 $\mu$m.}
\label{figure_8}
\end{figure}

Figure \ref{figure_8} shows the effect of clouds on temperature profiles, using Exo-REM with the simple cloud microphysics. Simulations were performed for a brown dwarf with log(g)=5 and T$_{eff}$=700, 900, 1300 and 1600 K. For T$_{eff}$=900, 1300 and 1600 K, we only included Fe and Mg$_2$SiO$_4$ clouds. For T$_{eff}$=700 K, we also included Na$_2$S and KCl clouds. For T$_{eff}$=1300 and T$_{eff}$=1600 K, iron and silicate clouds produce a strong warming ($\sim$200 K), which is visible for all pressures higher than 10$^{-3}$ bar. This warming is due to the greenhouse effect of clouds (absorption and backscattering) and is maximum close to the condensation level.
For T$_{eff}$=900 K, iron and silicate clouds condense below the photosphere and have almost no effect on the thermal structure.
For T$_{eff}$=700 K, Na$_2$S and KCl clouds condense in the photosphere and produce a moderate warming ($\sim$100 K).
In all cases, the photosphere goes up with the presence of clouds. This trend is due to the increase of atmospheric opacity by clouds. In addition, the maximal photosphere temperature (i.e. at the bottom of the photosphere) decreases while the minimal photosphere temperature (i.e. at top of the photosphere) increases.

\subsubsection{Atmospheric composition}

We can expect that the strong warming produced by clouds impacts atmospheric composition by shifting chemical equilibria. A warming should shift CO-CH$_4$ equilibrium toward CO and N$_2$-NH$_3$ toward N$_2$. Indeed, Figure \ref{figure_annexe2} in the Appendix and Figure \ref{figure_9}  show a decrease of the abundances of CH$_4$ and NH$_3$ with the presence of clouds. This effect is particularly strong concerning CH$_4$ for T$_{eff} >$1200 K, corresponding to L dwarfs with thick iron and silicate clouds. We found that the effect is even stronger for low gravity, since silicate and iron clouds are optically thicker and disappear at lower temperature. For T$_{eff}$=1300 K, the presence of clouds reduces CH$_4$ abundance by a factor of 2.5 for log(g)=5 and by a factor of 10 for log(g)=4. 
According to Figure \ref{figure_9}, the impact of gravity on non-equilibrium chemistry remains the dominant effect leading to methane depletion for low-gravity objects. The presence of clouds tends to significantly enhance this effect. Our model therefore predicts a strong methane depletion for cloudy low-gravity objects compared to field brown dwarfs of similar T$_{eff}$. This may help to explain the apparent methane depletion for the young giant exoplanets 2M1207b and HR8799bcd \citep{currie11, barman11, barman11b, konopacky13}, with effective temperature $\sim$1000-1200, while brown dwarfs of similar temperature show strong methane bands.
For NH$_3$, the depletion caused by clouds is moderate, generally a factor of 2-3 for low and high gravity (see Figure \ref{figure_annexe2} in the Appendix). Finally, with radiatively active clouds, the disappearance of TiO, VO, atomic Na and K (through respectively TiO, VO and Na$_2$S condensation and KCl formation) occurs higher in the atmosphere.

\begin{figure}[h] 
\begin{center} 
	\includegraphics[width=8.5cm]{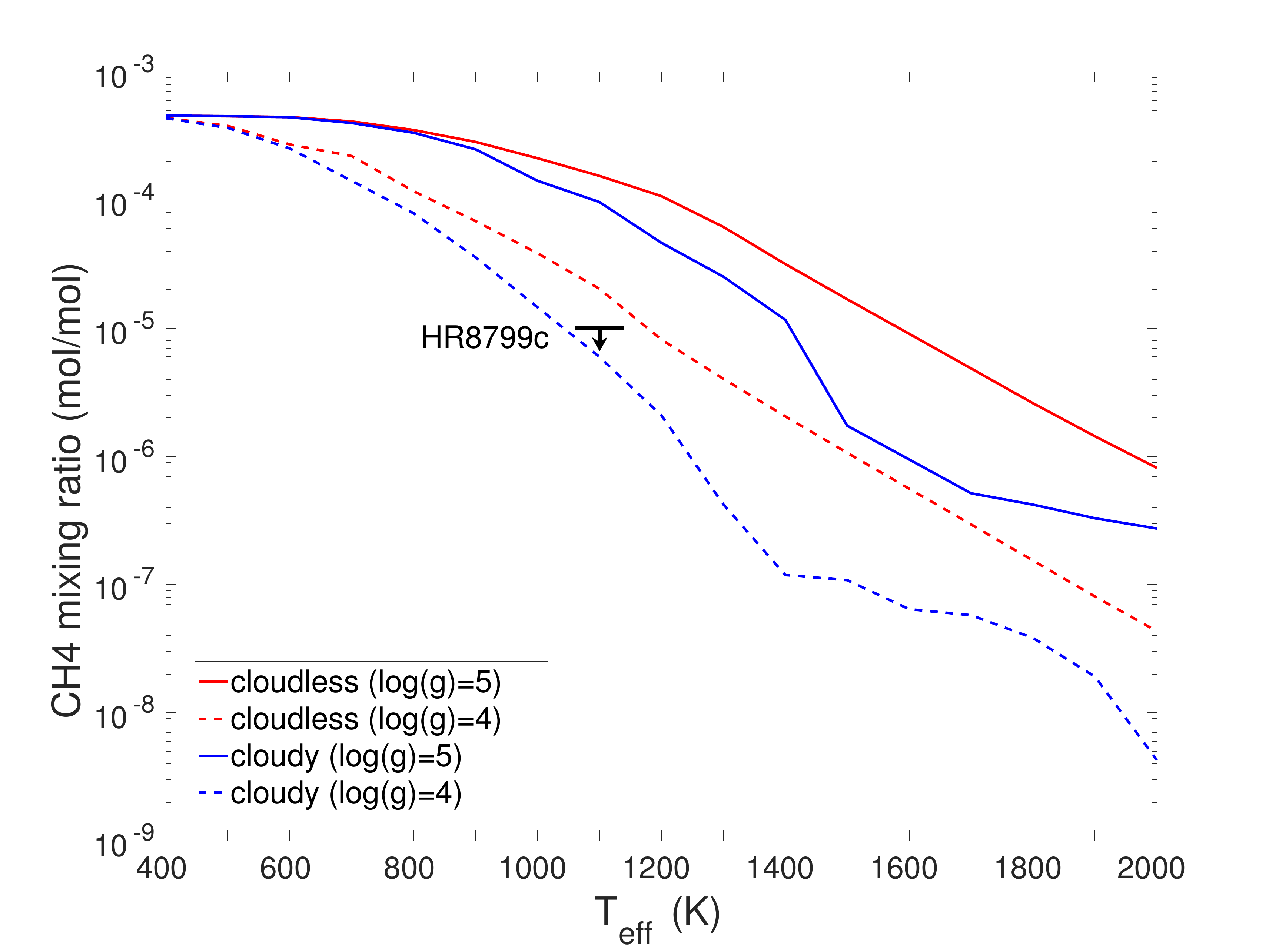}
\end{center}  
\caption{CH$_{4}$ mixing ratio in the upper atmosphere as a function of effective temperature for log(g)= 5 (solid line) and log(g)=4 (dashed line) without (red) or with clouds (blue), computed with non-equilibrium chemistry. The black line indicate the upper CH$_4$ abundance retrieved by \cite{konopacky13} for HR8799c.
}
\label{figure_9}
\end{figure}

\subsubsection{Spectra}

\begin{figure*}[!h] 
\begin{center} 
	\includegraphics[width=8.5cm]{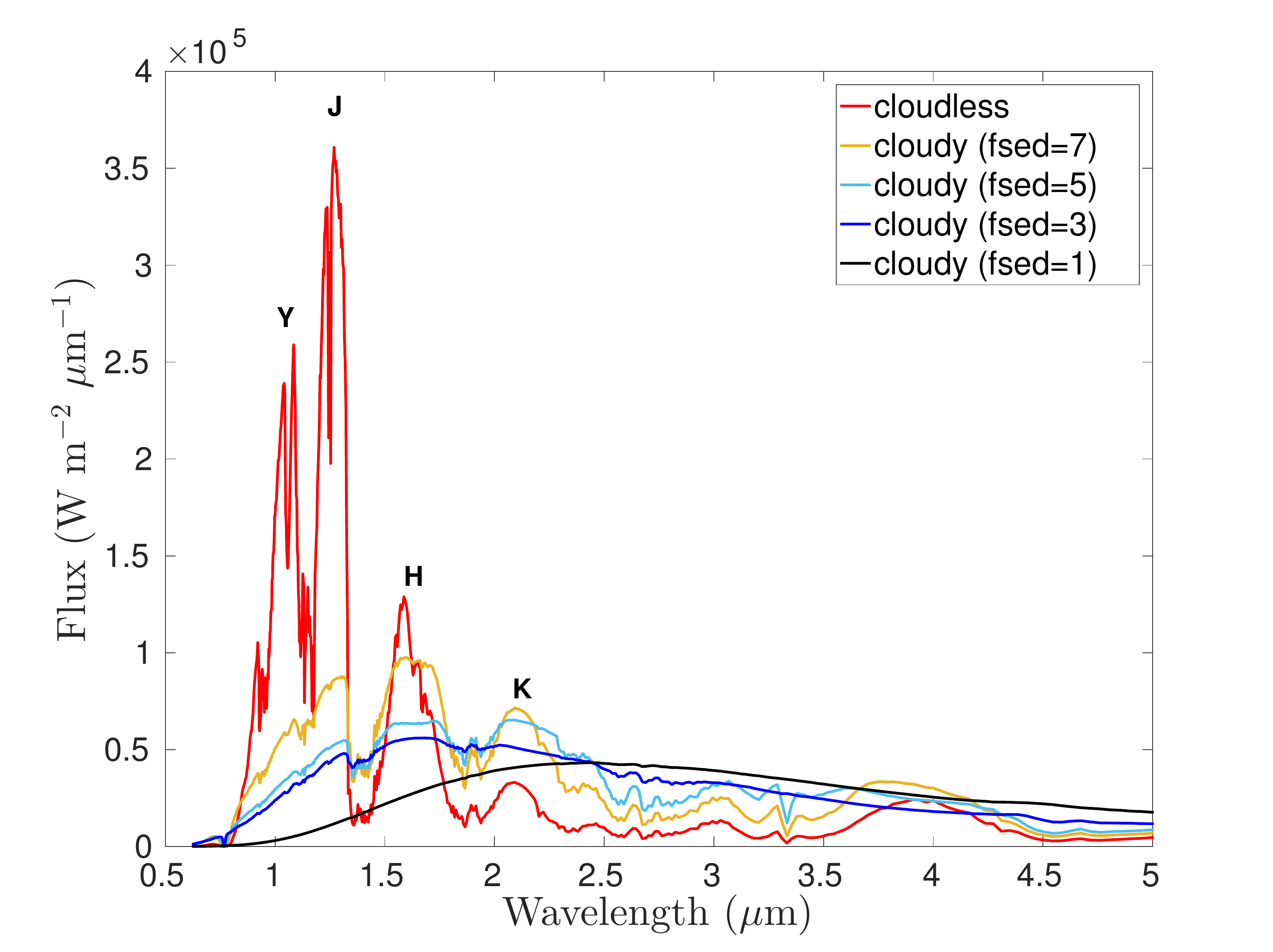}
	\includegraphics[width=8.5cm]{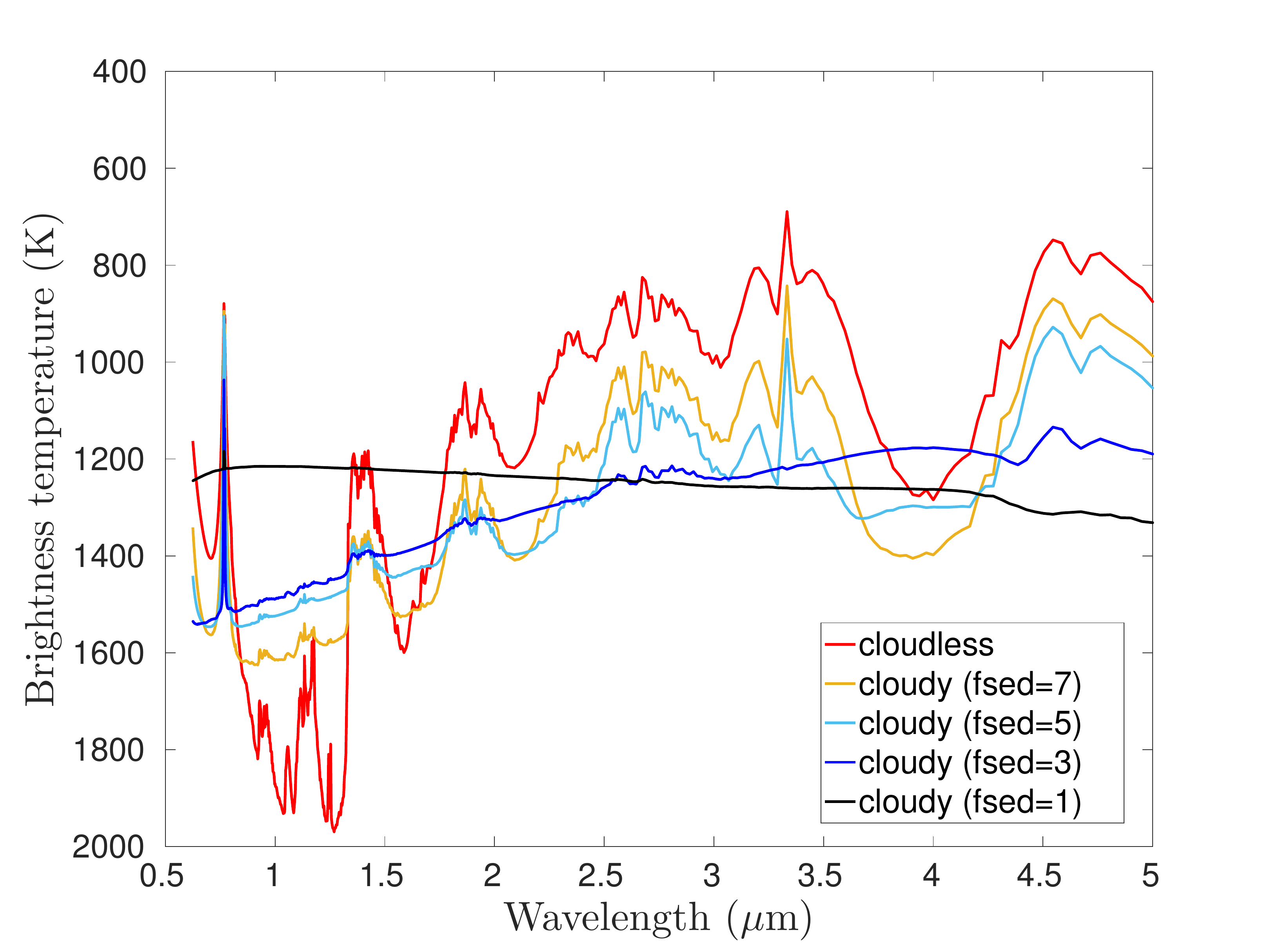}
\end{center}  
\caption{Emission spectra (left) and brightness temperature (right) computed for log(g)=5 and T$_{eff}$=1300 K. The cloudless case is shown in red. Cloudy cases are computed with $f_{sed}$=1-5.}
\label{figure_10}
\end{figure*}

Figure \ref{figure_10} shows the effect of clouds on emission spectra (left panels) and brightness temperatures (right panels) from 0.5 to 5 $\mu m$, using Exo-REM with fixed $f_{sed}$ (from 1 to 7), and for log(g)=5 and T$_{eff}$=1300 K. When clouds are included, thermal emission flux from atmospheric spectral windows at short wavelengths is reduced, while thermal emission at long wavelengths or from atmospheric bands is increased. In addition, variations of the brightness temperature with wavelength are reduced, as expected according to the change of the photosphere extension (see previous paragraph).

With $f_{sed}$=1, clouds are optically thick. The spectrum becomes very close to a blackbody, and the brightness temperature is almost constant and equal to 1300 K. For $f_{sed}$=1, the brightness temperature slightly decreases with wavelength. The object would thus appear even redder than a blackbody at 1300 K. This is due to Mie scattering and to the iron imaginary optical index, which make cloud optically thicker at short wavelengths.

\begin{figure*}[!h] 
\begin{center} 
	\includegraphics[width=8cm]{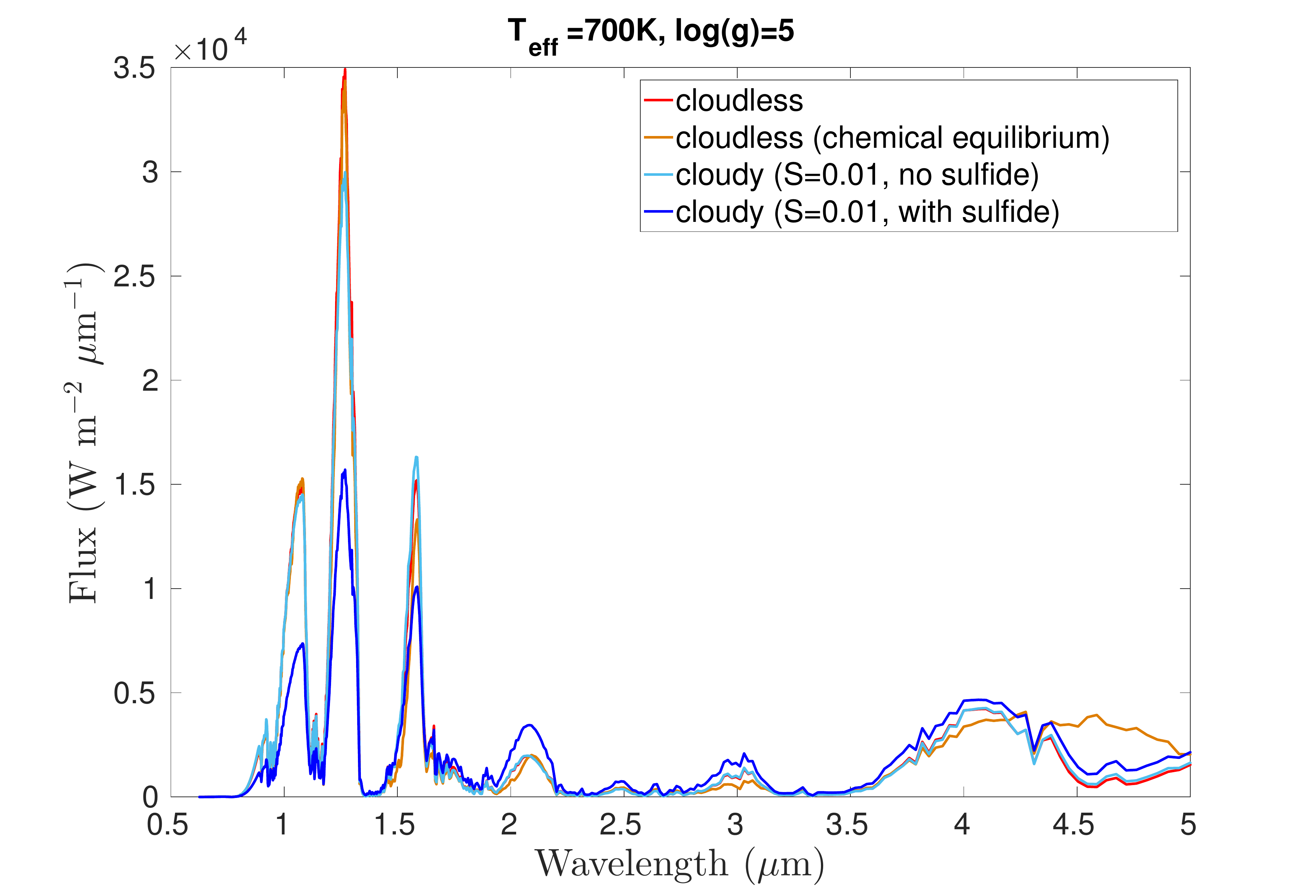}
	\includegraphics[width=8cm]{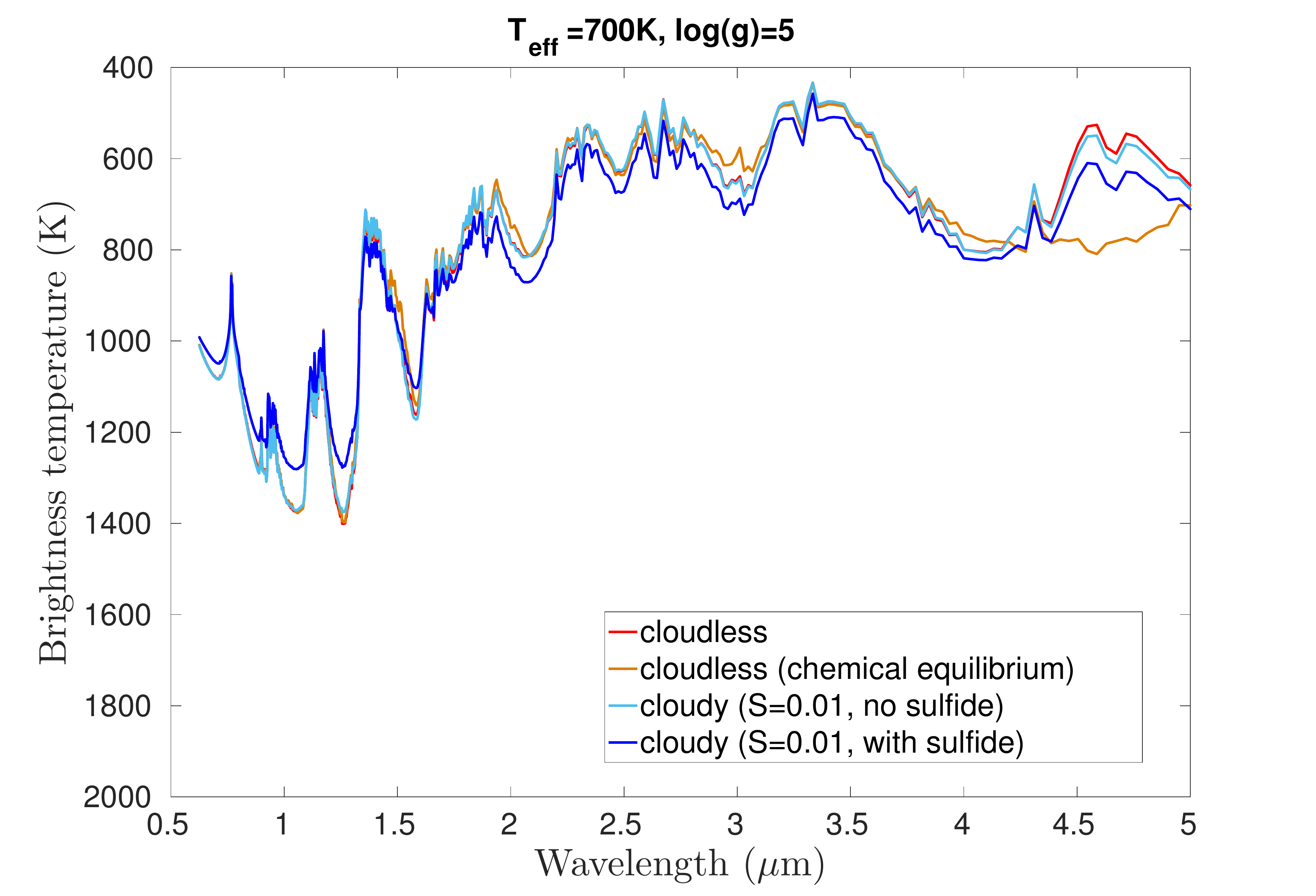}
	\includegraphics[width=8cm]{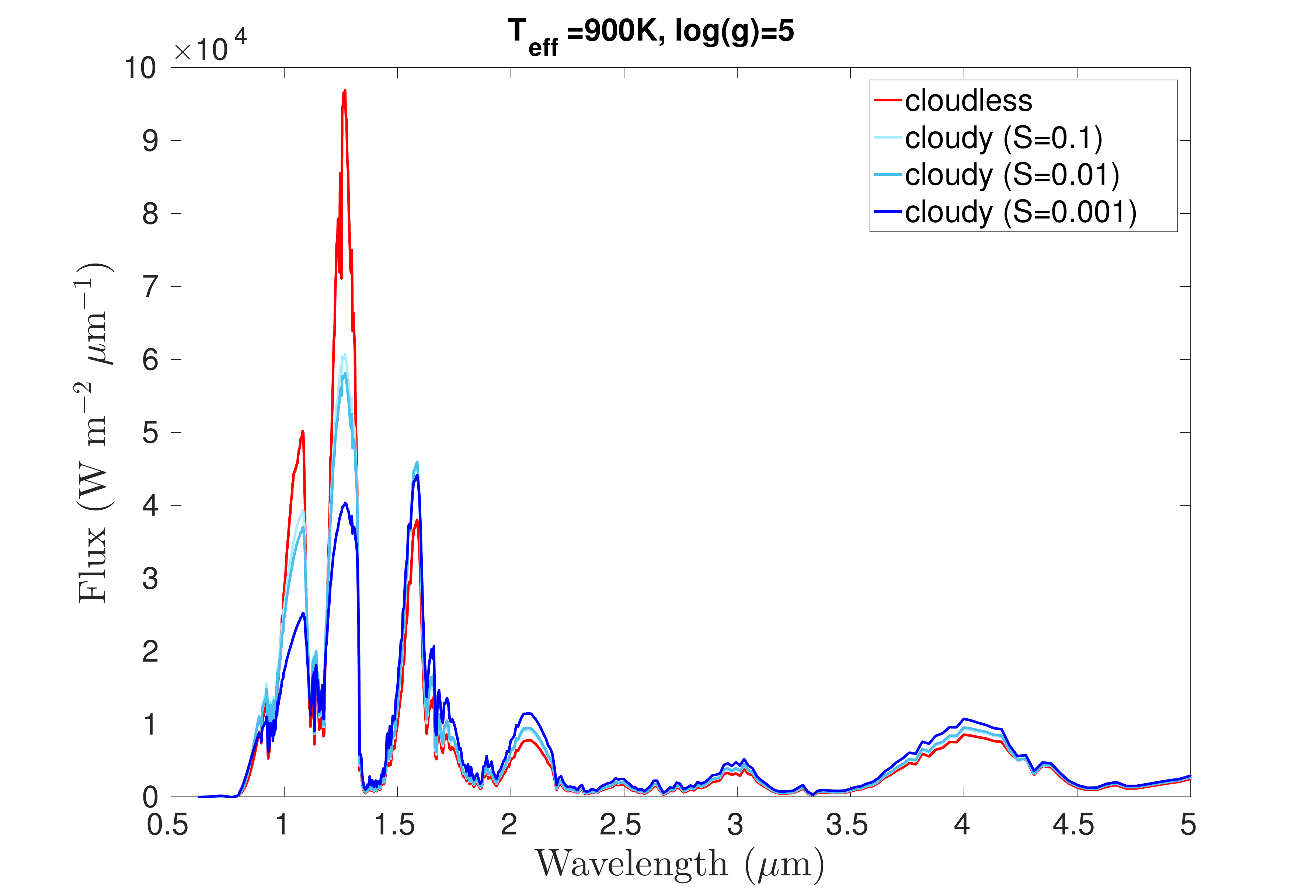}
	\includegraphics[width=8cm]{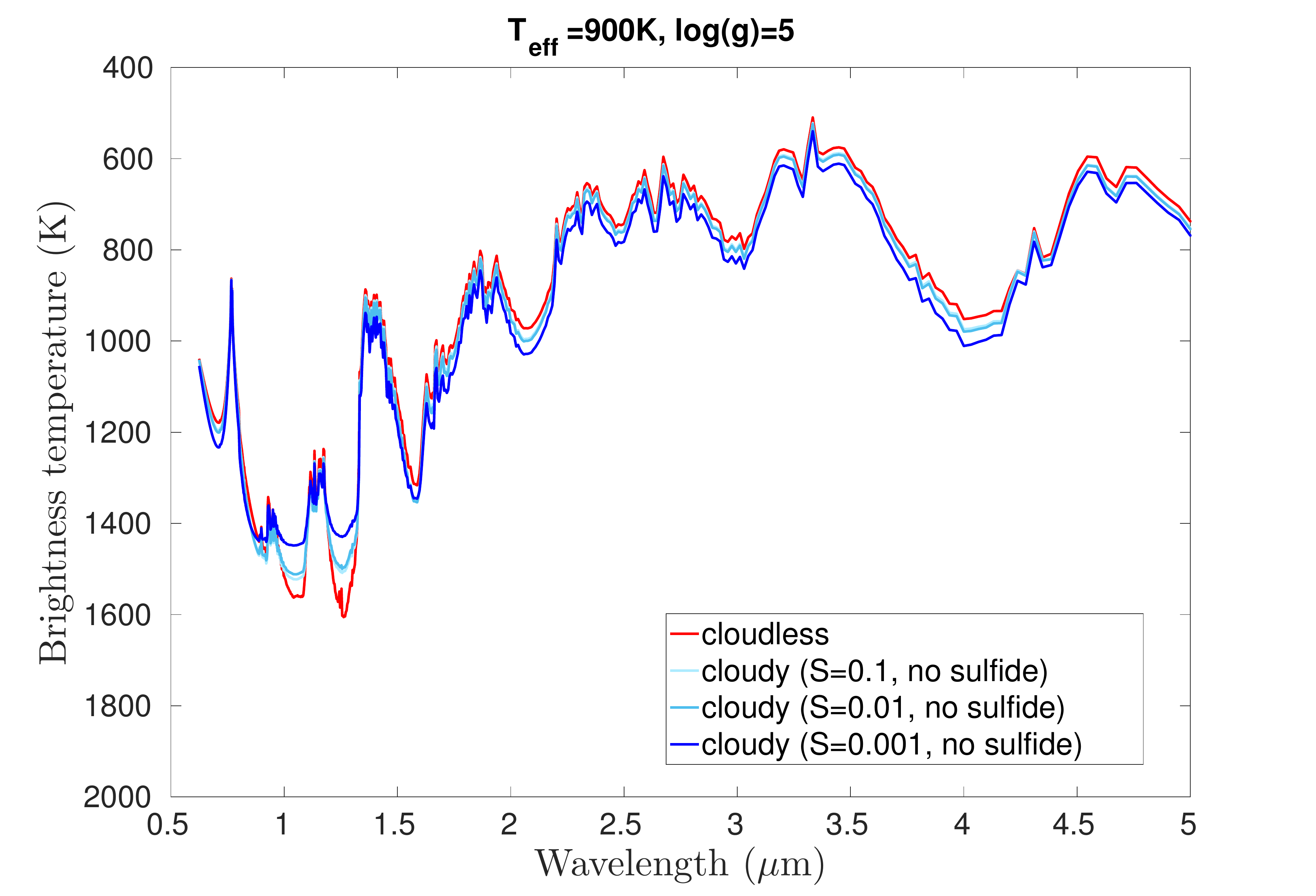}
	\includegraphics[width=8cm]{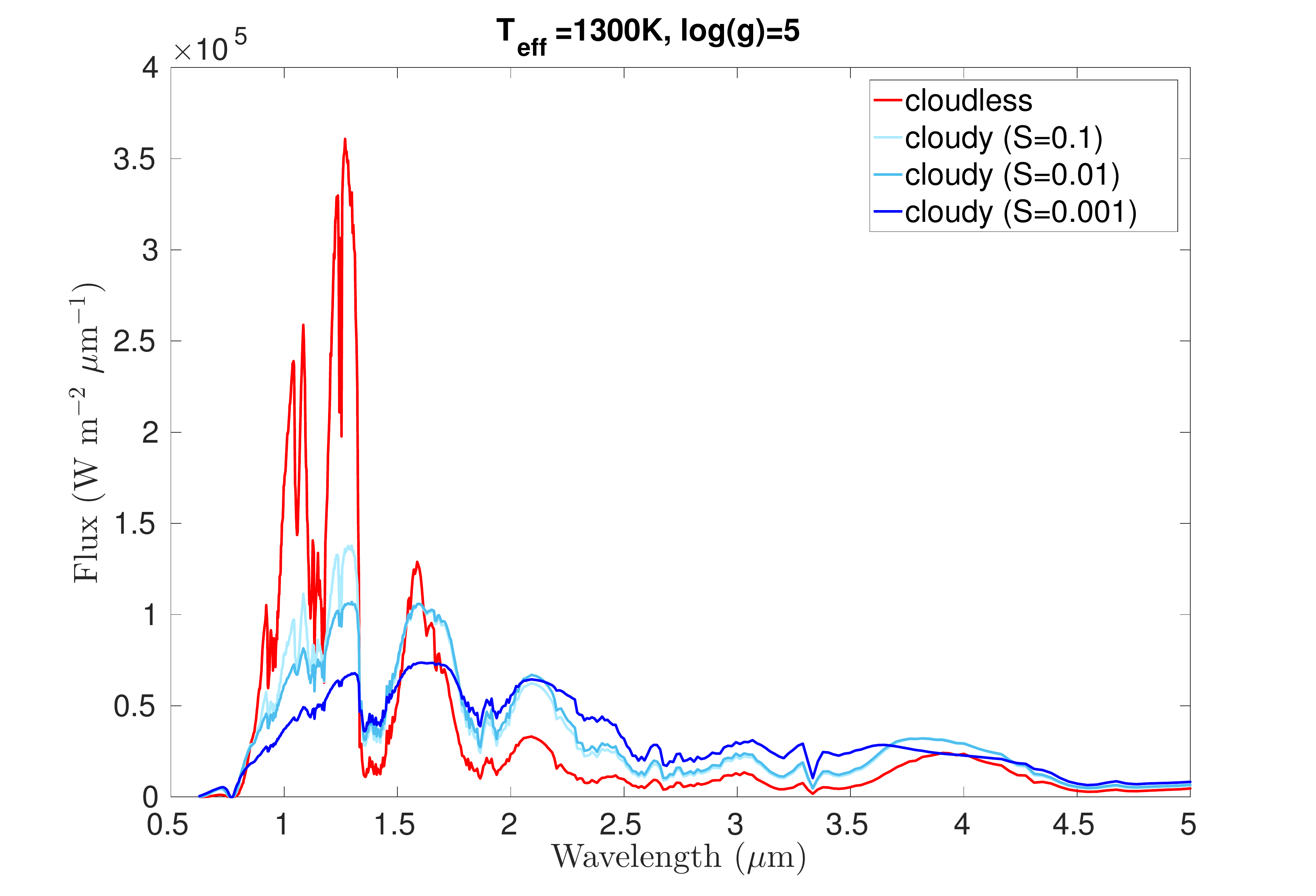}
	\includegraphics[width=8cm]{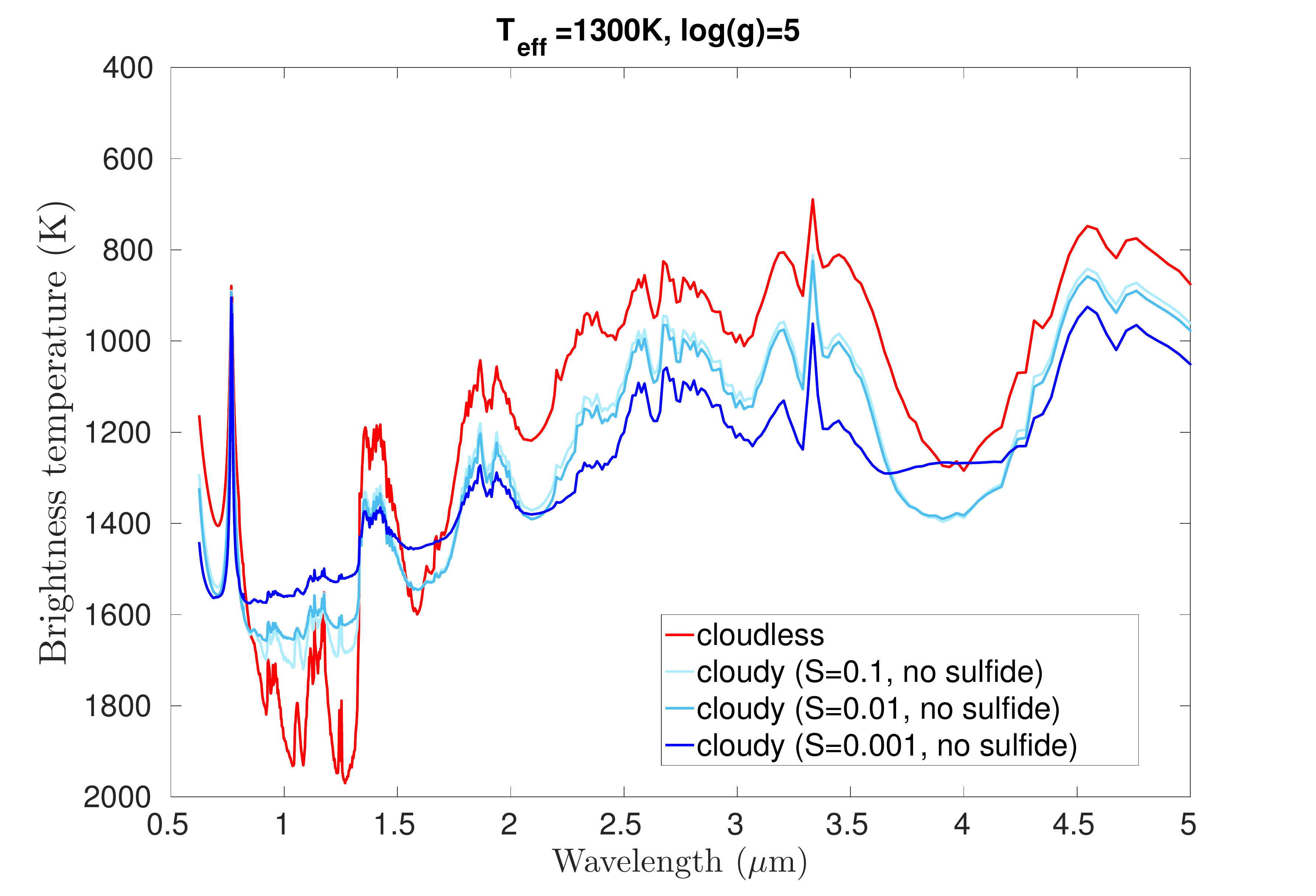}
	\includegraphics[width=8cm]{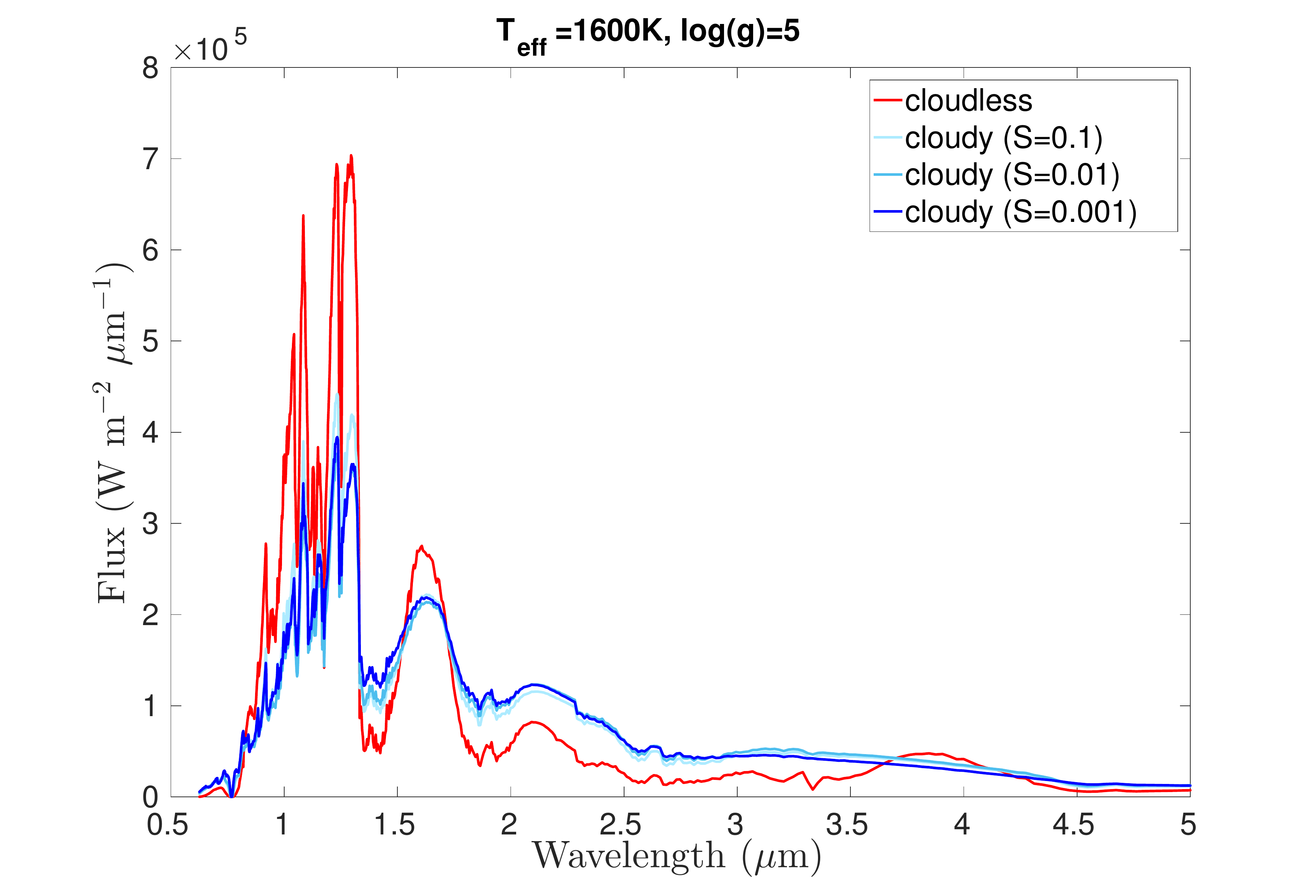}
	\includegraphics[width=8cm]{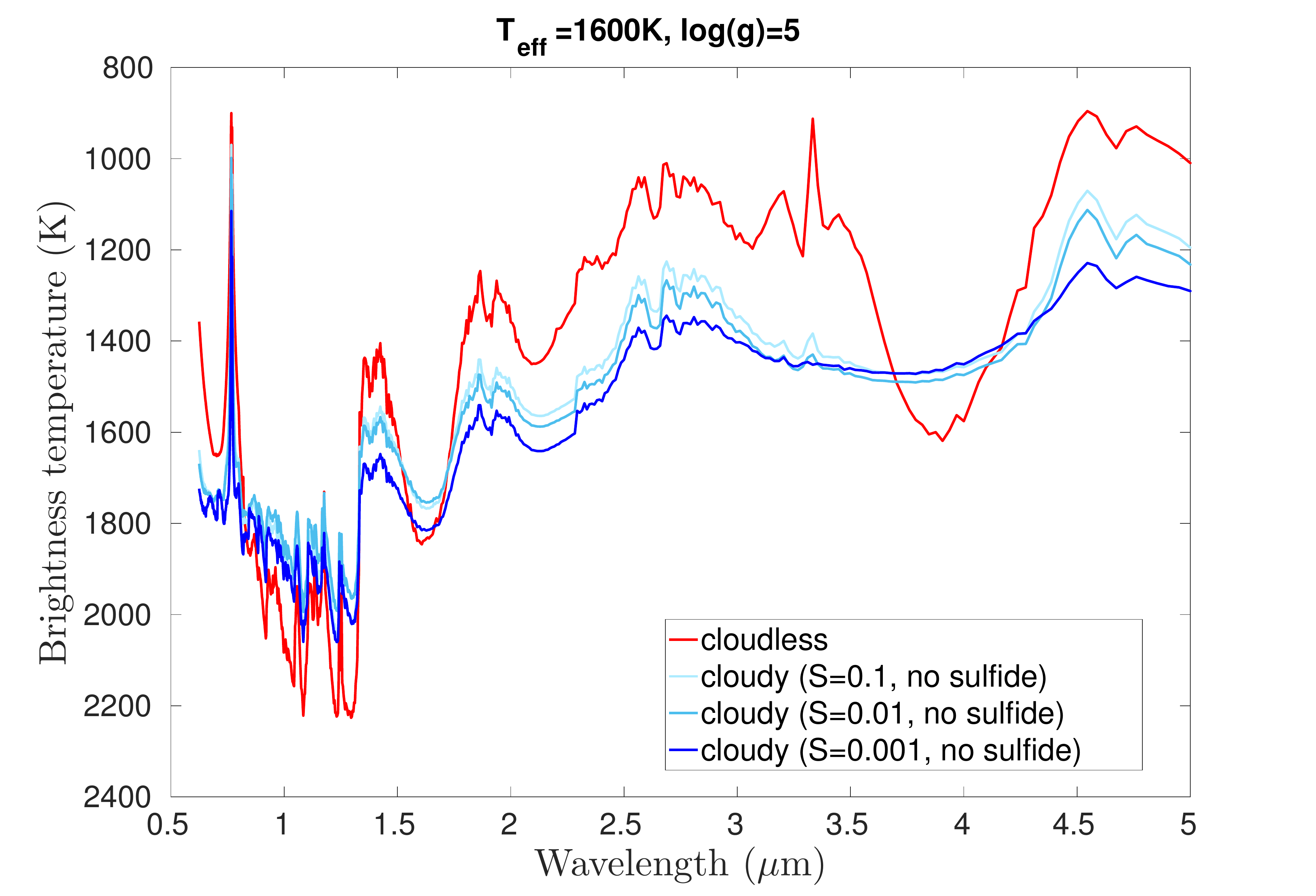}
\end{center}  
\caption{Emission spectra (left) and brightness temperature (right) computed for log(g)=5 and T$_{eff}$=700, 900, 1300 and 1600 K from top to bottom. For each effective temperature, the cloudless case is shown in red. Here, we used the cloud model with simple microphysics, with iron and silicate clouds and a supersaturation S=0.1, 0.01 and 0.001. For top panels (T$_{eff}$=700 K), the case with Na$_2$S and KCl clouds is shown in blue.}
\label{figure_11}
\end{figure*}

Figure \ref{figure_11} is similar to Figure \ref{figure_10}, but with the simple microphysics and for T$_{eff}$=700, 900, 1300 and 1600 K. Simulations were performed with Fe and Mg$_2$SiO$_4$ clouds at all temperatures and also Na$_2$S and KCl clouds for the coldest case (T$_{eff}$=700 K). As for the thermal structure (Figure \ref{figure_8}), the effect of clouds is maximal for L dwarfs (i.e. T$_{eff}$=1300-1600 K).
For the case at 700 K, we also included a simulation assuming chemical equilibrium and with no cloud. For this effective temperature, the presence/absence of silicate and iron clouds has a lower impact on spectra than chemical equilibrium/disequilibrium. In particular, the CO abundance in the photosphere is increased by a factor of $10^5$ when chemical disequilibrium is taken into account, leading to a strong reduction of the emitted flux at $\sim$4.5 $\mu m$.

\begin{figure}[!h] 
\begin{center} 
	\includegraphics[width=8.5cm]{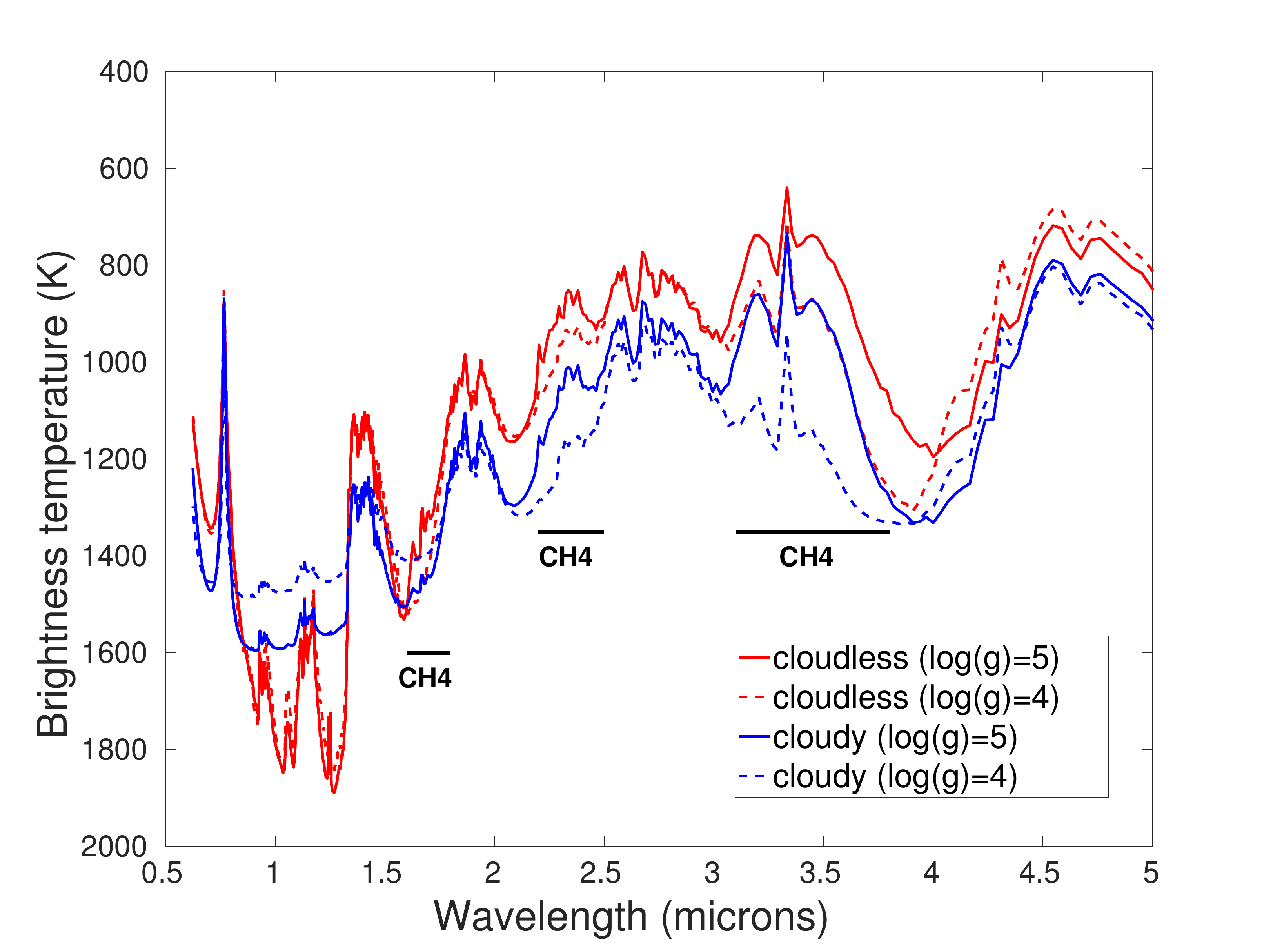}
\end{center}  
\caption{Brightness temperature computed for log(g)=4 (dashed lines) 5 (solid lines) and T$_{eff}$=1200 K. For each case, the cloudless case is shown in red and the cloudy case is shown in blue. Here, we used the cloud model with simple microphysics with S=0.01. Methane bands are indicated with black lines.
}
\label{figure_12}
\end{figure}

CH$_4$ has strong absorption bands at 1.7, 2.3 and 3.3 $\mu$m. The methane depletion due to cloud greenhouse effect and described in the previous section, is noticeable in spectra and brightness temperature with T$_{eff}$$>$1100 K in Figures \ref{figure_10} and \ref{figure_11}. The increase of the cloud thickness (by a lower $f_{sed}$ or a lower supersaturation S) leads to a decrease of the CH$_4$ absorption band depth compared to the surrounding water bands. In addition, the depth of the K line at 0.75 $\mu$m increases relatively to other absorption bands with the presence of clouds. Figure \ref{figure_12} illustrates the stronger methane depletion for cloudy atmospheres with low surface gravity.

\subsubsection{Case of inhomogeneous cloud cover}

Our model is 1D and simulate a homogeneous atmospheric column. In order to simulate patchy clouds, we included the parametrisation from \cite{marley10} and \cite{morley14}. For such cases, the atmosphere is supposed to possess a clear part with no cloud, and a cloudy part, corresponding to a surface fraction $f$. We also assume that the thermal structure is the same in both parts, meaning that the horizontal heat redistribution is very efficient. For each iteration, we run radiative transfer for both atmospheric columns and compute flux for the clear part (F$_{\rm clear}$) and for the cloudy part (F$_{\rm clear}$). The total flux is then: 
\begin{equation} 
\rm F_{total}=(1-\it f) \rm F_{clear} + \it f  \rm F_{cloudy}
\end{equation} 

\begin{figure}[!h] 
\begin{center} 
	\includegraphics[width=8.5cm]{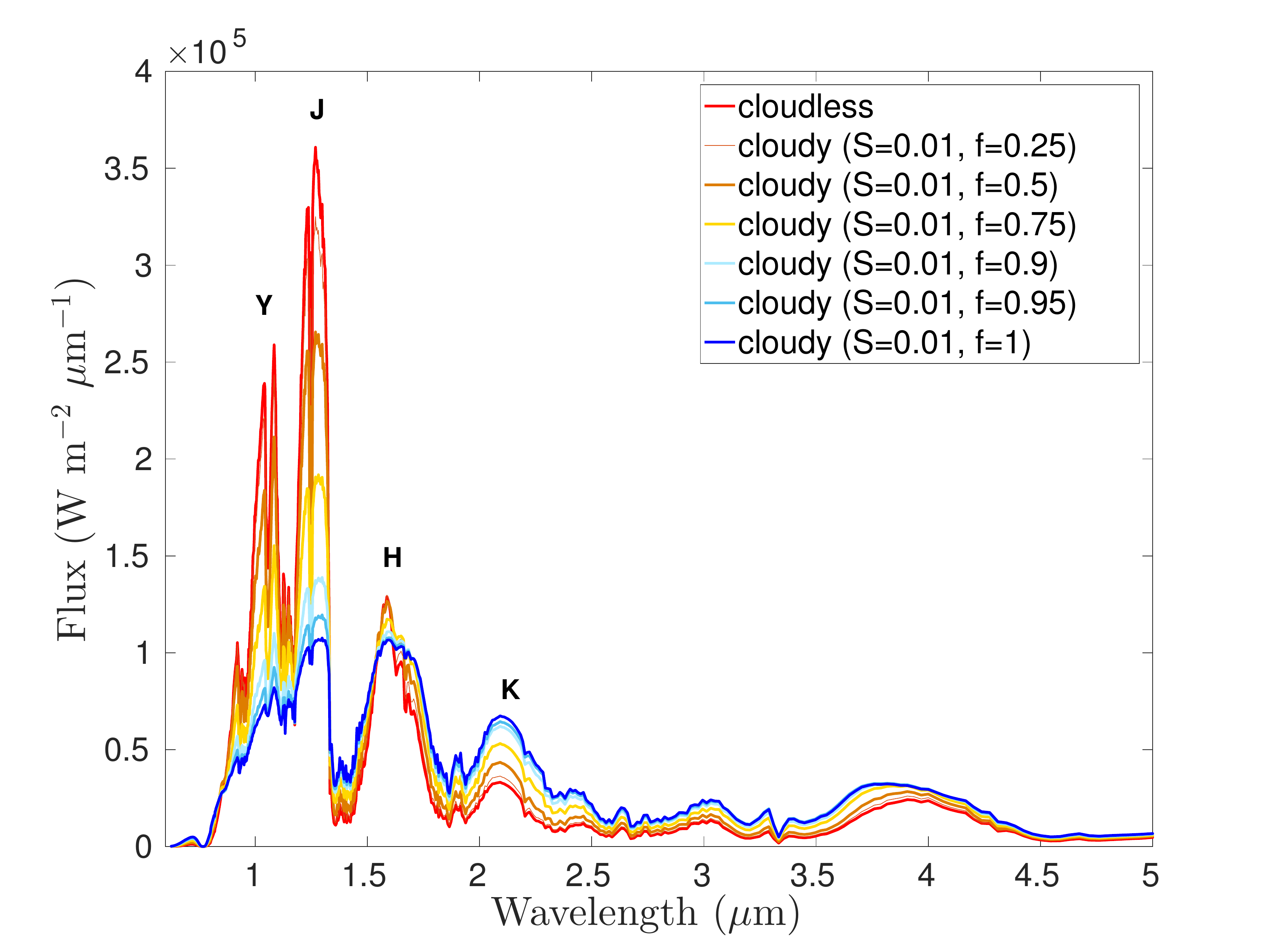}
	\includegraphics[width=8.5cm]{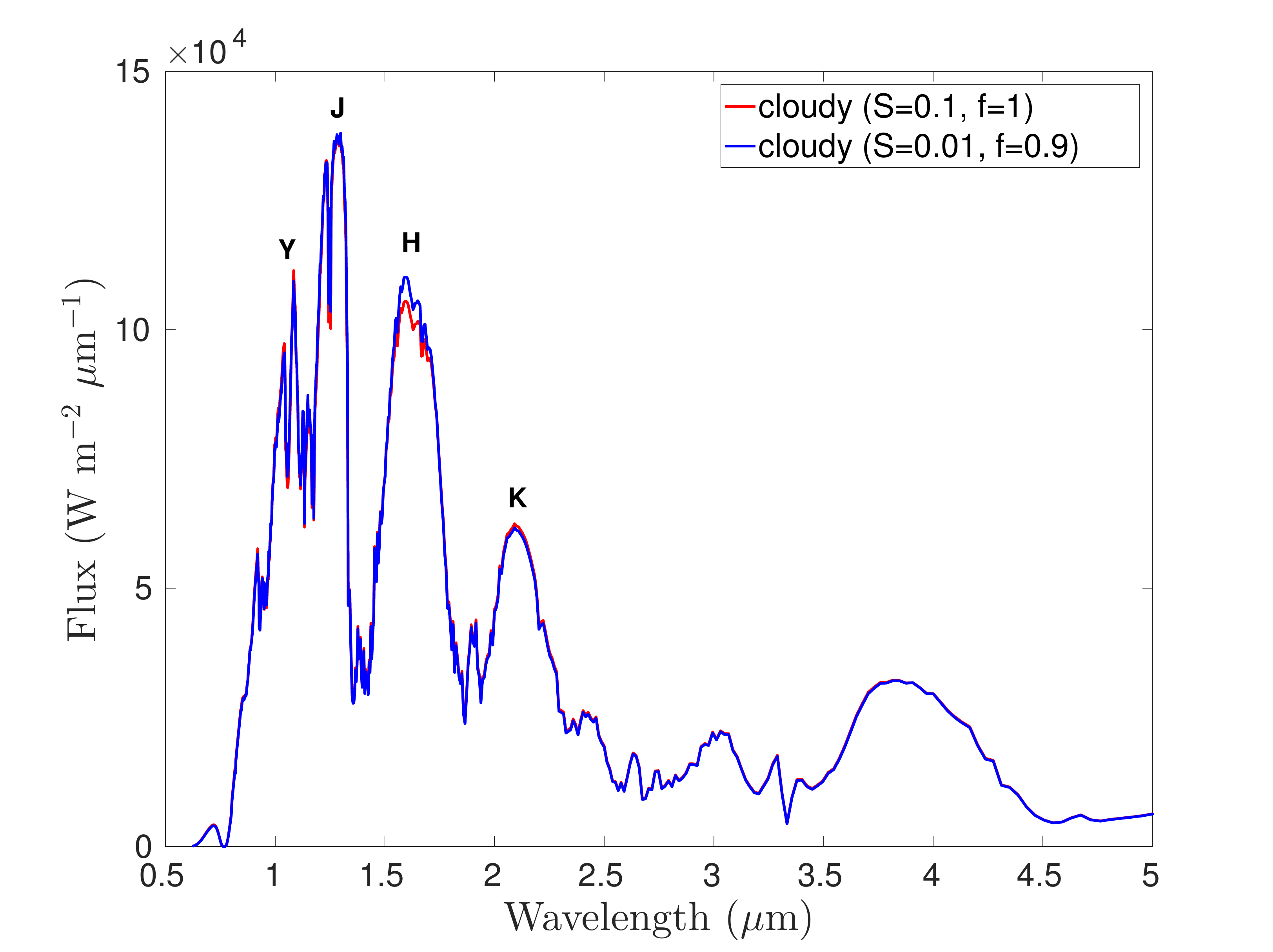}
\end{center}  
\caption{Emission spectra with cloud cover from 0 to 100$\%$ (parameter f), for log(g)=5 and T$_{eff}$=1300 K. We used the cloud model with simple microphysics, with iron and silicate clouds and a supersaturation S=0.01 in the top panel. In the bottom panel, we used S=0.1 with 100$\%$ cloud cover and S=0.01 with 90$\%$ cloud cover.
}
\label{figure_13}
\end{figure}

The top panel in Figure \ref{figure_13} shows the effect of increasing the cloud cover, which is similar to making clouds optically thicker. The emission spectrum progressively passes from that of a cloud-free atmosphere to a fully cloudy atmosphere. The bottom panel in Figure \ref{figure_13} shows spectra with a fully cloudy atmosphere with thin clouds ($f$=1 and $S$=0.1) compared to a partially cloudy atmosphere with thicker clouds ($f$=0.9 and $S$=0.01). These spectra are almost identical (there is just a small difference in H band) leading to potentially degenerated fitting solutions. 

For objects covered by water clouds as Y dwarfs, we found that an inhomogeneous cloud cover generally is necessary to allow model convergence (see also Morley et al. 2014). Water clouds can be extremely opaque for $T_{eff}$$<$400 K and all surface gravity, producing strong warming and triggering numerical instabilities. A 50$\%$-cloud cover generally is sufficient to avoid such instabilities.

\subsection{Comparison with brown dwarf photometry}

In this section, we compare our cloud models to photometric observations of brown dwarfs and young giant exoplanets, to investigate 1) the L-T transition, 2) the apparent reddening of exoplanets and low-gravity brown dwarfs and 3) colors of T dwarfs.

For studying the evolution of colors and magnitudes of brown dwarfs and exoplanets, we produced grids of spectra with Exo-REM, in effective temperature (T$_{eff}$=400-2000 K, step=100 K), gravity (log(g)=3-5, step=0.1) and metallicity (M=-0.5, 0, +0.5; in log scale). We used the three options for cloud modelling (fixed $f_{sed}$, fixed radius and simple microphysics). From these spectra, we computed colors in bands J, H and K (MKO filters). We derived the absolute magnitude in J band using radii of brown dwarfs from the Ames-Dusty evolutionary model. We interpolated the points from Ames-Dusty grid into our temperature-gravity grid. 
For our range of temperature and gravity, the radius is $\sim$1 R$_{J}$ excepted for objects with high temperature and low gravity, for which it can exceed 4  R$_{J}$. 
The choice of the evolution model has a relatively small impact in a color-magnitude diagram. For brown dwarfs, we found very similar results using Ames-Dusty, Ames-Cond and power-law relations from \cite{burrows01}. Evolution models have discrepancies mostly for young giant planets, but these uncertainties should not modify our conclusions.

In the next three sections, we only consider the effect of iron and silicate clouds with 100$\%$ cloud covering and use the three options for computing the cloud particle radii. We discuss the effect of sulfide and salt clouds with inhomogeneous cloud cover in section 3.2.4 for T dwarfs.

\subsubsection{Fixed radius}

\begin{figure*}[!h] 
\begin{center} 
    \textbf{Color-magnitude diagrams with fixed radius}\par
	\includegraphics[width=8.5cm]{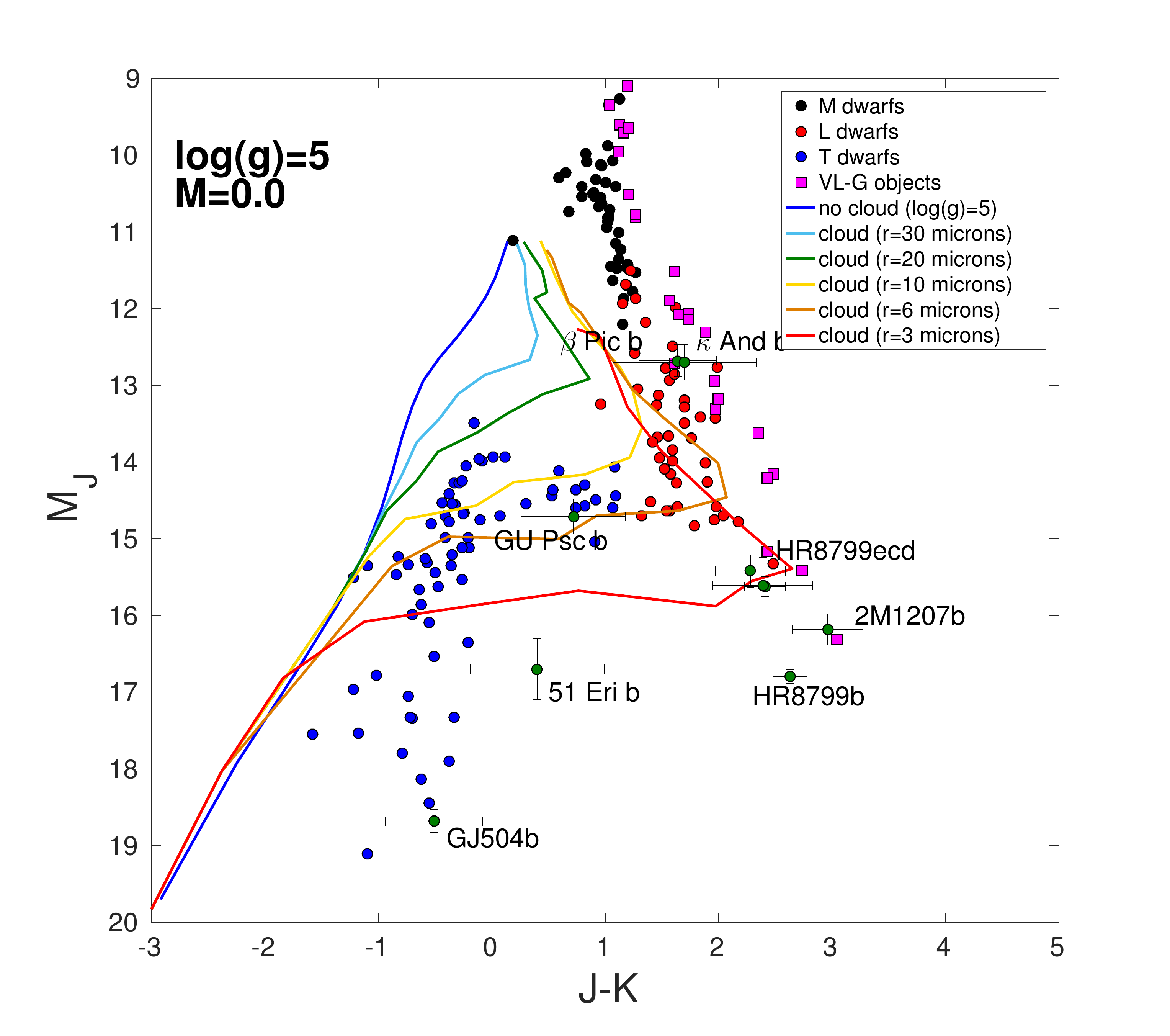}
	\includegraphics[width=8.5cm]{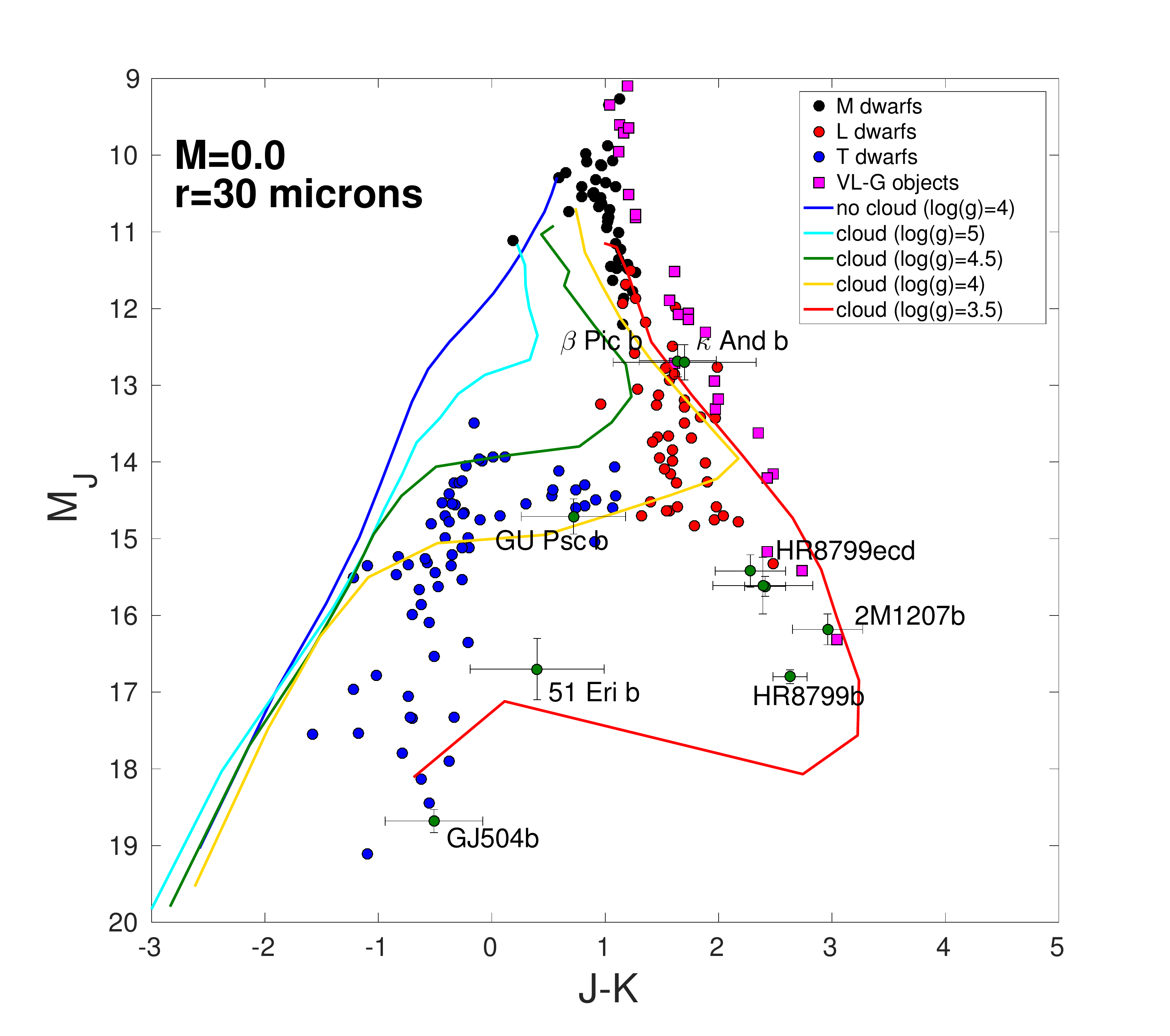}
\end{center}  
\caption{Same format as Figure \ref{figure_1} with fixed radius. For the left panel, curves are computed with Exo-REM, with log(g)=5 and particle radii equal to 3, 6, 10, 20 and 30 microns. For the right panel, curves are computed with 30 micron particle radii and log(g)=5, 4.5, 4, 3.5. M=[Fe/H] is the metallicity (in log scale). M=0 corresponds to the solar metallicity.}
\label{figure_14}
\end{figure*}

Figure \ref{figure_14} shows color-magnitude diagrams in J magnitude and J-K color assuming fixed values of particle radius. The left panel shows the effect of radius for log(g)=5 and the right panel the effect of log(g) for r=30 $\mu$m. The photometric curves are extremely sensitive to the value of particle radius (left panel). The size of particles impacts both cloud vertical distribution and optical thickness. With large particles (e.g. 30 $\mu$m), clouds are optically thin and colors are close to the cloudless case. With smaller particles, the branch of M and L dwarfs can be well reproduced and the colors of HR8799cde can be reproduced with 3 micron particles. For log(g)=4, the colors of HR8799bcde and 2M1207b can be reproduced with 10-20 micron particles. These particle sizes are lower than in the model of \cite{madhusudhan11} who obtained a best fit with particle size of 60 $\mu$m, but this model does not solve cloud vertical distribution self-consistently.
The colors of brown dwarfs at the L-T transition are reproduced with 6 $\mu$m particles for log(g)=5 (left panel) and with 30 $\mu$m particles for log(g)=4 (right panel). The model produces a sharp L-T transition, which is partly due to the evolution of the convective region. Indeed, the vertical extent of the convective region decreases when T$_{eff}$ is reduced, passing below the photosphere at T$_{eff}$$\sim$1200 K for log(g)=5 and at T$_{eff}$$\sim$1100 K for log(g)=4, similarly to the silicate condensation level. This evolution of the vertical mixing enhances the L-T transition for such a cloud parametrization. When the size of particle is fixed equal to 30 $\mu$m (right panel of Figure \ref{figure_14}), a lower surface gravity leads to redder colors, consistent with observations of directly imaged exoplanets. With this cloud model, colors are extremely sensitive to gravity/particle size. Moreover, one single value of particle radius cannot reproduce both brown dwarfs and exoplanets, meaning that the size of cloud particle should depend on gravity and potentially also on effective temperature.

\subsubsection{Fixed $f_{sed}$}

\begin{figure*}[!h] 
\begin{center} 
    \textbf{Color-magnitude diagrams with fixed $f_{sed}$}\par
	\includegraphics[width=8.5cm]{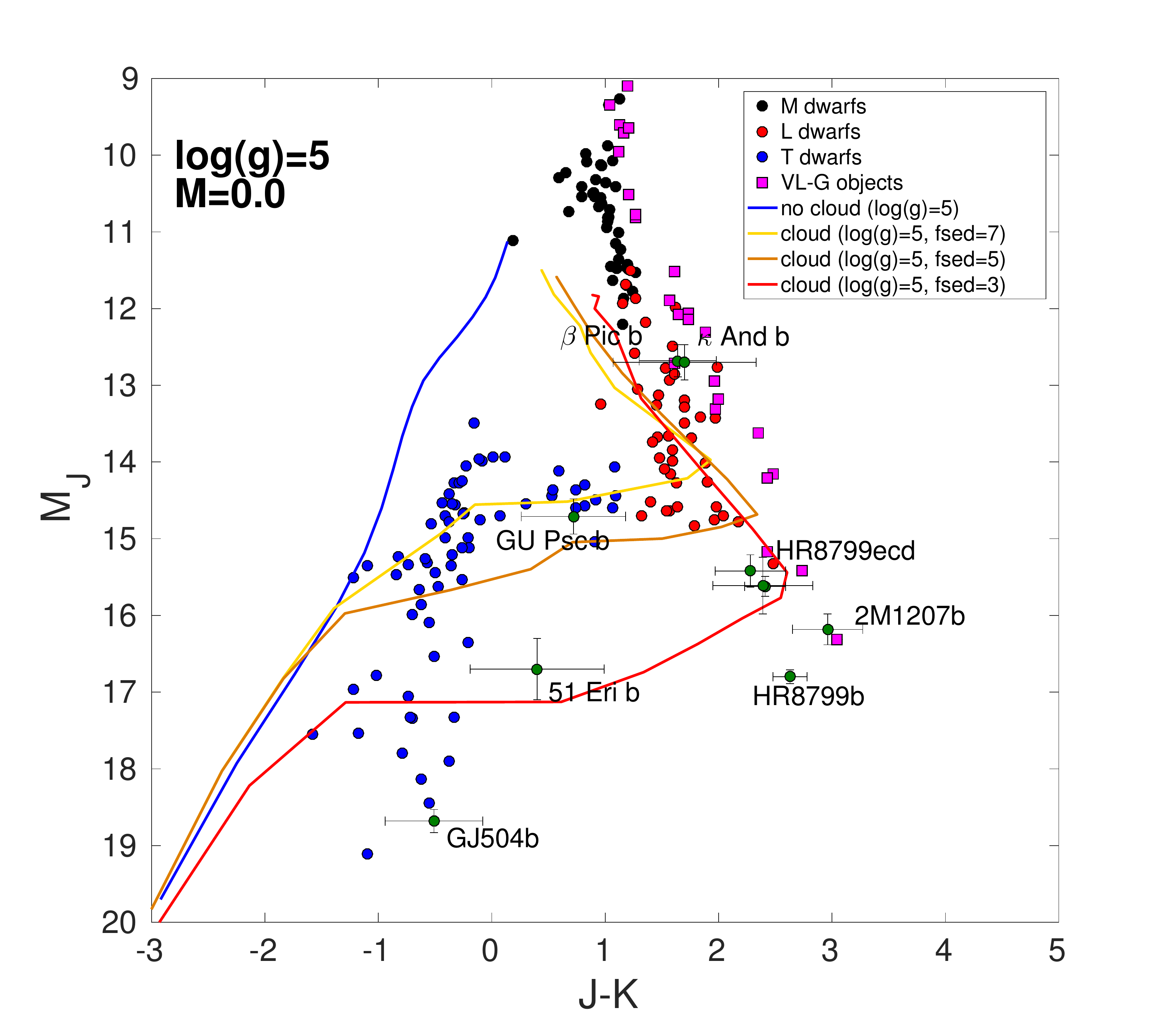}
	\includegraphics[width=8.5cm]{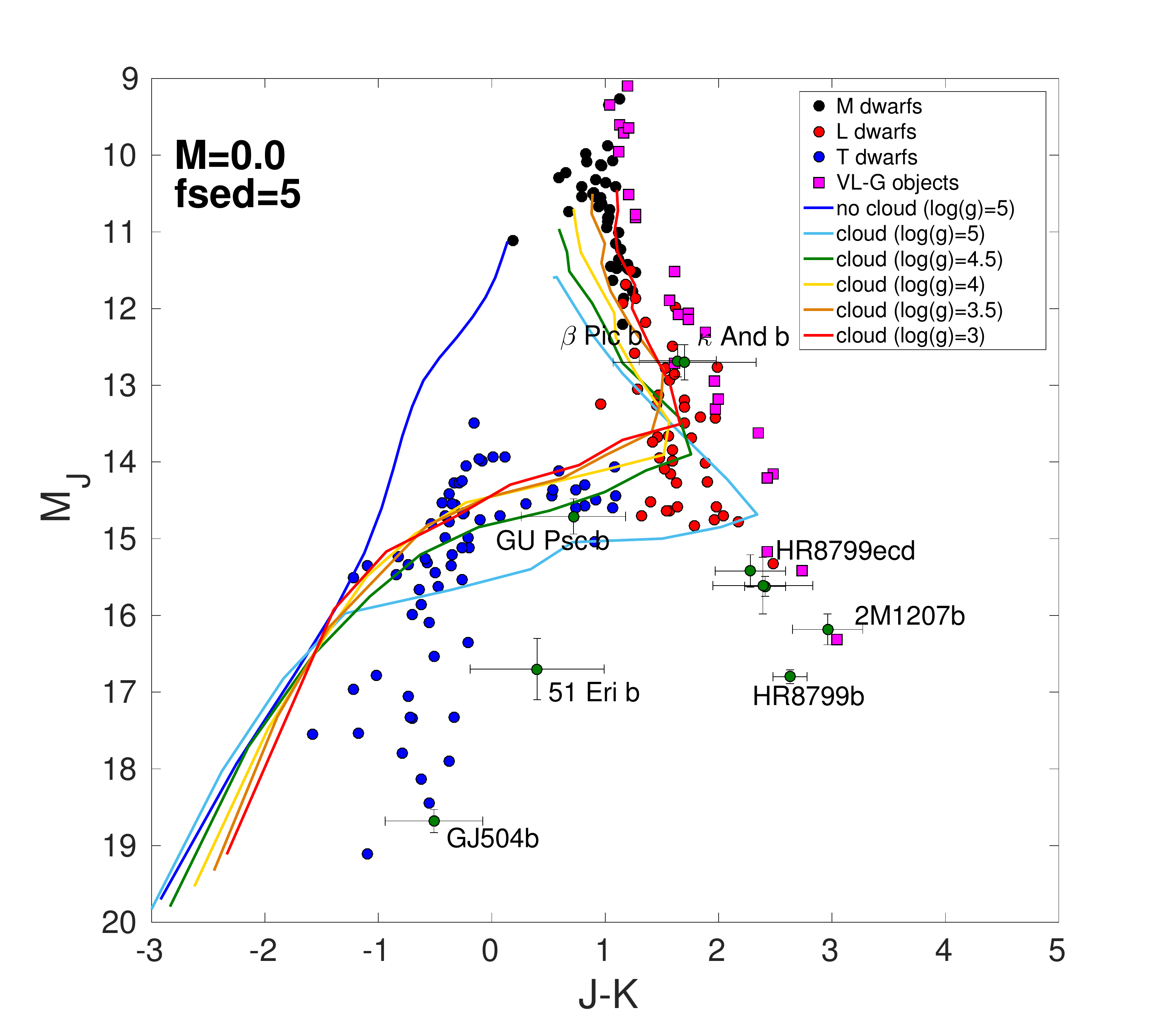}
\end{center}  
\caption{Same format as Figure \ref{figure_1} with fixed $f_{sed}$. For the left panel, curves are computed with Exo-REM, with log(g)=5 and $f_{sed}$=3 (red line), $f_{sed}$=5 (orange line) and $f_{sed}$=7 (yellow line). For the right panel, curves are computed with $f_{sed}$=5 and log(g)=5 (light blue line), 4.5 (green line), 4 (yellow line), 3.5 (orange line) and 3 (red line).}
\label{figure_15}
\end{figure*}

Figure \ref{figure_15} shows color-magnitude diagrams in J magnitude and J-K color assuming a fixed value of $f_{sed}$. The left panel shows the effect of $f_{sed}$ for log(g)=5 and the right panel the effect of log(g) for $f_{sed}$=5. With $f_{sed}$=3 and log(g)=5, the model reproduces the colors of most L dwarfs, including very red L dwarfs, but the L-T transition is delayed and the model does not reproduce the photometry of T dwarfs. With a larger $f_{sed}$, the L-T transition is sharp and occurs at a lower effective temperature. In particular with  $f_{sed}$=7, the L-T transition is well reproduced but the color is too blue for mid- and early-L dwarfs. This may suggest that $f_{sed}$ increases as brown dwarfs cool from early-L to early-T dwarfs \citep{saumon08}. Another possible explanation is that some absorbing species and condensates are missing for high temperatures, such as FeH gas and Al$_2$O$_3$ clouds. Most cloud-free models converge toward J-K=1 at 2000 K (see for instance \cite{marley10}) while our cloud-free model only reaches J-K=0. This strongly suggests that some opacity sources are missing in our model. Correcting this issue may allow us to match colors of early-L to early-T dwarfs with the same value of $f_{sed}$.
For a given $f_{sed}$, our model produces clouds optically thicker than the model by \cite{saumon08} and \cite{marley10}. This is certainly due to our parametization for $K_{zz}$, which gives values typically one order of magnitude lower than in \cite{ackerman01}. For a given $f_{sed}$, a lower value of $K_{zz}$ leads to smaller particles and consequently to optically thicker clouds. We thus need a higher $f_{sed}$ than \cite{marley10} to reproduce the photometry and spectra of L dwarfs. In that case, clouds are less extended, leading to a sharper L-T transition.

When $f_{sed}$ is fixed equal to 5 (right panel in Figure \ref{figure_15}), changing surface gravity has a relatively modest effect on photometry. 
However, for objects at the L-T transition, a lower surface gravity produces a shift towards blue color. In the convective region, the vertical mixing velocity (i.e. $K_{zz}/H$) is roughly independent of gravity. This implies that for a fixed $f_{sed}$, the particle sedimentation speed is almost independent of gravity as well. In that case, we have approximately $r_{cloud}\propto \left( \frac{K_{zz}}{g H} \right)^{-1/2} \propto g^{-1/2}$ and the cloud optical depth $\tau_{cloud} \propto \sqrt{g}$ \citep{marley12}. A decrease in surface gravity thus leads to slightly thinner clouds. This effect is reinforced with our $K_{zz}$ parametrization by the fact that, as gravity is decreasing, the top of the convective zone moves upward more rapidly than the condensation level of silicate clouds (see Figure \ref{figure_2}). In addition, in our parametrization for $K_{zz}$, the scale height of mixing by overshoot above the convective region varies with gravity as $g^{-1/2}$ (see \cite{ludwig03} and Figure \ref{figure_3}).
Therefore $K_{zz}/H$ increases when surface gravity is reduced at the level of condensation, making cloud particles larger and consequently clouds optically thinner.
This gravity trend is opposed to observations that suggest a reddening for low-gravity objects. One single value of $f_{sed}$ cannot thus reproduce both brown dwarfs and exoplanets.

\subsubsection{Simple microphysics}
The two previous parametrizations correspond to limit cases for the vertical size distribution of cloud particles: 1) when there is no evolution at all (fixed radius) and 2) when there is a very efficient size evolution with altitude (fixed $f_{sed}$).
Fixed radii seem to better reproduce the variation with temperature (i.e. sharp L-T transition) and with gravity (reddening for low surface gravity). This suggests that the size of particles does not vary much with altitude, in order to capture the effect due to the transition between the convective region and the radiative region. Since the parametrization with fixed radii is too sensitive to gravity, it also suggests that the size of particles should increase for lower gravity, but not as efficiently as with fixed $f_{sed}$.
We developed the parametrization with simple microphysics in order that it may behave as an intermediate case between the two previous ones. With this parametrization, the size of particles does not evolve much above the level of condensation, making the cloud mass distribution sensitive to vertical mixing and to the extent of the convective region. In addition, the size of particles varies approximately as $r_{cloud} \propto g^{-1/2}$ when condensation growth is limited by sedimentation, and as $r_{cloud} \propto \left( g\frac{K_{zz}}{H} \right)^{-1/2}$ when condensation growth is limited by vertical mixing. A decrease in surface gravity thus leads to large particles, but the dependence with $K_{zz}$ is either null or reverse compared to the case with fixed $f_{sed}$, making clouds vertically more extended and optically thicker for low gravity.

\begin{figure*}[!h] 
\begin{center} 
    \textbf{Color-magnitude diagrams with simple microphysics}\par
	\includegraphics[width=8cm]{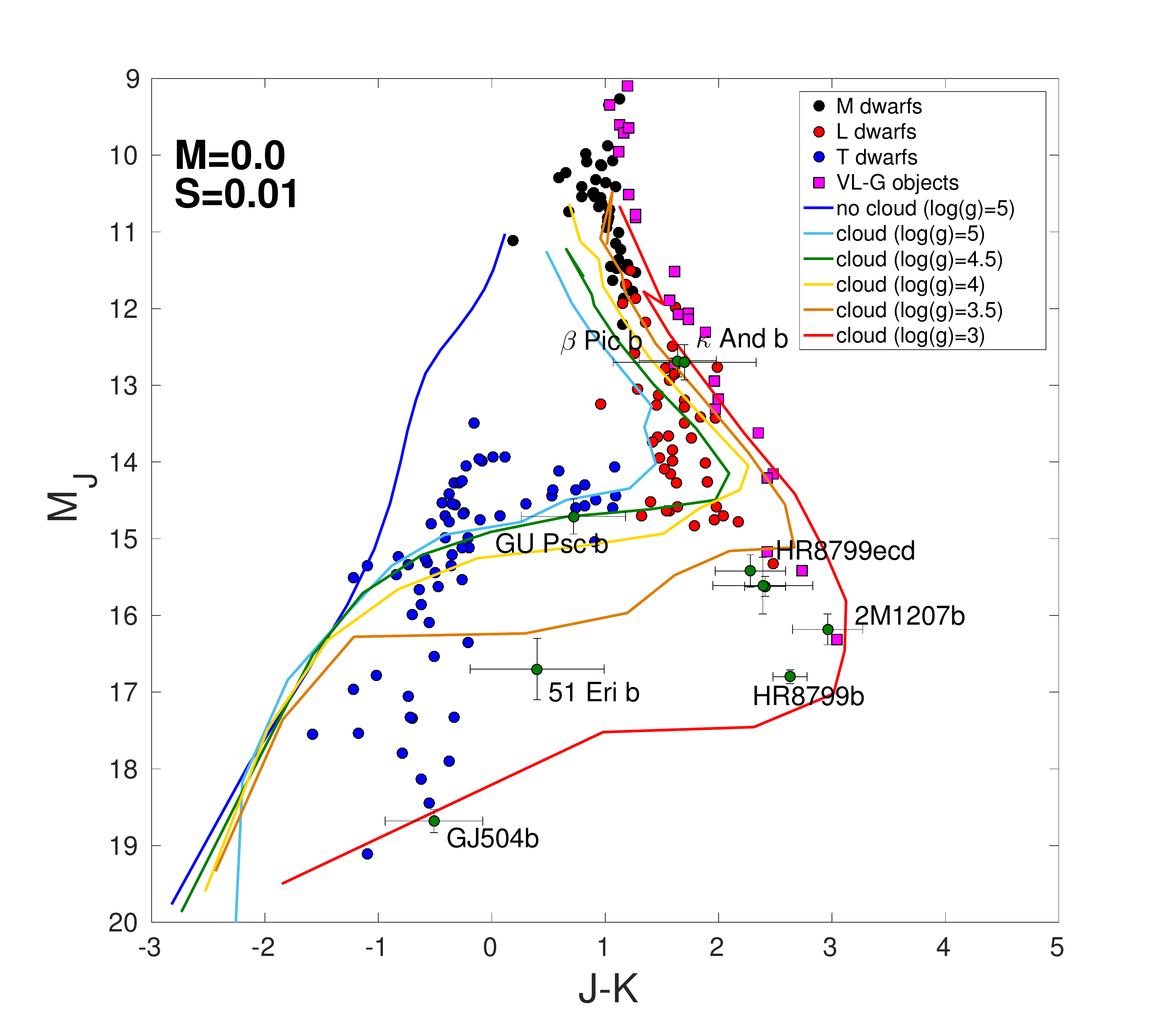}
	\includegraphics[width=8cm]{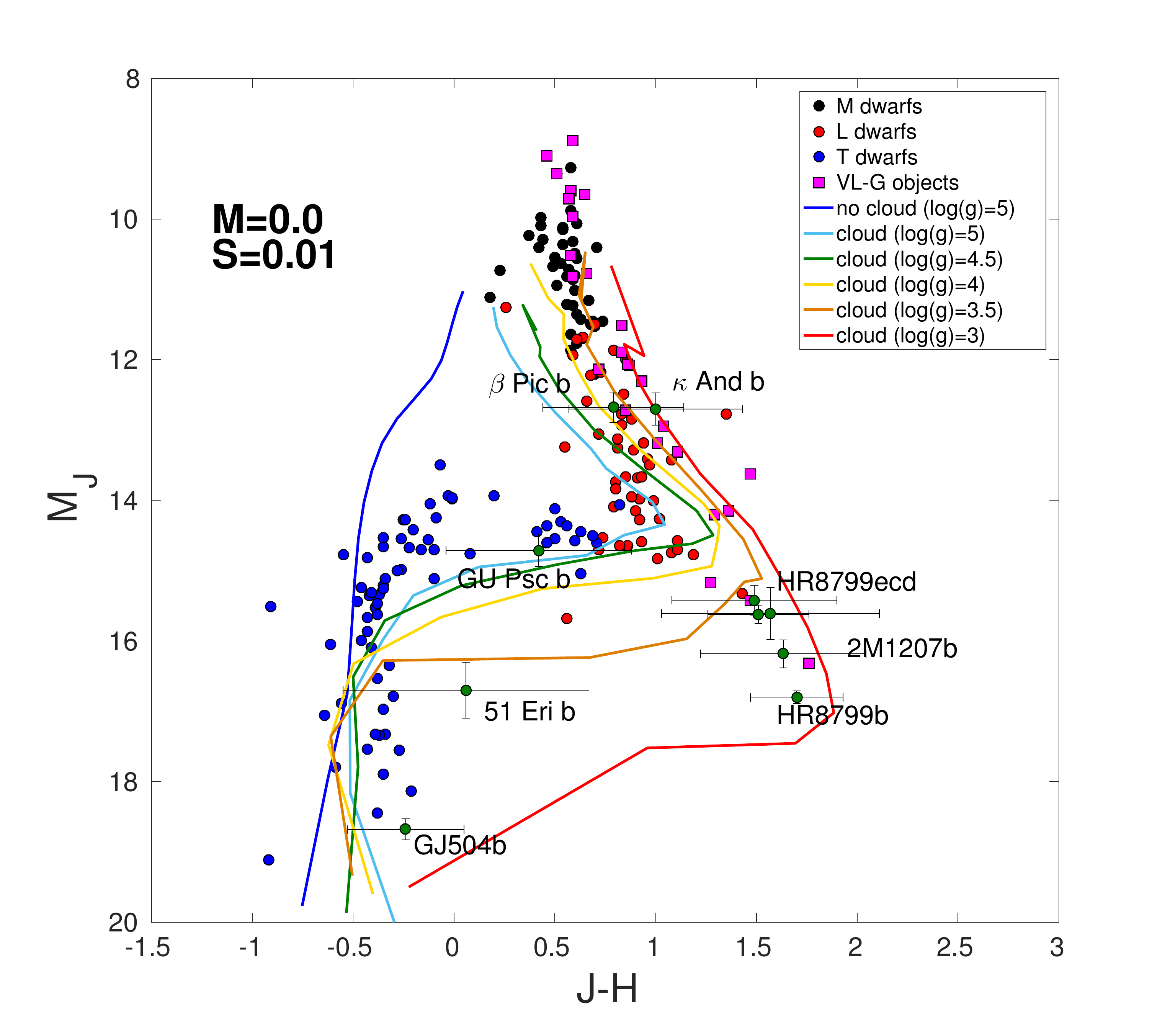}
	\includegraphics[width=8cm]{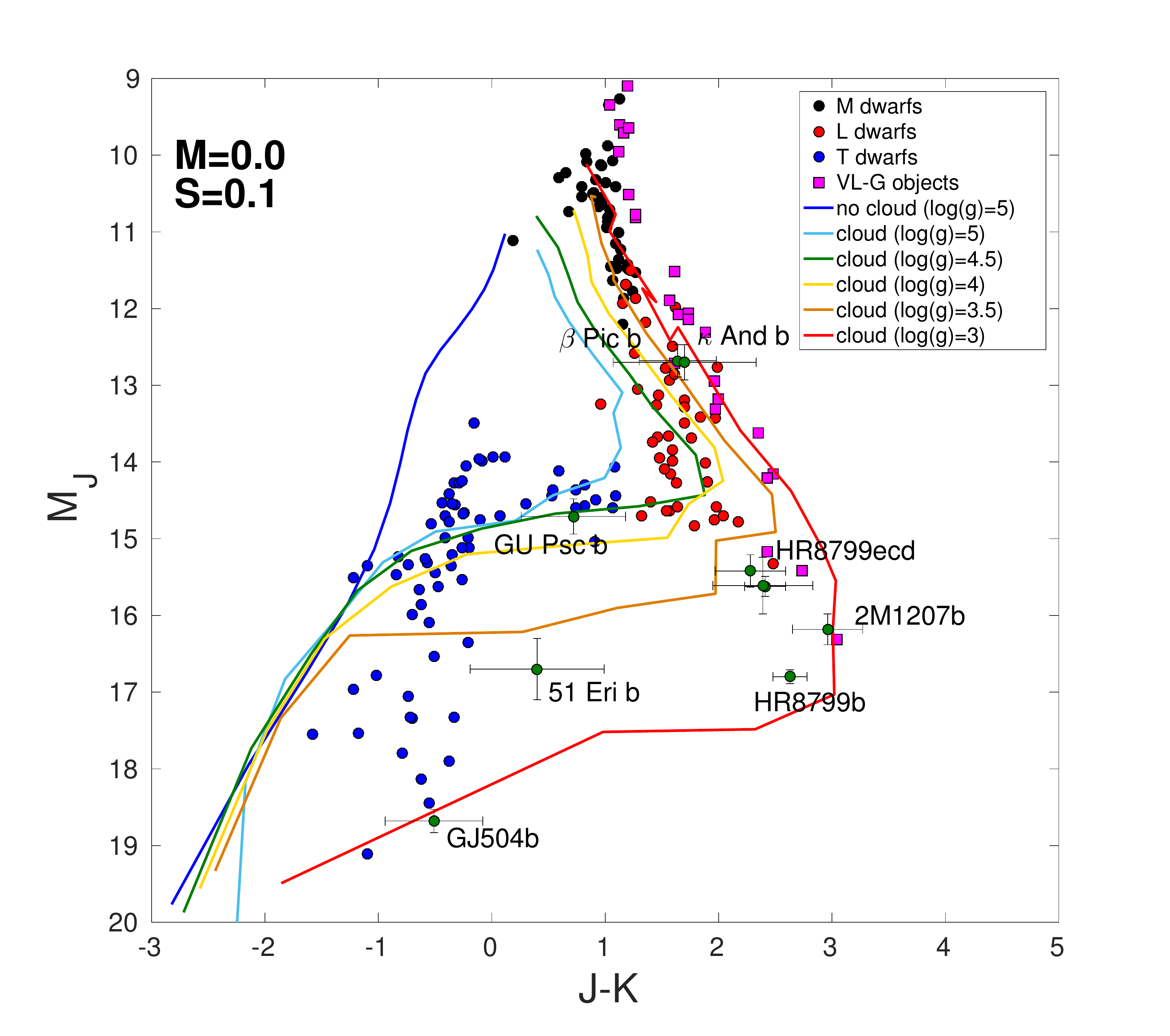}
	\includegraphics[width=8cm]{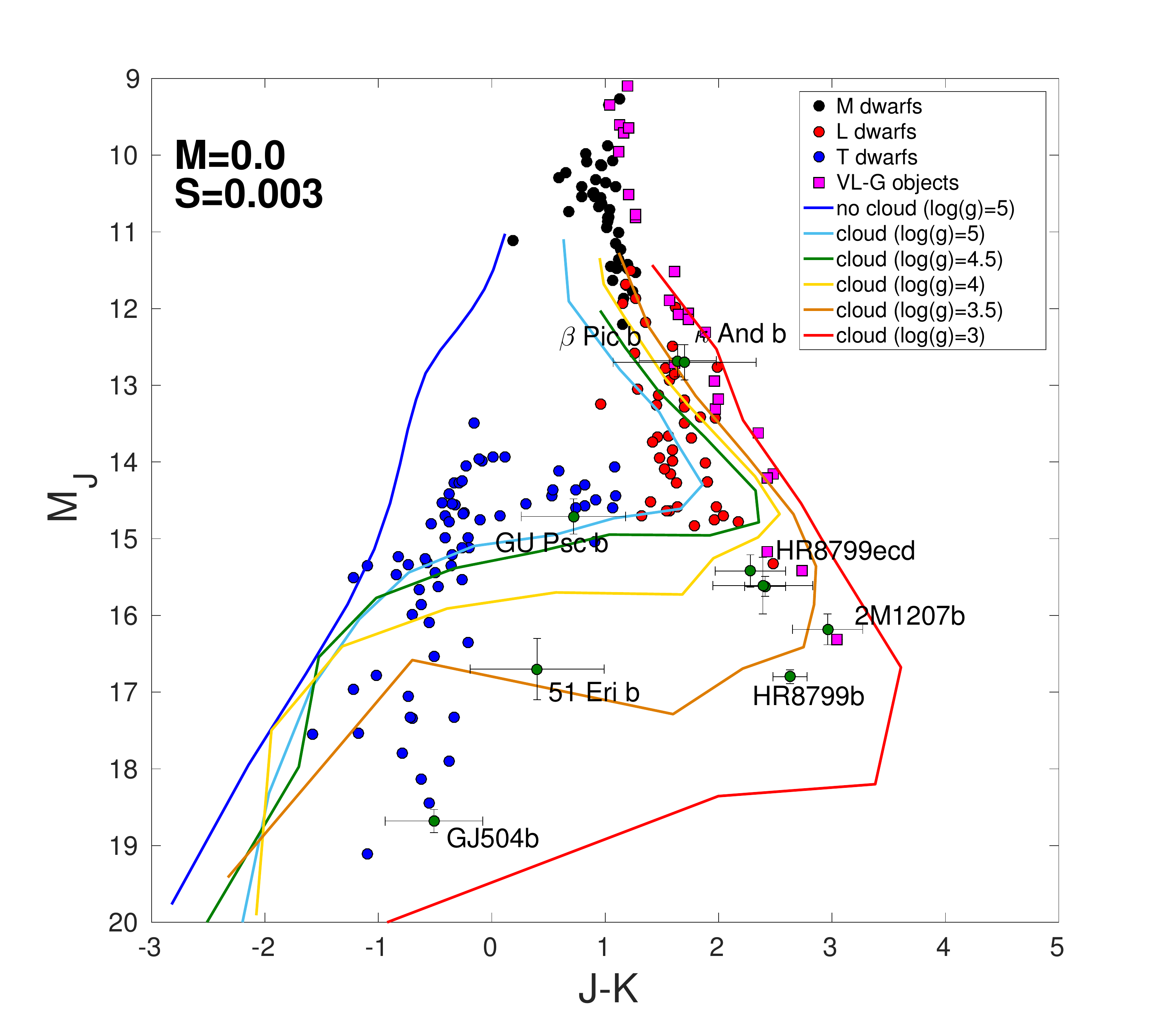}
	\includegraphics[width=8cm]{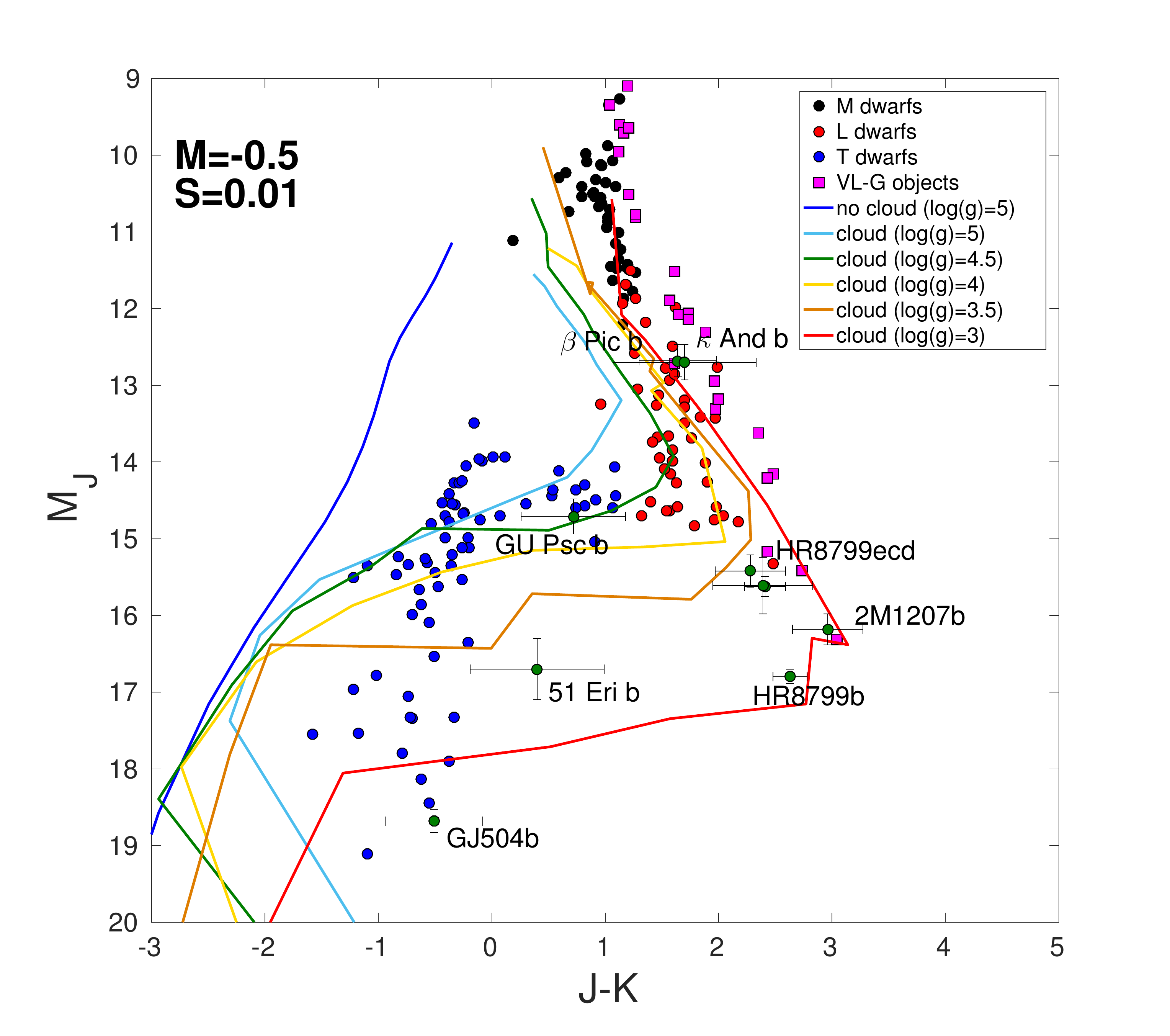}
	\includegraphics[width=8cm]{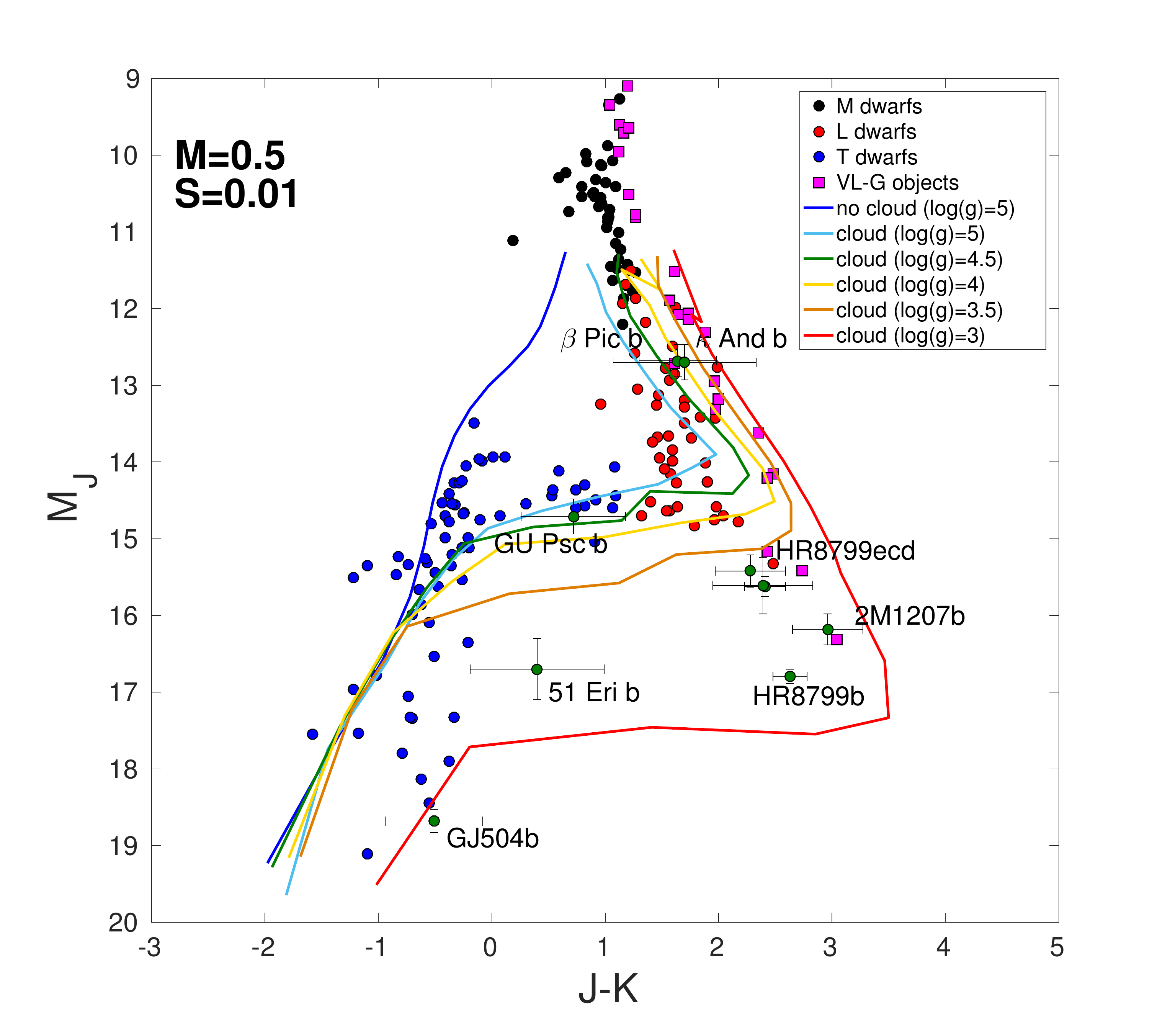}
\end{center}  
\caption{Same format as Figure \ref{figure_1} with microphysical model . For top panels, curves are computed with Exo-REM, with a supersaturation S=0.01 and log(g)=3.5, 4, 4.5, 5. The top left panel is with J-K colors and the top right panel is with J-H colors. Middle panels show J-K colors with S=0.1 (left) and S=0.003 (right). Bottom panels show J-K colors with S=0.01 and 0.3$\times$solar (M=-0.5, left) and 3$\times$solar (M=0.5, right) metallicity.
}
\label{figure_16}
\end{figure*}

Figure \ref{figure_16} shows color-magnitude diagrams in J magnitude and J-K or J-H colors, using supersaturation values of 0.1, 0.01 and 0.003, with 0.3, 1 or 3 $\times$ solar metallicity. For all cases, there is a sharp L-T transition, a reddening and a delayed transition for low-gravity objects. As in the previous section, colors are too blue for high effective temperatures, probably due to missing clouds and absorbing gas.
For low gravity, colors of L dwarfs are significantly shifted toward red, which is consistent with observations of young objects. Decreasing the value of supersaturation leads to redder colors (see the middle panels in Figure \ref{figure_16}). However, the supersaturation parameter has a much weaker impact than the particle radius and $f_{sed}$ in Figure \ref{figure_14} and \ref{figure_15}. Field brown dwarfs are well reproduced with S=0.01 and log(g)=4.5-5. For this supersaturation value, HR8799bcde and 2M1207b can be reproduced with log(g)=3-3.5, lower than expected. For S=0.003, their colors can however be reproduced with log(g)=3.5-4, which is more compatible with evolutionary models \citep{marley12}.
An enhanced metallicity tends to slightly shift curves toward red (see bottom panels in Figure \ref{figure_16}).
For all cases, photometry curves for high gravity (log(g)=4.5-5) tend be be clustered in the branches of observed L and T field dwarfs. Curves for low gravity  (log(g)=3-4) are more spread, implying more variations in the colors of low-gravity objects.

\subsubsection{T dwarfs}

\begin{figure}
\begin{center} 
    \textbf{Color-magnitude diagrams with simple microphysics}\par
	\includegraphics[width=8cm]{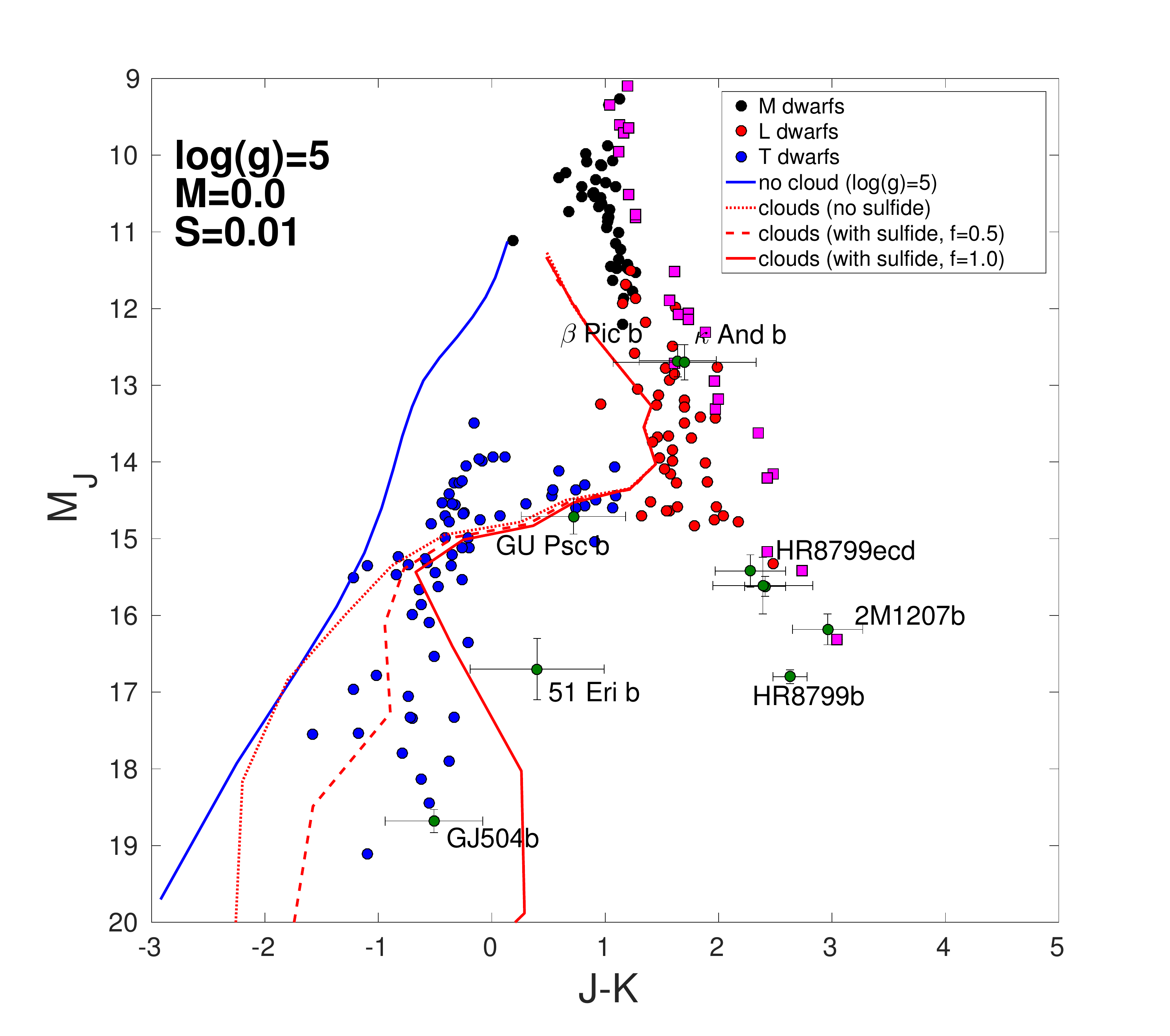}\\
	\includegraphics[width=8cm]{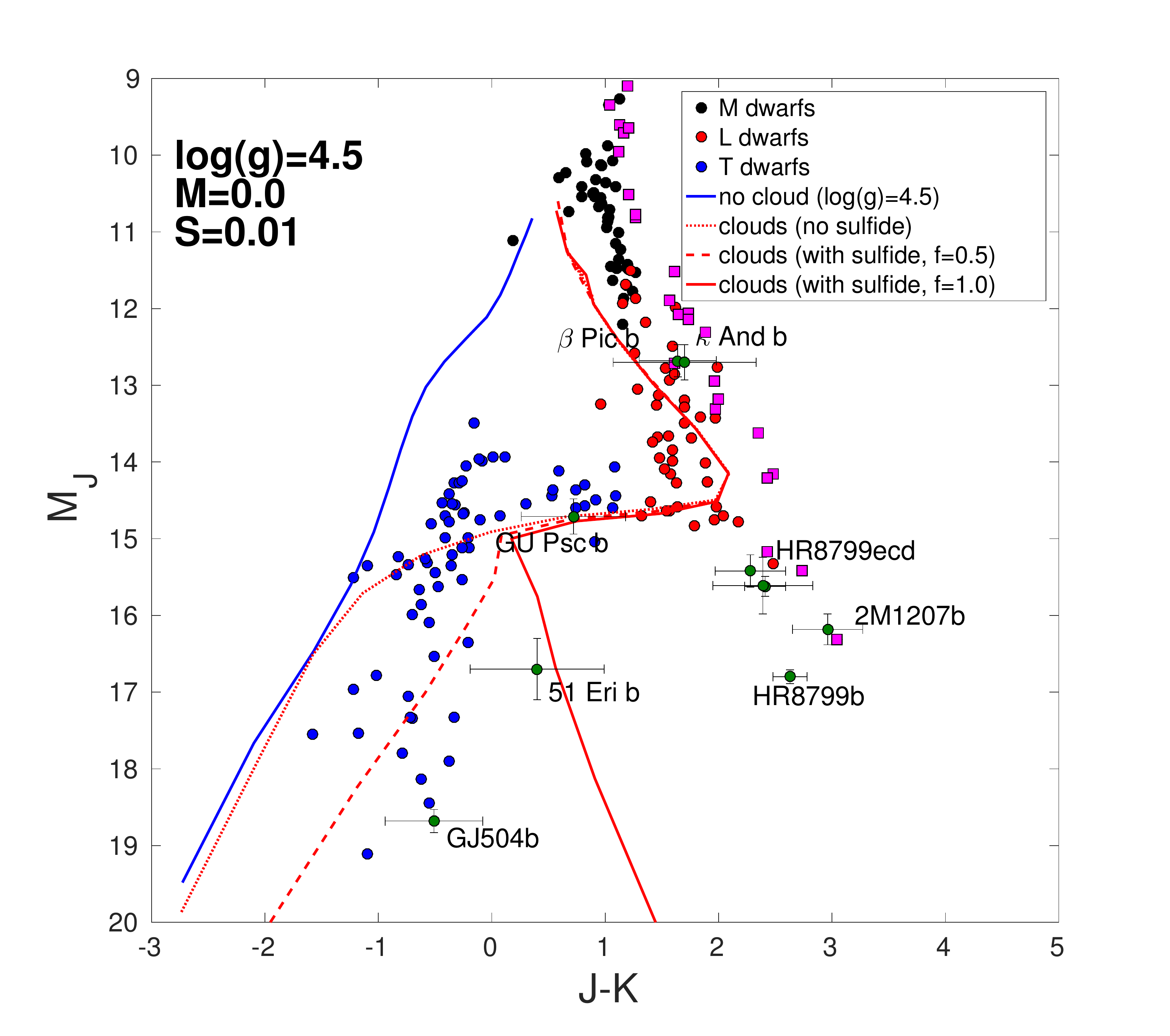}\\
	\includegraphics[width=8cm]{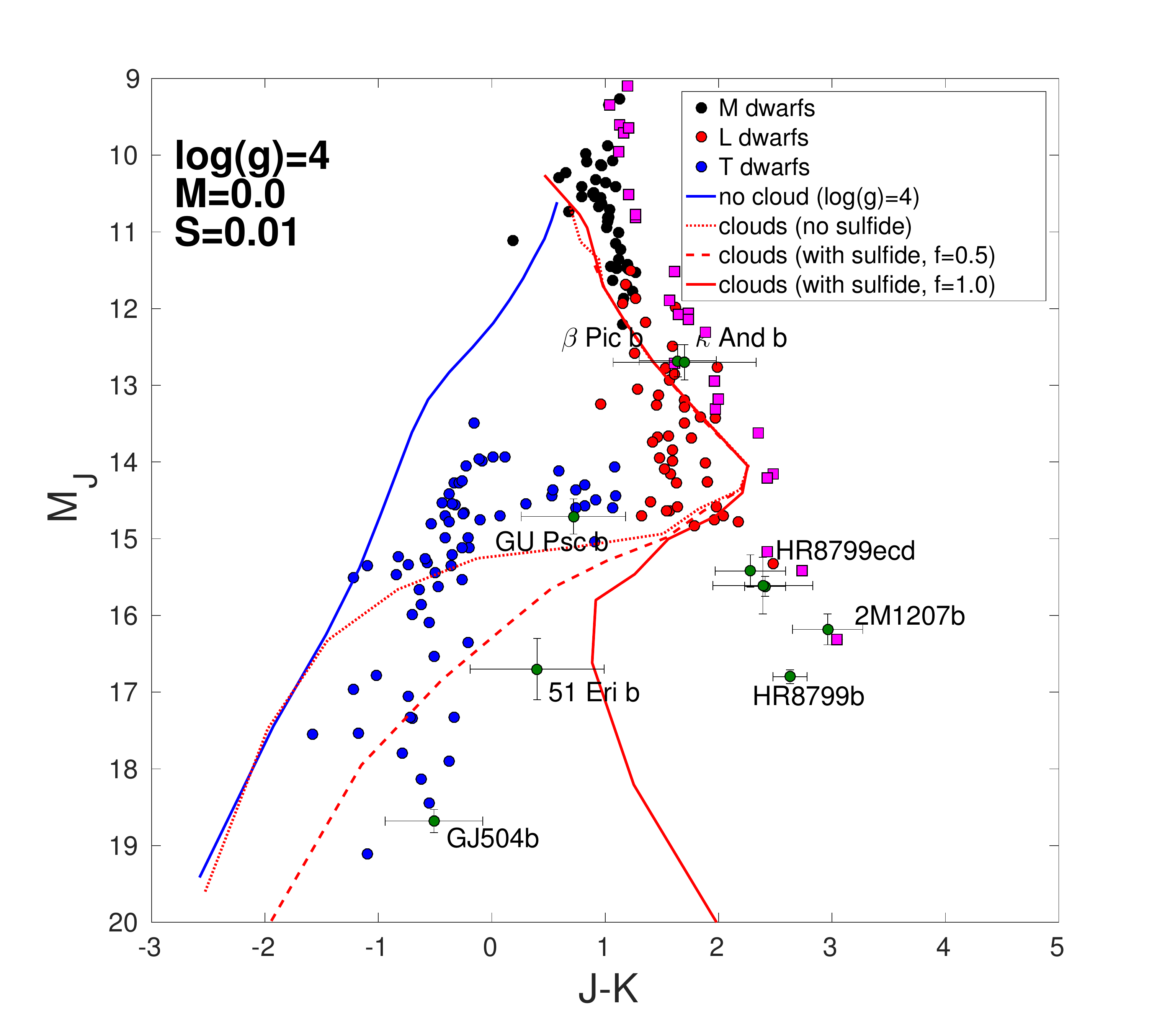}
\end{center}  
\caption{Same data as Figure \ref{figure_1} with microphysical model including sulfide and salt clouds. Cloudy cases are computed with log(g)=5, 4.5 and 4 from top to bottom, with a supersaturation S=0.01 for iron and silicate clouds and S=0.1 for sulfide and salt clouds. For sulfide and salt clouds, we use a cloud cover of 0, 50 and 100 $\%$ (dotted, dashed and solid red lines).}
\label{figure_17}
\end{figure}

In the previous section, we discussed about the photometry of L dwarfs and early-T dwarfs by including only silicate and iron clouds, for solar metallicity. With these clouds, our model generally fails  to reproduce the J-K color of mid-T and late-T dwarfs. Additional reddening is required. Sulfide and salt condensates are expected to form in the photosphere at these temperatures \citep{morley12}. In particular, Na$_2$S should be the dominant source of cloud opacity for effective temperature between 800 K and 400 K. Cr and MnS could have optical thickness larger than Na$_2$S in the deep atmosphere, but they should be dominated by silicate and iron clouds in this region \citep{morley12}. We therefore neglected them and considered only Na$_2$S and KCl clouds, even if the latter has quite a negligible effect.
When we include these clouds, the model with simple microphysics produces a too strong reddening even using S=0.1. We succeed in matching T dwarf photometry by using an inhomogeneous cloud cover of around $50\%$ for Na$_2$S and KCl clouds (see Figure \ref{figure_17}). \cite{morley12} also had to reduce the cloud thickness of sulfide clouds (using $f_{sed}$=5 for sulfide clouds instead of $f_{sed}$=3 for iron and silicate clouds) to match T dwarf photometry.
This suggests either a sodium depletion, a condensation growth more efficient than expected from our formula (maybe due to a weaker vertical mixing), or an inhomogeneous sulfide cloud cover. We remark that sulfide clouds have a much stronger effect for low gravity. For log(g)=4, sulfide clouds produce a noticeable effect for T$_{eff}<1000$ K, while this only occurs for T$_{eff}<800$ K for log(g)=5. The formation of sulfide clouds occurs in parallel to the disappearance of iron and silicate clouds for low-gravity objects making the L-T transition less sharp in that case. We also remark that a 3$\times$solar metallicity with no cloud can match the photometry of several T dwarfs (see the bottom right panel in Figure \ref{figure_16}).

\subsection{Comparison with brown dwarf spectra}

\begin{figure}[!h] 
\begin{center} 
	\includegraphics[width=8.5cm]{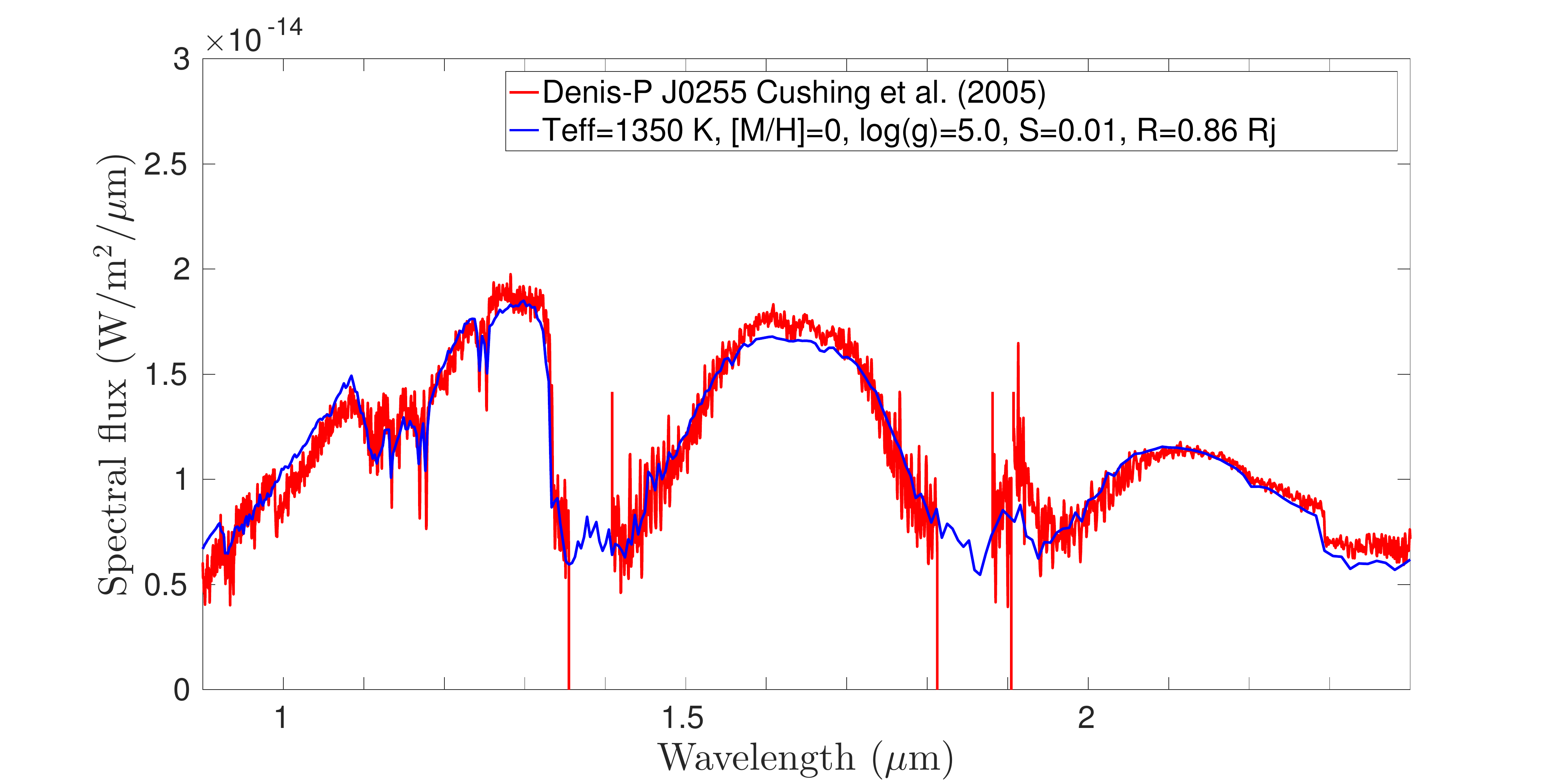}
	\includegraphics[width=8.5cm]{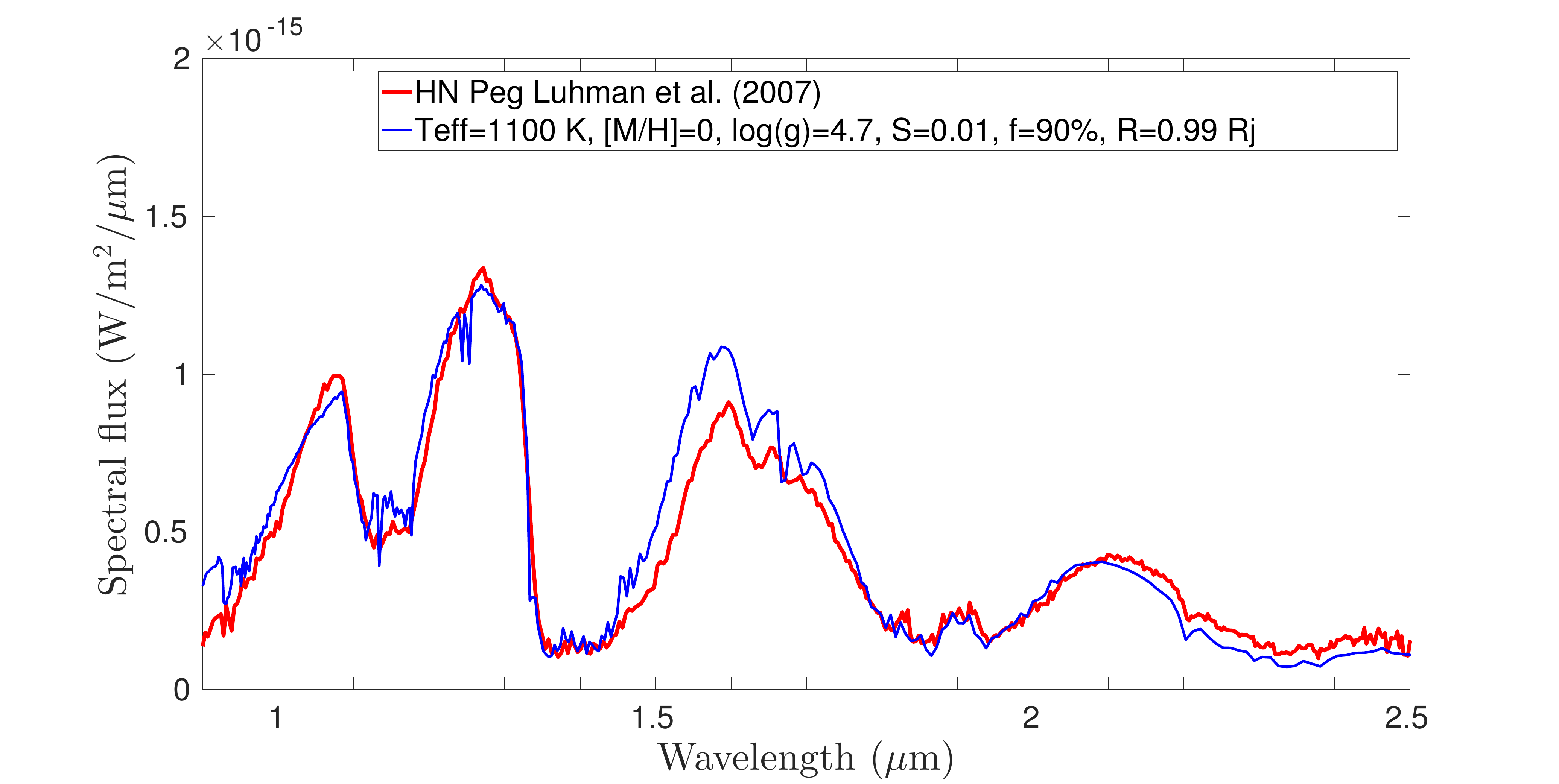}
	\includegraphics[width=8.5cm]{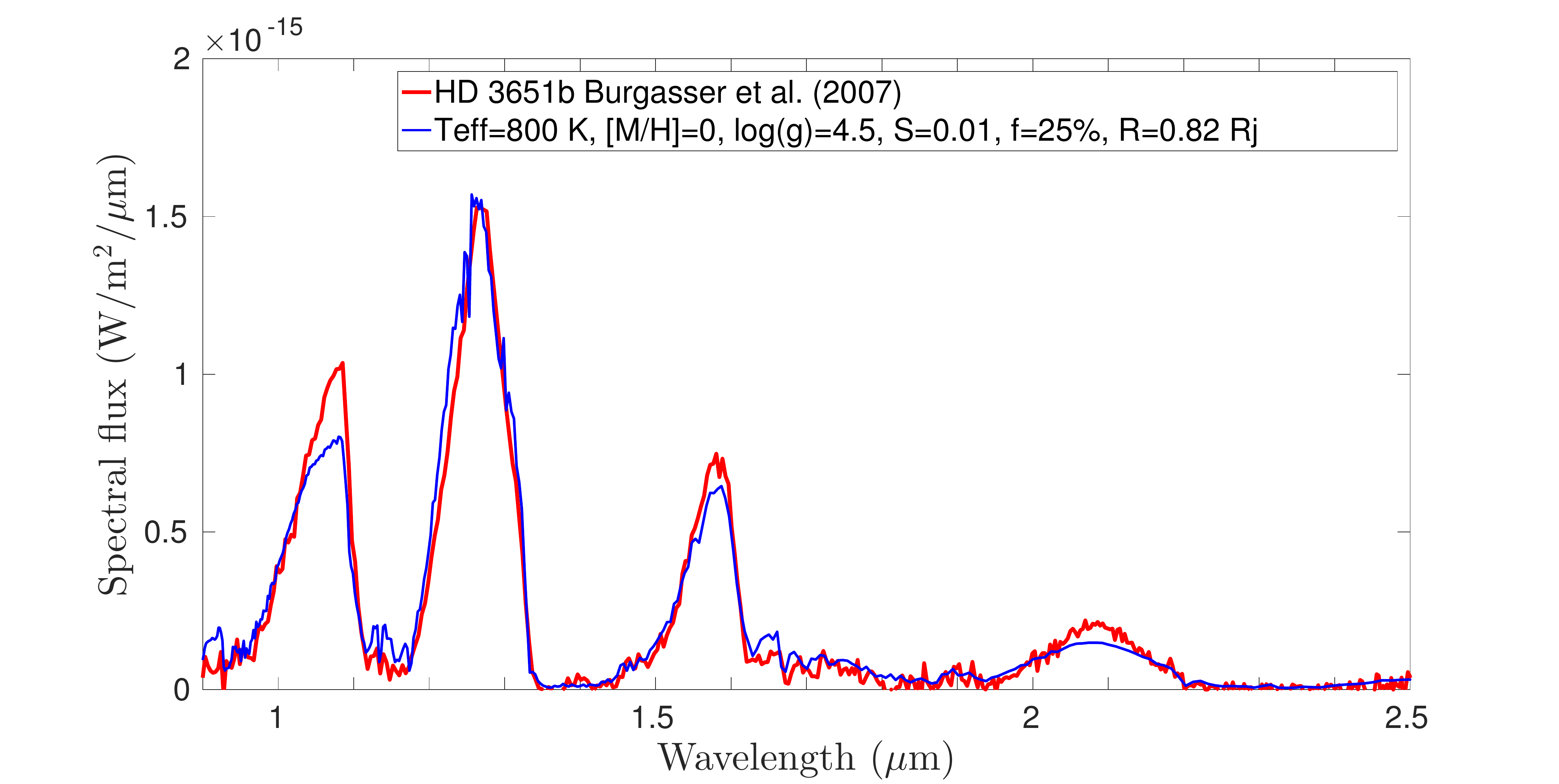}
\end{center}  
\caption{Spectra of Denis-P J0255 (top),  HN Peg b (middle) and HD 3651 b (bottom). Red lines correspond to observations and blue lines to the best fits from Exo-REM.}
\label{figure_18}
\end{figure}

After having validated the cloud model with the photometry of brown dwarfs and exoplanets, we compared the model to spectra of well-characterized brown dwarfs. We chose one late-L dwarf (Denis-P J0255) and two T dwarfs (HN Peg b and HD 3651 b) to cover the L-T transition. The observed spectra are shown in red in Figure \ref{figure_18} while simulated spectra are in blue. For all models, we used the cloud scheme with simple microphysics (S=0.01) and the formation of iron, silicate and sulfide clouds. The comparison (detailed below) is globally satisfying and derived parameters are consistent with evolutionary models. For this spectral fitting, we only changed surface gravity, radius, effective temperature and cloud cover, when needed. Despite this relatively small number of free parameters, the model can match observations of very different objects and appears suitable for atmospheric retrievals.

\subsubsection{Denis-P J0255}
Denis-P J0255 is a typical late-L dwarf at a distance of 4.97 pc. The top panel in Figure \ref{figure_18} shows its SpeX Prism spectrum from \cite{cushing05}, corresponding to spectral type L8.
The best fit is obtained for T$_{eff}$=1350 K, log(g)=5.0, [M/H]=0.0 and R=0.86 R$_J$. 
These results are similar to \cite{tremblin16} who obtained a best fit with T$_{eff}$=1400 K, log(g)=5, [M/H]=0.0 and R=0.76 R$_J$, using the cloud-free model ATMO.

\subsubsection{HN Peg b}
HN Peg b is a T dwarf companion at a distance of 18.4 pc discoverd by \cite{luhman07}. The middle panel in Figure \ref{figure_18} shows its SpeX Prism spectrum from \cite{luhman07}, corresponding to spectral type T2. The age of its host star is estimated  in the range 0.2-0.5 Gyr, younger than typical field brown dwarfs. Near-IR photometric observations with SpeX gives absolute magnitudes of 14.54 in J band and 13.80 in K band (MKO filters). 
Our model with simple microphysics is slightly too cloudy in that case and we have to use inhomogeneous cloud cover to match the photometry or the spectrum. Doing so, the best fits are obtained with T$_{eff}$=1100 K, log(g)=4.7, [M/H]=0, R=0.99 R$_J$ and a cloud fraction of 90 $\%$. These parameters compatible with evolutionary model by \cite{baraffe03} and \cite{burrows97}, which predict T$_{eff}$=1130 $\pm$70 K and R=1.08 $\pm$0.06 R$_J$ for a 0.5 Gyr old T dwarf (see Figure 10 in \cite{luhman07}). Our model produces slightly too much emission in H band and not enough in K band.

\subsubsection{HD 3651 b}
HD 3651 b is another T dwarf companion at a distance of 11.0 pc discovered by \cite{luhman07}. The bottom panel in Figure \ref{figure_18} shows its SpeX Prism spectrum from \cite{burgasser07}, corresponding to spectral type T8. The age of its host star is poorly constrained and estimated in the range 2-12 Gyr.
Near-IR photometric observations with SpeX gives absolute magnitudes of 15.95 in J band and 16.66 in K band \citep{luhman07}.
To match the spectrum of this object, we need to reduce the cloud cover as for the photometry of T dwarfs. In that case, the best fit is obtained for T$_{eff}$=800 K, log(g)=4.5, [M/H]=0.0 and R=0.82 R$_J$ and a sulfide cloud fraction of 25 $\%$. These parameters are compatible with evolutionary models which predict T$_{eff}$=810 $\pm$50 K and R=0.82 $\pm$0.06 R$_J$ (see Figure 10 in \cite{luhman07}).

\section{Summary and Discussion}

We developed a new cloud model for Exo-REM, which simulates the formation of silicate, iron, sulfide, alkali salt and water clouds. The cloud distribution is computed taking into account sedimentation and vertical mixing with realistic K$_{zz}$ profiles based on the mixing length theory.
By using first fixed particle radii, we found that the photometry of brown dwarfs and directly imaged exoplanets can be reproduced only if the size of cloud particles decreases with gravity, implying a feedback by microphysics. By using the parametrization with $f_{sed}$, we found that a sharp L-T transition can be obtained with high values of $f_{sed}$ (i.e. 5-7) but that the size of particles decreases too rapidly with gravity, leading to reddening for high gravity, contrary to observations.
The model with simple microphysics is intermediate between these two options and solves these issues. It reproduces well the photometry and spectra of brown dwarfs and young giant exoplanets, in particular the L-T transition and the reddening of low-gravity objects. 

This self-consistent model with simple microphysics also has a high potential for observational predictions. We found that photometry curves for high gravity are clustered in the branches of observed L and T field dwarfs. But they are more spread for low gravity. This implies more variations for colors of young brown dwarfs and young giant exoplanets. Based on these curves, we also predict the existence of a continuum of low-gravity objects in the apparent gap right to the T dwarf branch, corresponding to magnitude J=15-18 and J-K=0-3. Such objects would have effective temperature around 700-900 K. In this region, colors vary strongly with gravity and effective temperature (J-K decreases from $\sim$2 to $\sim$0 for T$_{eff}$ decreasing from 1000 K to 700 K, for log(g)=3.5 and S=0.01). Therefore the exoplanet population should not be dense in this region of the color-magnitude diagram, and we will probably detect only a few objects there. The exoplanet 51 Eri b may be the first of this kind.

We revealed a strong feedback caused by the greenhouse effect of clouds on the chemistry, leading to a decrease of the methane abundance for cloudy objects. The effect is particularly strong before the L-T transition and for low-gravity objects. It contributes significantly to the observed methane depletion in low-gravity objects, such as 2M1207b and HR8799bcd. This feedback is a fundamental consequence of cloud formation. It implies a strong correlation between colors and the CH$_4$/CO ratio, which help to explain the sudden appearance of strong methane bands at the L-T transition as clouds disappear.
Cloudless models lead to a smoother variation of the CH$_4$/CO ratio with effective temperature than we predict. 
In particular, the model by \cite{tremblin16} predicts a cooling of the deep atmosphere by the triggering of fingering convection, opposite to the effect of clouds. For an effective temperature of 1400 K, we predict a temperature around 3000 K at 100 bars, while \cite{tremblin16} predict a temperature around 2000 K. Such a large temperature difference should have implications for the chemical composition and the radius of the object. A statistical analysis of the chemical composition of brown dwarfs may reveal which mechanism dominates.

We also noticed that the top of the convective region passes below the photosphere at the L-T transition. This leads to a decrease of the vertical mixing in the cloud forming region and so to a decrease of the cloud optical thickness when using fixed particle radii or microphysics. To our knowledge, this effect has not been discussed before. In our numerical simulation, the evolution of the convective region extension makes the L-T transition slightly steeper but it is not critical for producing the L-T transition, which remains controlled by the evolution of the condensation level. The evolution of the L-T transition with gravity is linked to changes of vertical mixing, particularly the extension of the convective region and the mixing by convective overshoot. 2D or 3D simulations of convection and mixing, similar as the work by \cite{freytag10}, are required to validate our parametrization for $K_{zz}$ in this regime.
An other limitation in our modelling is that we use the eddy mixing coefficient from cloud-free simulations, to avoid numerical instabilities. Cloud warming should stabilize the atmosphere below the cloud deck and should trigger convection inside and above the cloud deck. Such a feedback should impact the cloud vertical structure and potentially the L-T transition. If the cloud warming tends to increase vertical mixing, we could expect a steeper L-T transition than simulated here.
2D-3D simulations including the radiative effect of clouds are needed to analyse these cloud feedbacks and to parametrize more precisely the vertical mixing.

Finally, this work suggests that microphysics and vertical mixing play important roles in the L-T transition for brown dwarfs and exoplanets. A cloud model taking into account these elements can produce a L-T transition without requiring other physical effects.

\section{Perspectives}

The natural next step is to apply this new cloud model to observations of young exoplanets and brown dwarfs imaged by instruments like SPHERE and GPI, as well as JWST in the near futur. It has already been applied to three objects discovered by SPHERE: HIP 65426b \citep{chauvin17}, HD206893b \citep{delorme17} and HR2562b \citep{mesa17}. The J-K colors for these objects are represented in Figure \ref{figure_1}. The spectra of HIP 65426b and HD206893b were successfully reproduced with the $f_{sed}$ parameter (the parametrization with microphysics was developed later). They require low values of $f_{sed}$ suggesting the presence of thick clouds. It will be interesting to fit these spectra with grids computed with the simple microphysics model to improve the estimation of the gravity and radius of these objects. For HR2562b, we used the grid computed with the simple microphysics (S=0.01) and found that the spectrum can be matched with only three free parameters (i.e. gravity, effective temperature and radius) as in the previous section.

Many brown dwarfs, in particular at the L-T transition, show temporal variability in near-IR, interpreted as the presence of inhomogeneous cloud covers combined with fast rotation rates. These longitudinal inhomogeneities in the cloud cover could be produced by atmospheric waves (e.g. gravity waves and planetary waves), cloud convection triggered by cloud radiative heating \citep{freytag10}, latent heat release \citep{tan17} or other physical processes. Information about the horizontal and vertical cloud distribution can be obtained by analysing this variability at different wavelengths. Our cloud model with inhomogeneous cloud cover and various options for modelling clouds is well suited to test hypotheses for the brown dwarf variability.

Finally, the self-consistent cloud model that we developed is very generic and other condensates can easily been added, in particular NH$_{3}$ and NH$_{4}$HS for cold exoplanets. A future development will be to include stellar heating and to compute reflected spectra. Grids of reflected spectra for cloudy giant exoplanets would be very useful to predict and to interpret observations by future planet imager telescopes such as WFIRST or for luminosity curves of close-in planets.
\\

\acknowledgments

We thank Pascal Tremblin, Tobias Schmidt and S\'{e}bastien Fromang for useful discussions. B. C. acknowledges support from the PSL Fellowship. We acknowledge financial support from the French National Research Agency (ANR-14-CE33-0018/GIPSE) and from the Programme National de Planétologie (PNP) of CNRS-INSU co-funded by CNES.

\appendix
\section{Saturation vapour pressures}

Monatomic Fe gas is the dominant Fe-bearing gas for L dwarfs and iron condensation occurs via:
\begin{equation} 
\rm  Fe = Fe(s,l)
\end{equation} 

Condensed iron generally is liquid for brown dwarfs (iron melts at 1809 K).
The equilibrium saturation vapour pressure of Fe ($p'_{\rm Fe}$, in bar) as a function of temperature ($T$, in K) is approximated by \citep{visscher10}:
\begin{equation} 
log (p'_{\rm Fe}) \approx 7.23 - 20995/T 
\end{equation} 

Monoatomic Mg is the most abundant Mg-bearing and condenses as forsterite (Mg$_2$SiO$_4$) and enstatite (MgSiO$_3$) via the net thermochemical reactions:
\begin{equation} 
\rm  2Mg + 3H_2O + SiO = Mg_2SiO4(s,l) + 3H_2
\end{equation}
\begin{equation} 
\rm  Mg + 2H_2O + SiO = MgSiO_3(s,l) + 2H_2
\end{equation} 

Enstatite forms at a slightly lower temperature. For simplicity, we assumed Mg only condenses as forsterite and its saturation vapour pressure ($p'_{\rm Mg}$) as a function of temperature and metallicity ($[Fe/H]$) is approximated by \citep{visscher10}:
\begin{equation} 
log (p'_{\rm Mg}) \approx 11.83 - 27250/T  - [Fe/H] 
\end{equation} 

Monoatomic Na is the most abundant Na-bearing gas and condenses as Na$_2$S via the net thermochemical reaction: 
\begin{equation} 
\rm H_2S+ 2Na = Na_2S(s) +H_2
\end{equation} 

Na saturation vapour pressure ($p'_{\rm Na}$) is approximated by \citep{visscher06}:
\begin{equation} 
log (p'_{\rm Na}) \approx 8.550 - 13889/T - 0.50 [\rm Fe/H]
\end{equation}

For T dwarfs, KCl is the most abundant K-bearing gas and condenses via the net thermochemical reaction: 
\begin{equation} 
\rm  KCl = KCl(s)
\end{equation} 

KCl saturation vapour pressure ($p'_{\rm KCl}$) is approximated by \citep{morley12}:
\begin{equation} 
log (p'_{\rm KCl}) \approx 7.611 - 11382/T
\end{equation} 

For Y dwarfs, H$_2$O condenses as water ice clouds and the saturation water vapour pressure ($p'_{\rm H_2O}$) is approximated by:
\begin{equation} 
log (p'_{\rm H_2O}) \approx 7.6116 - 2694.96/T+3040.1/T^2
\end{equation}

\begin{figure*}[h] 
\begin{center} 
	\includegraphics[width=6.5cm]{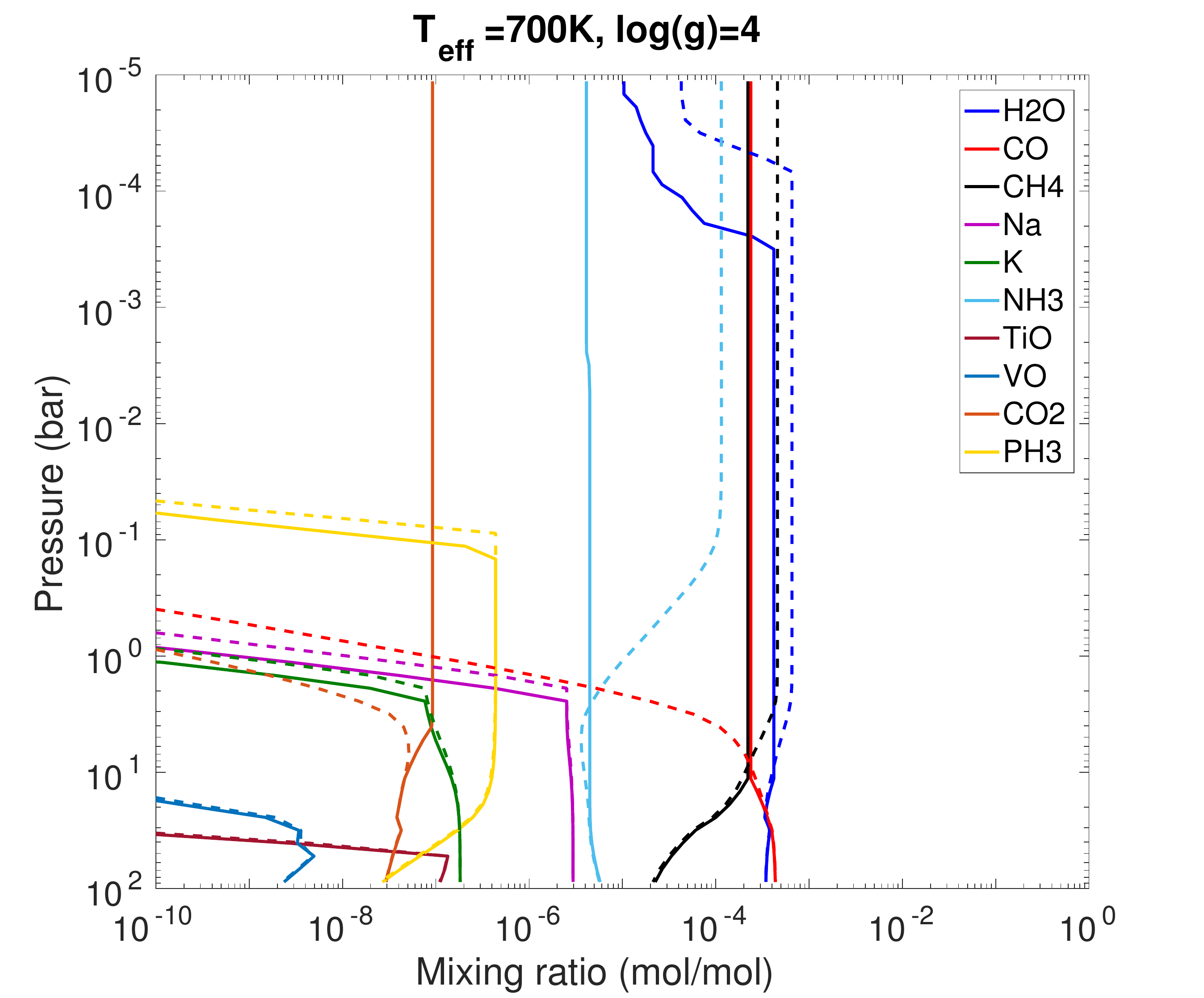}
	\includegraphics[width=6.5cm]{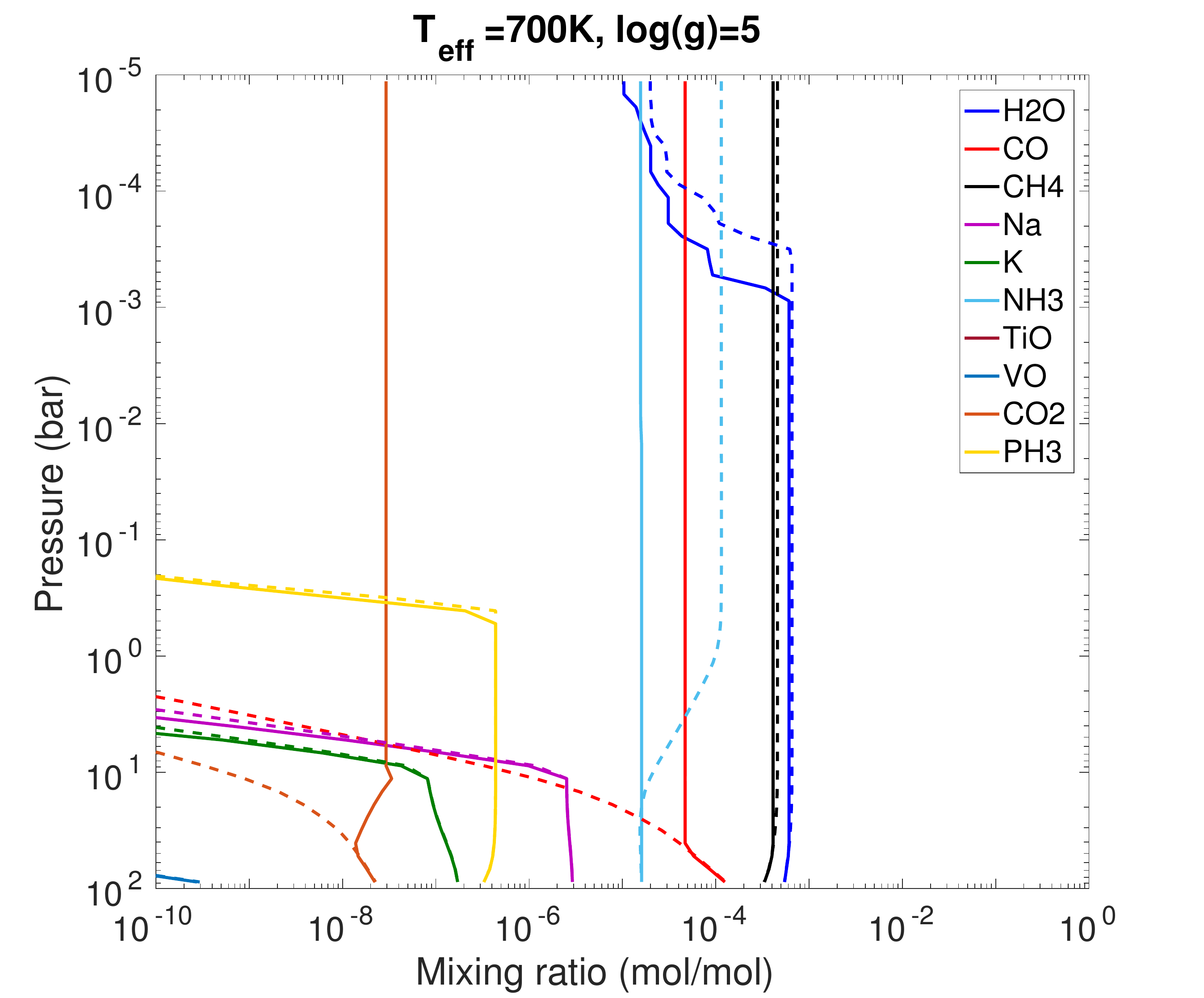}
	\includegraphics[width=6.5cm]{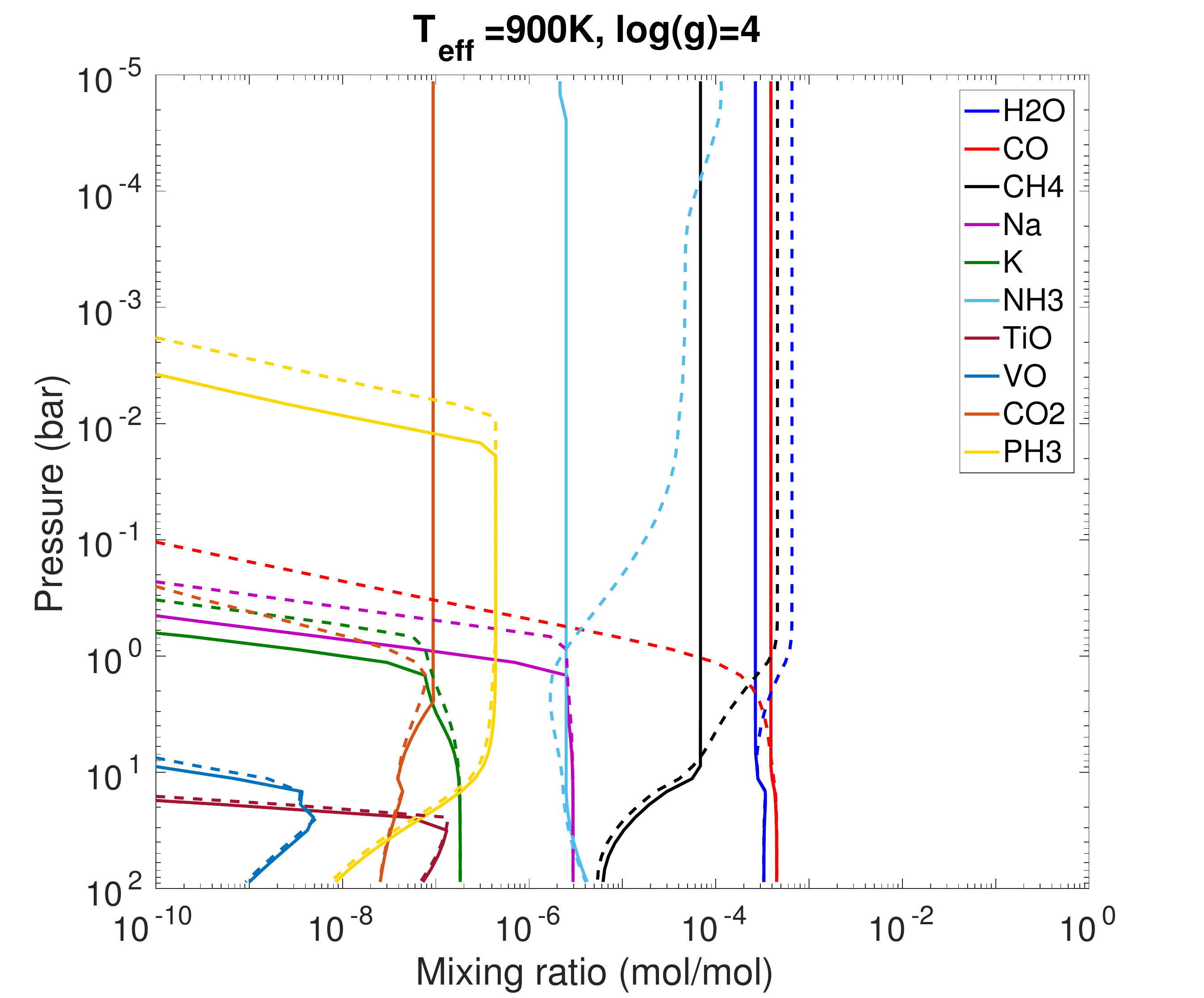}
	\includegraphics[width=6.5cm]{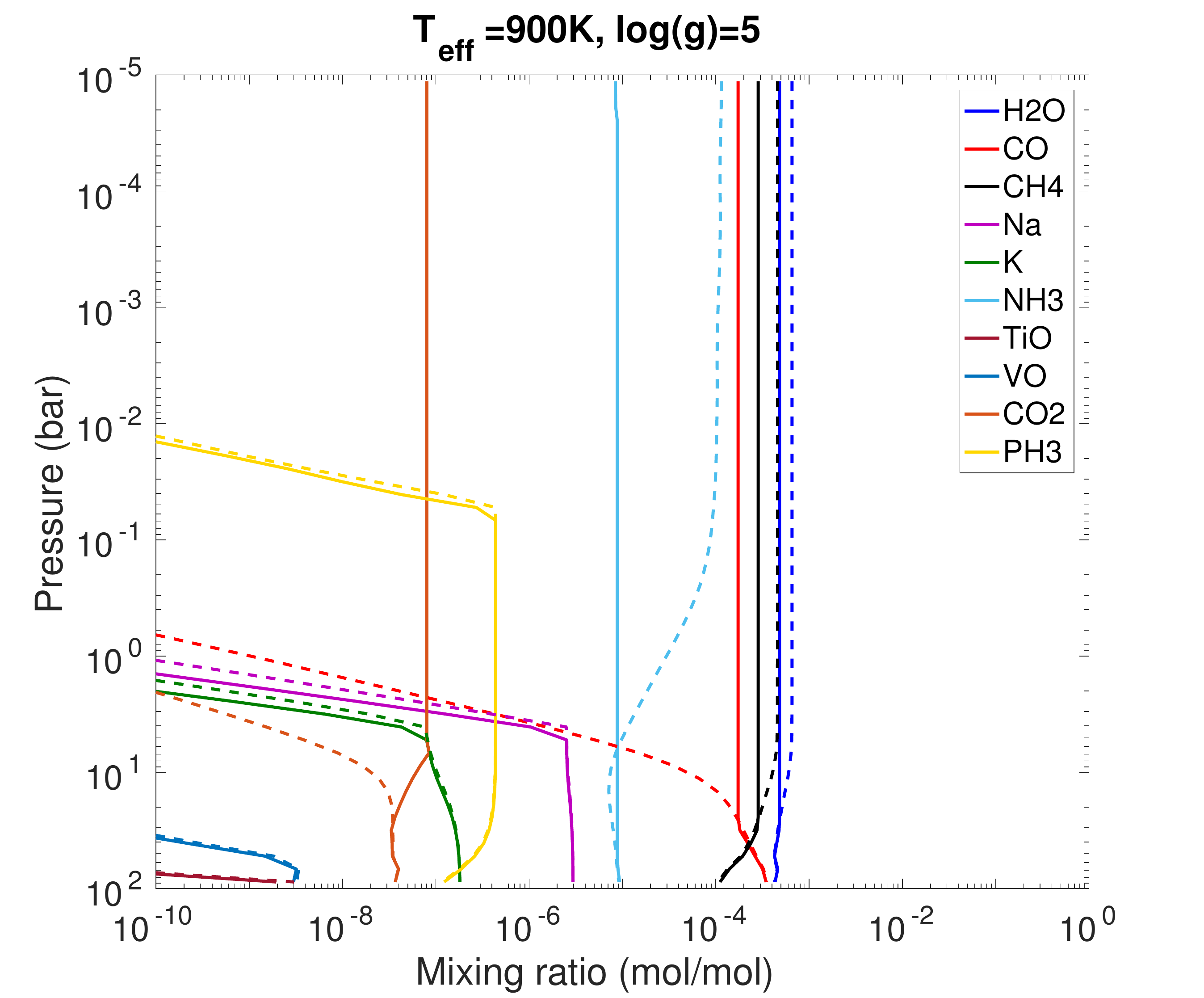}
	\includegraphics[width=6.5cm]{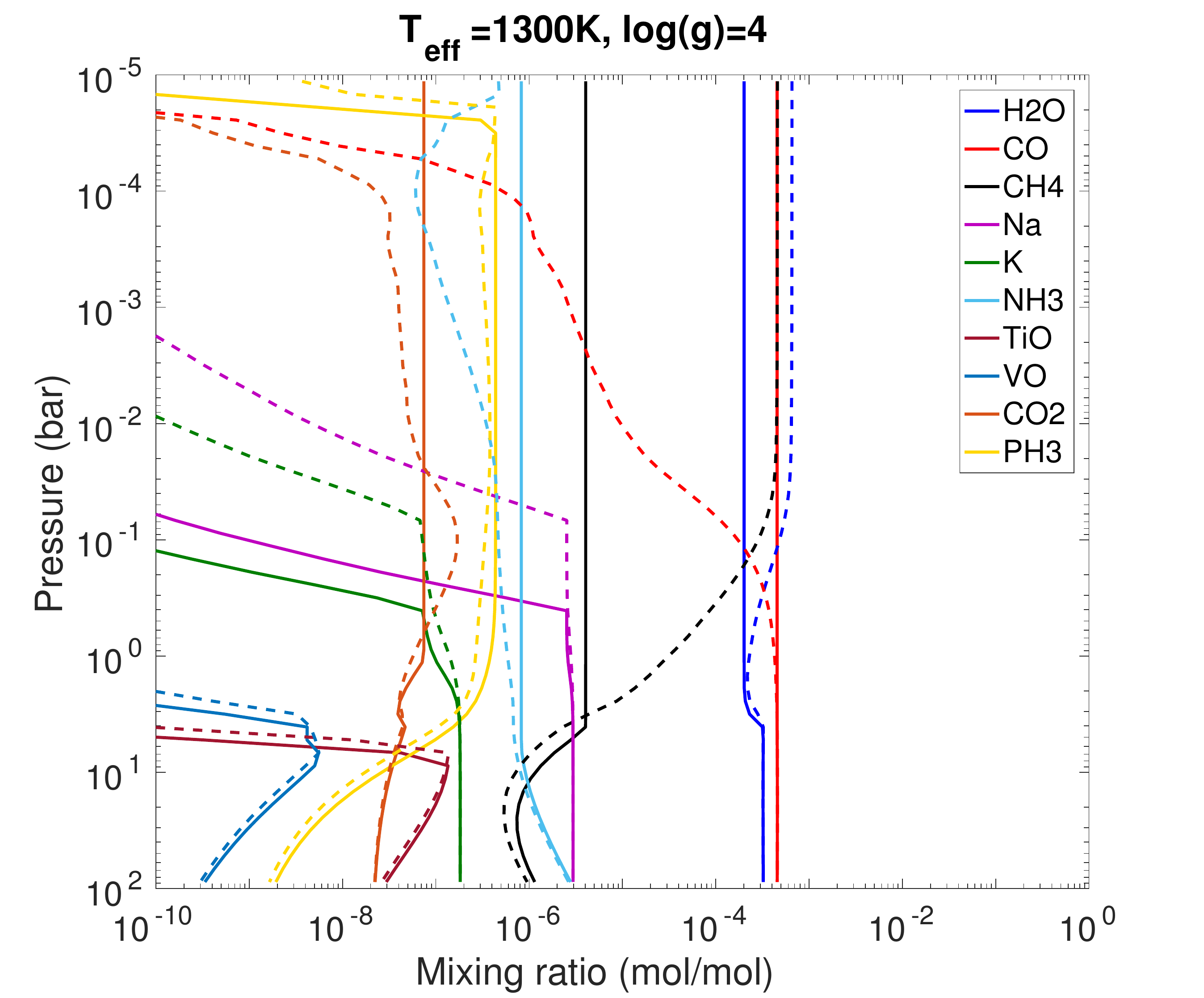}
	\includegraphics[width=6.5cm]{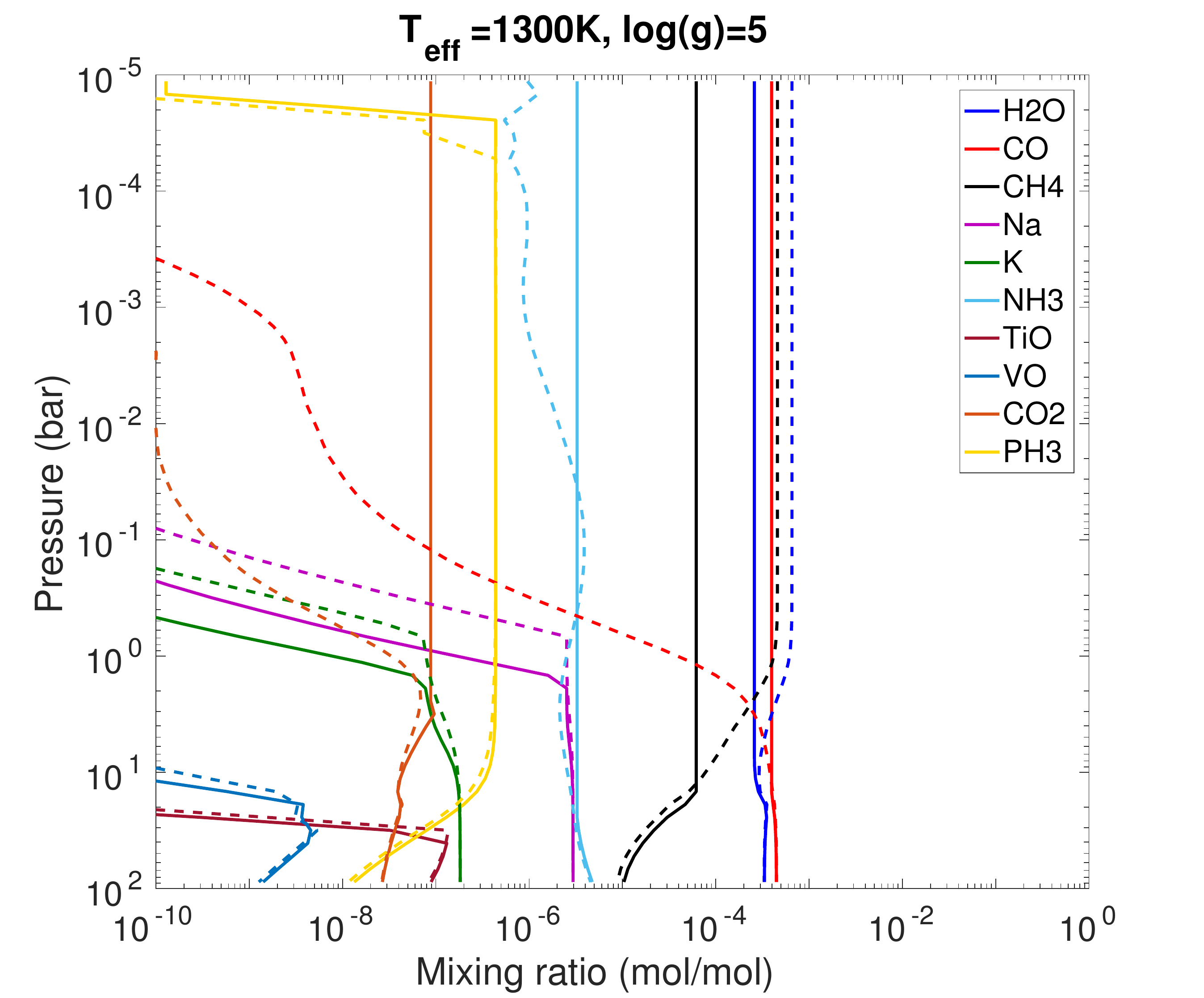}
	\includegraphics[width=6.5cm]{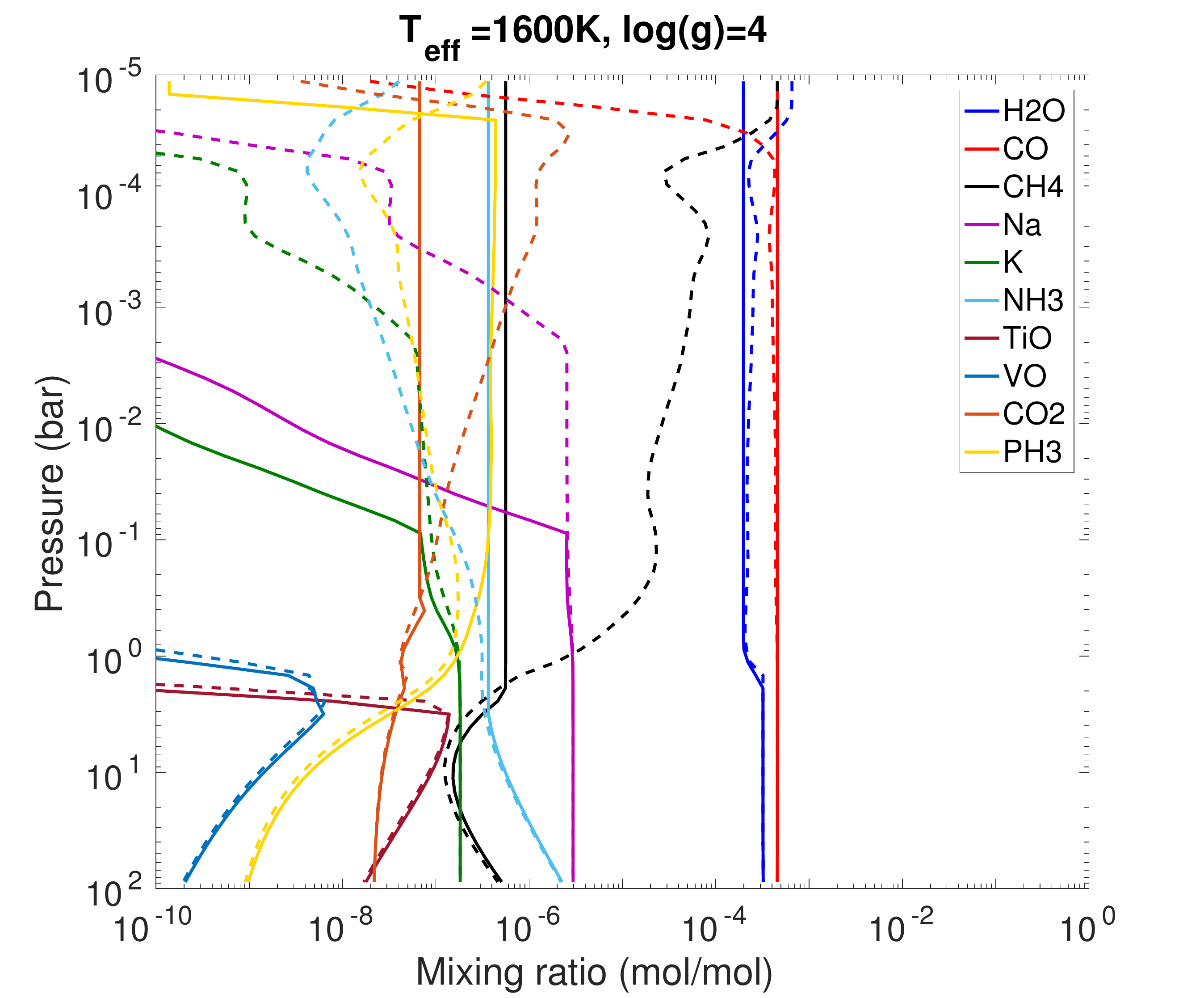}
	\includegraphics[width=6.5cm]{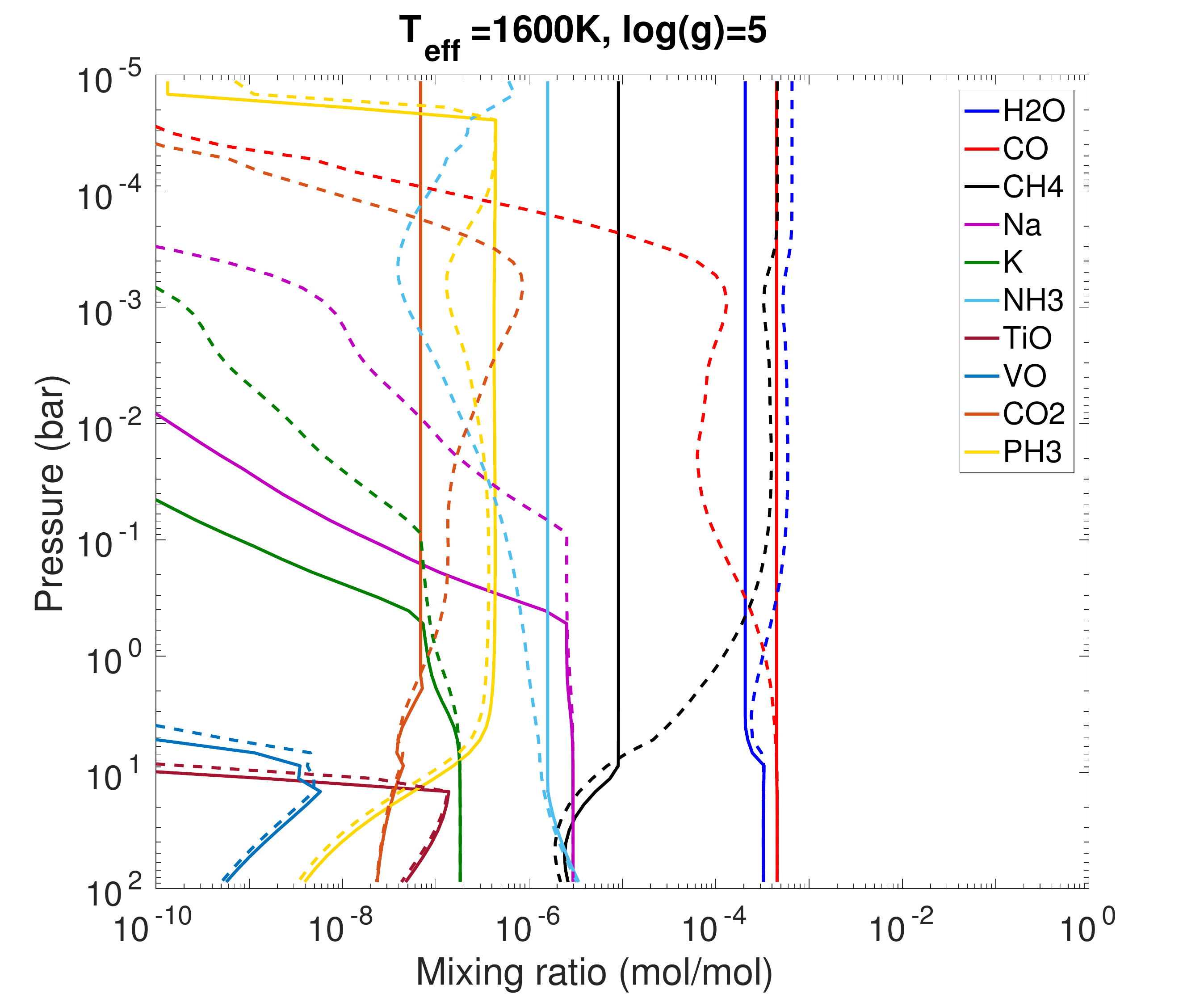}
\end{center}  
\caption{Atmospheric abundance profiles computed with Exo-REM. The panels show the mixing ratio (in mol/mol) of H$_2$O, CO, CH$_4$, Na, K, NH$_3$, TiO, VO, CO$_2$ and PH$_3$ assuming chemical equilibrium (dashed lines) or disequilibrium (solid lines). Left panels correspond to log(g)=4 and right panels  correspond to log(g)=5. Effective temperature is 700, 900, 1300 and 1600 K from top to bottom panels.
}
\label{figure_annexe1}
\end{figure*}

\begin{figure*}[h] 
\begin{center} 
	\includegraphics[width=6.5cm]{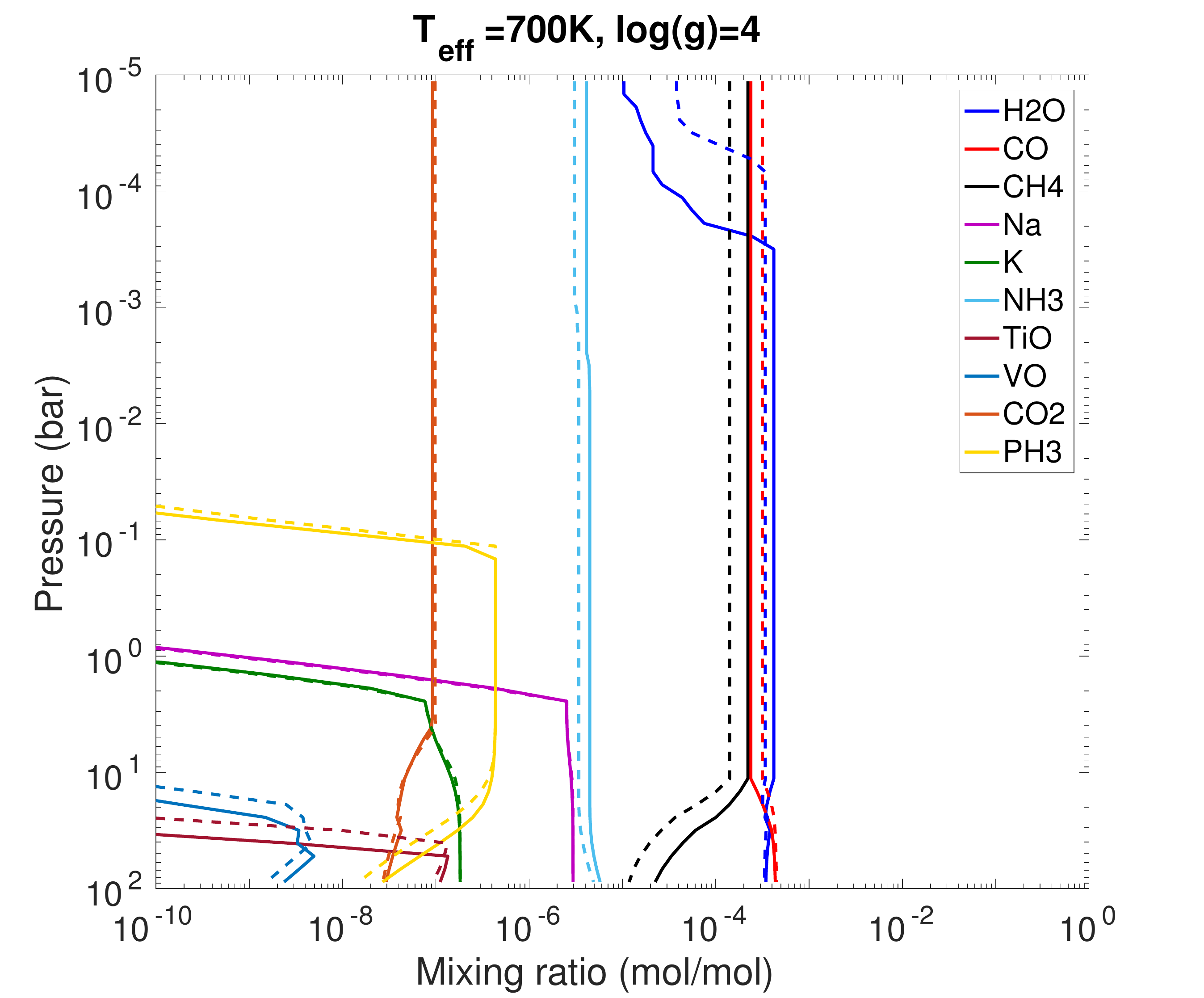}
	\includegraphics[width=6.5cm]{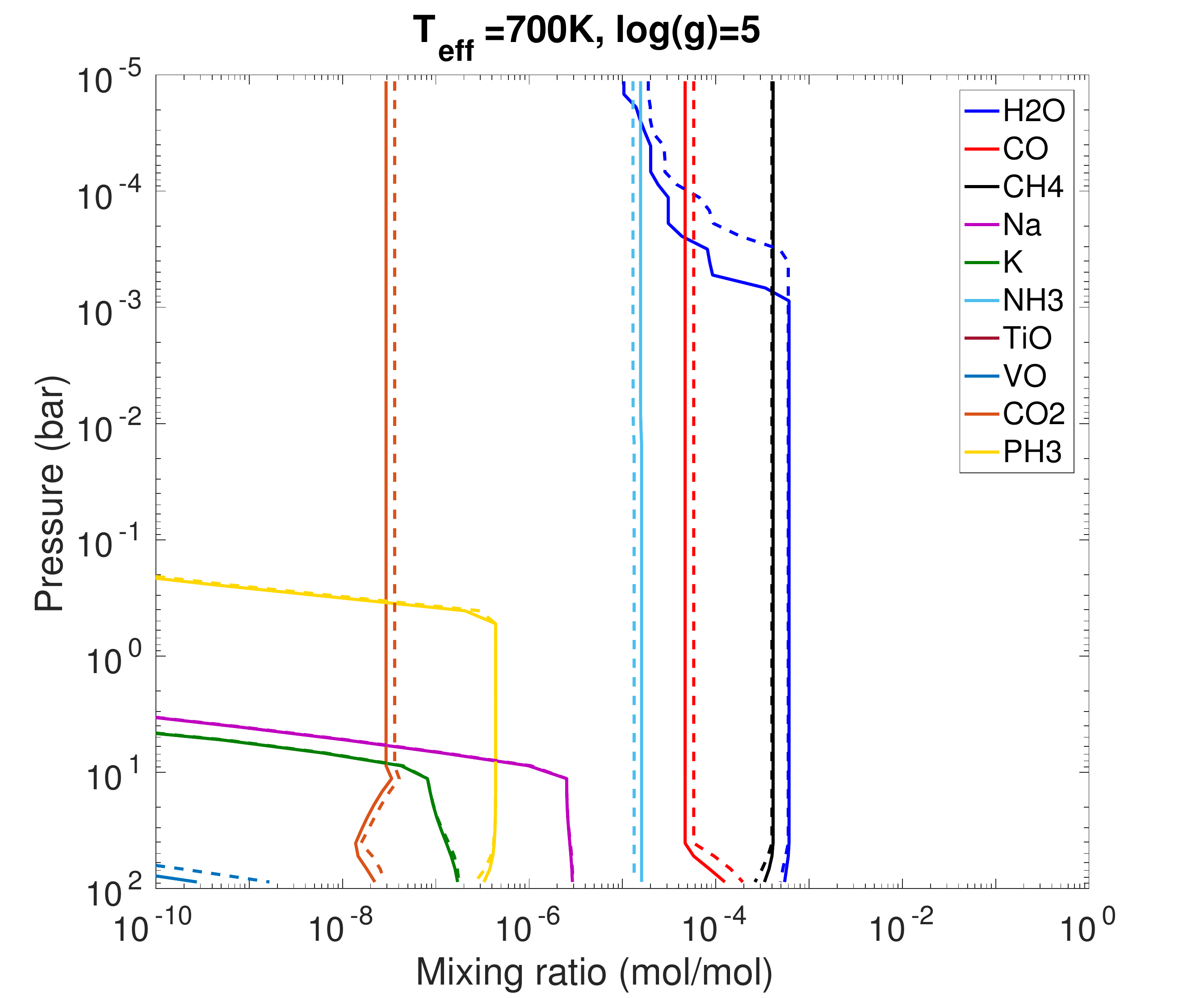}
	\includegraphics[width=6.5cm]{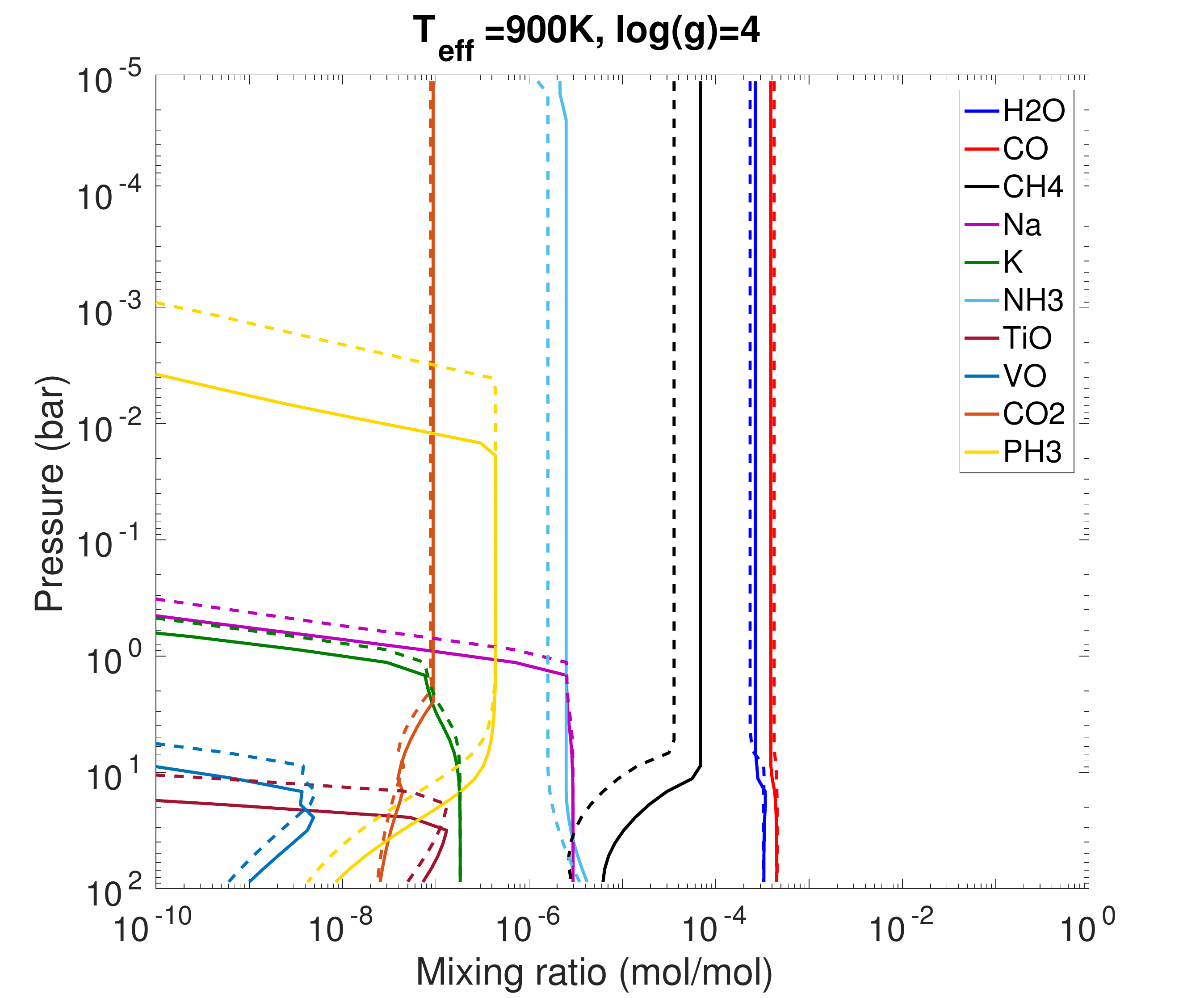}
	\includegraphics[width=6.5cm]{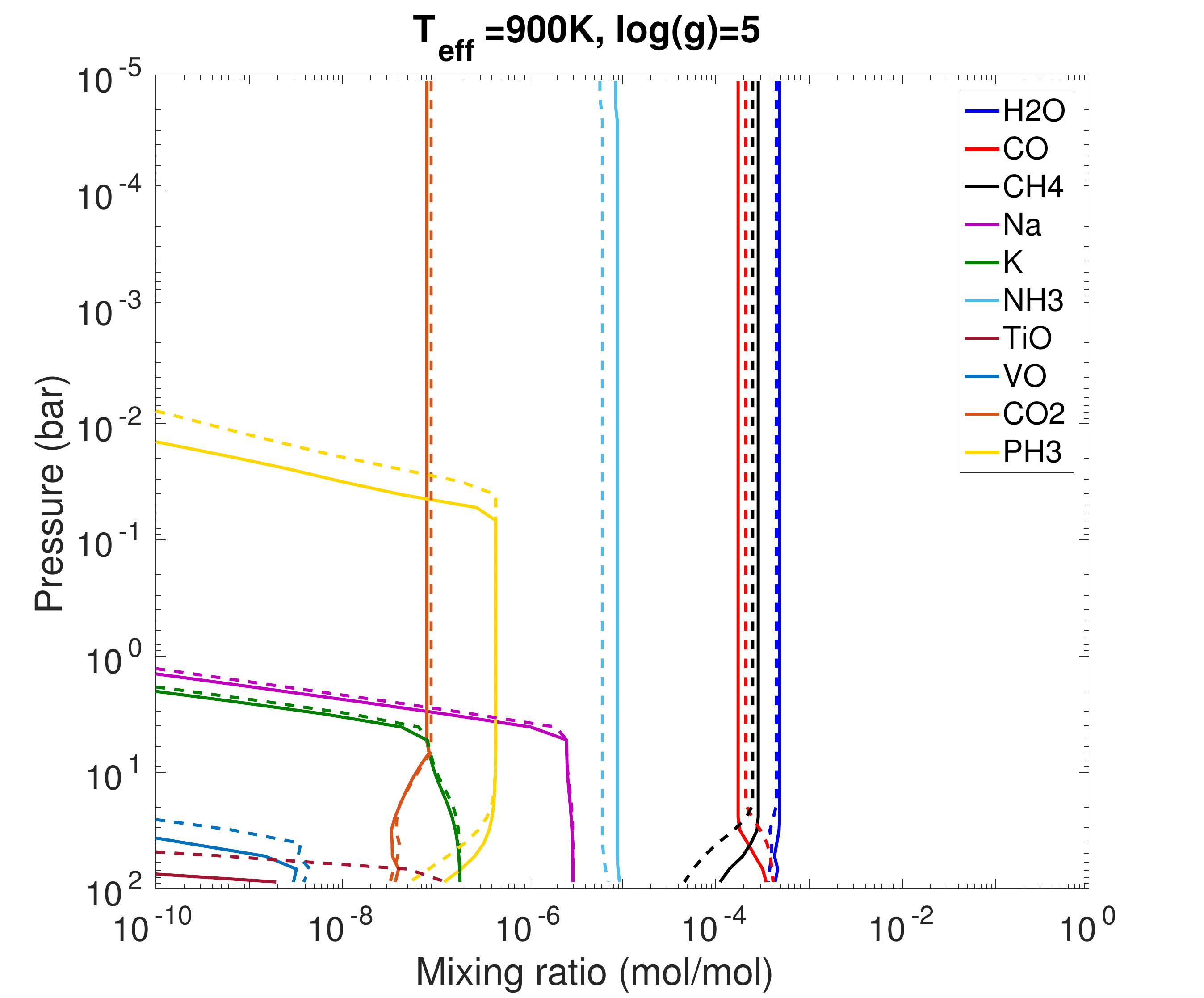}
	\includegraphics[width=6.5cm]{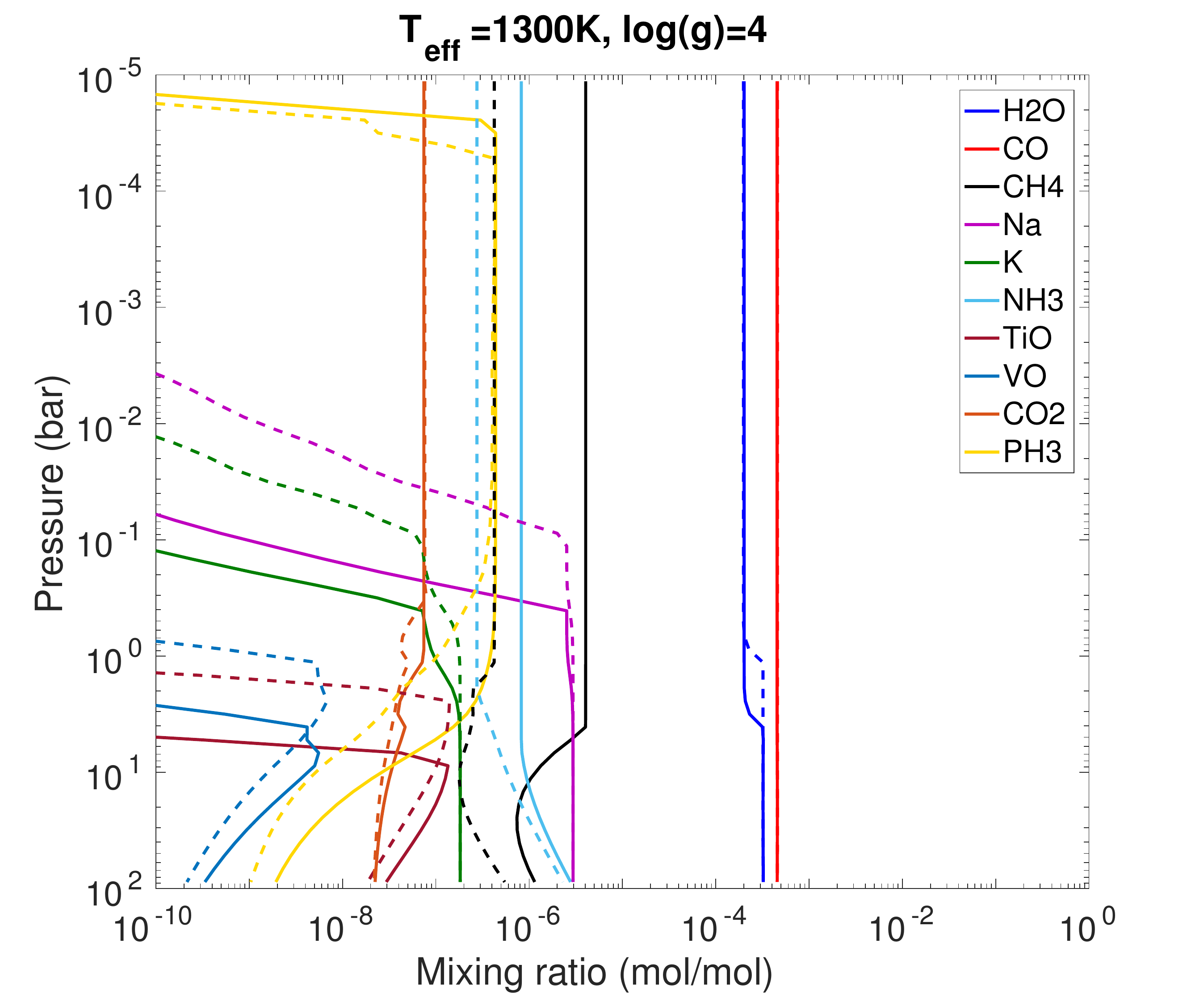}
	\includegraphics[width=6.5cm]{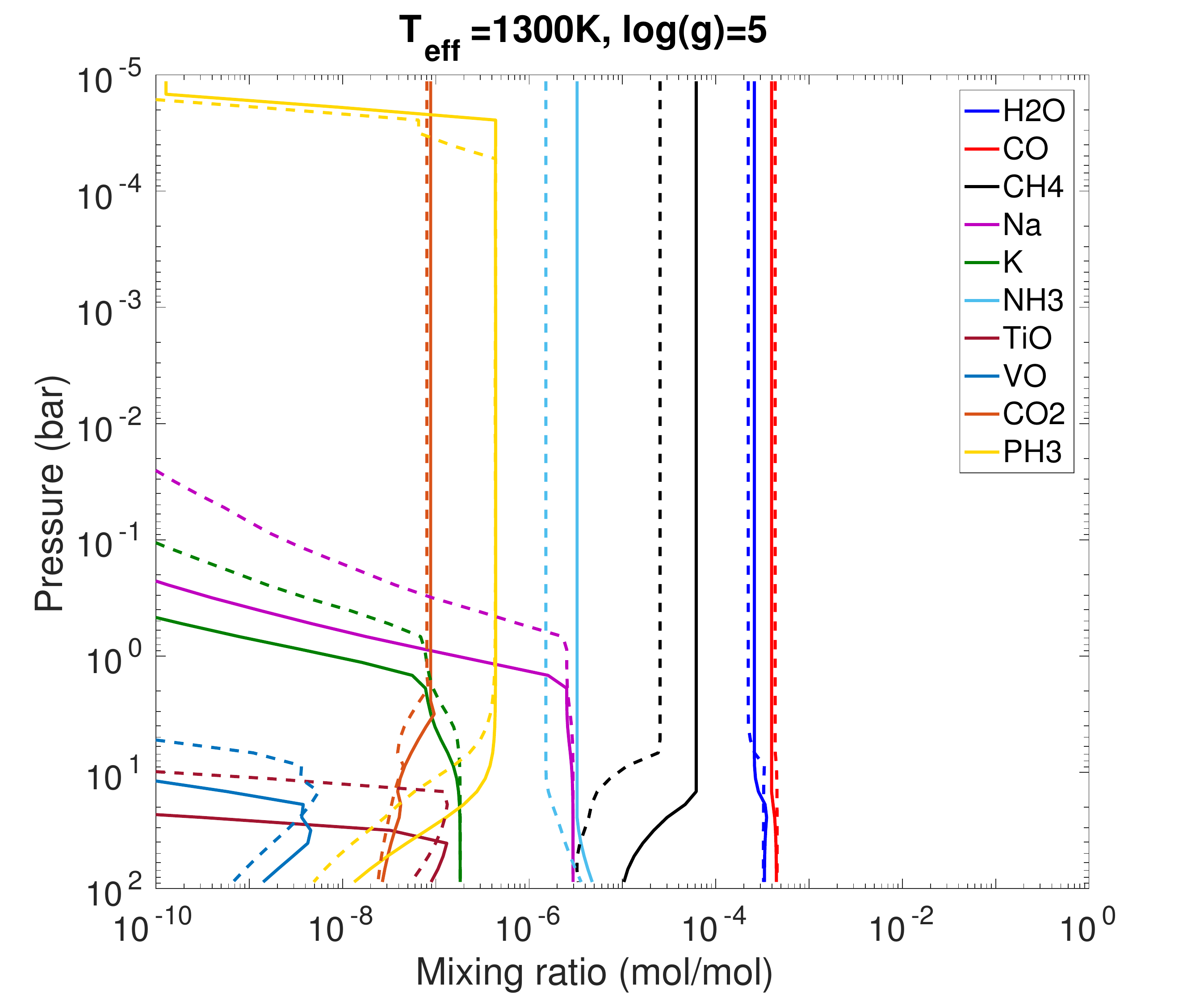}
	\includegraphics[width=6.5cm]{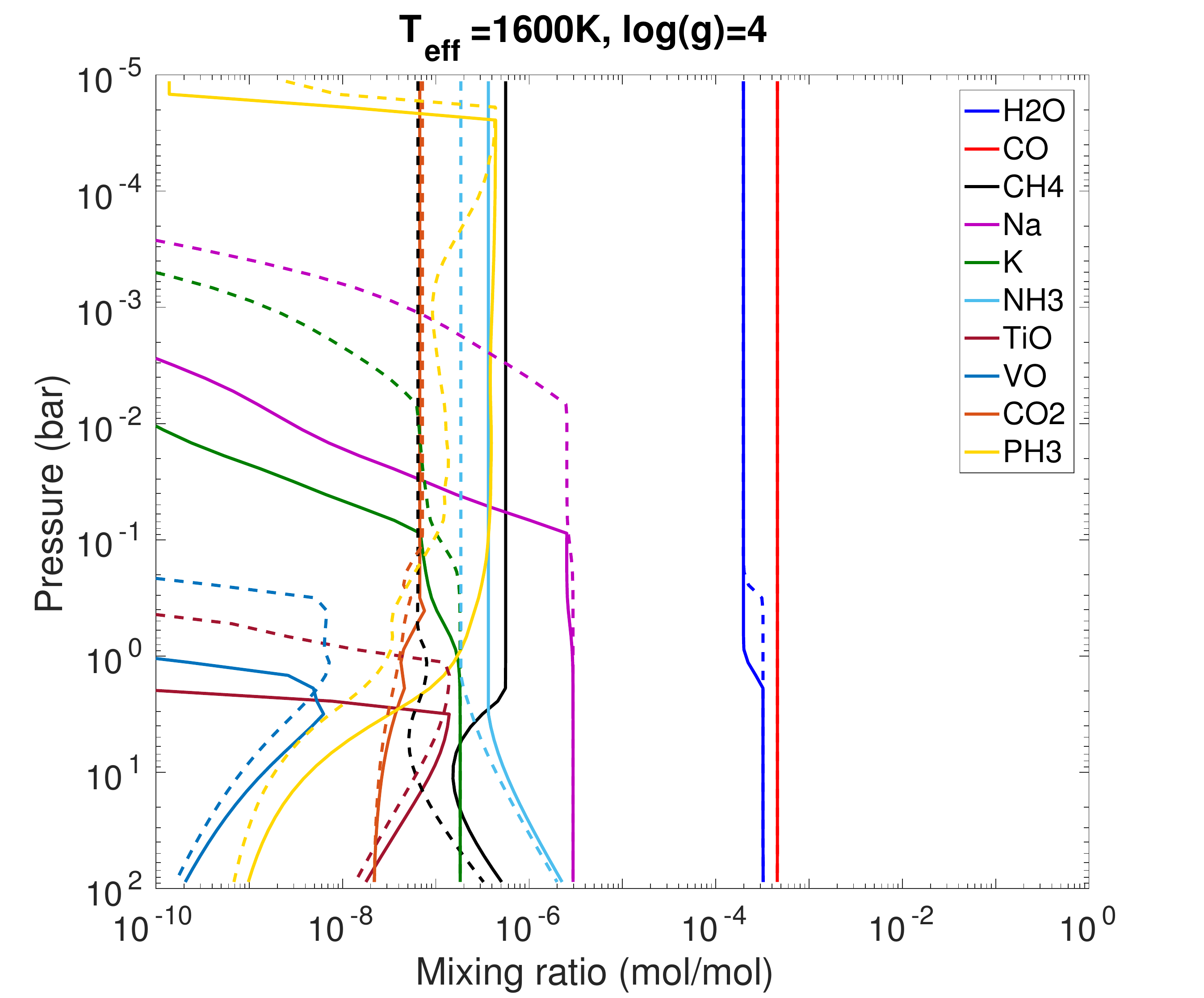}
	\includegraphics[width=6.5cm]{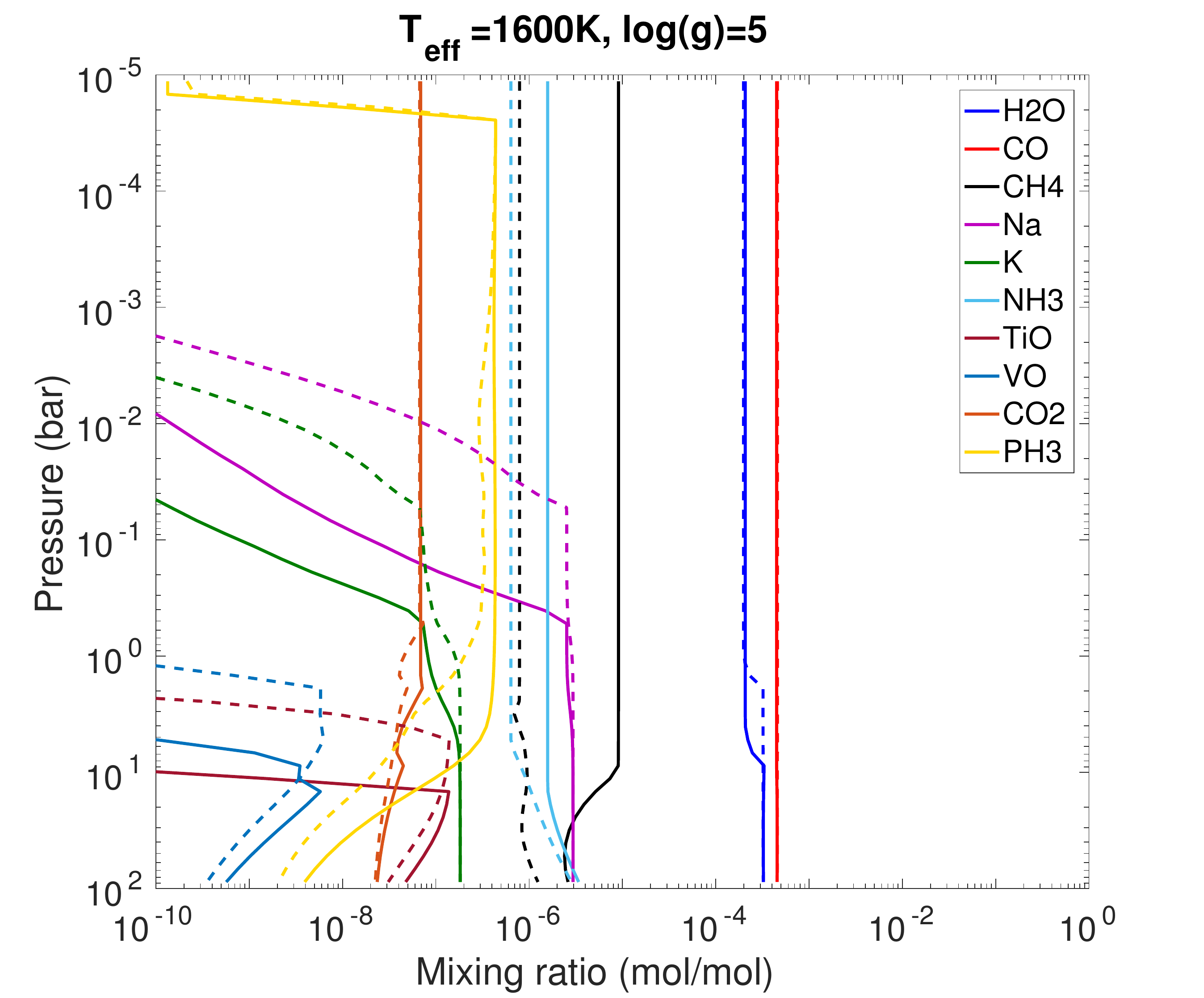}
\end{center}  
\caption{Same as Figure \ref{figure_annexe1} but dashed lines correspond to cloudy atmospheres  computed with simple microphysics (S=0.01) and solid lines correspond to cloud-free atmospheres. All cases were computed with chemical disequilibrium.
}
\label{figure_annexe2}
\end{figure*}

\clearpage



\end{document}